\documentclass[sn-nature]{sn-jnl}% Style for submissions to Nature Portfolio journals
\usepackage{upgreek}

\usepackage{graphicx}%
\usepackage{multirow}%
\usepackage{amsmath,amssymb,amsfonts}%
\usepackage{amsthm}%
\usepackage{mathrsfs}%
\usepackage[title]{appendix}%
\usepackage{xcolor}%
\usepackage{textcomp}%
\usepackage{manyfoot}%
\usepackage{booktabs}%
\usepackage{algorithm}%
\usepackage{algorithmicx}%
\usepackage{algpseudocode}%
\usepackage{listings}%
\usepackage{epstopdf}
\usepackage{subfiles}

\def\deg{\mbox{$^{\circ}$}}

\theoremstyle{thmstyleone}%
\theoremstyle{thmstyletwo}%

\theoremstyle{thmstylethree}%

\begin{document}

\title[Article Title]{Ground Calibration Result of the Wide-field X-ray Telescope (WXT) onboard the {\em Einstein Probe}}

%%=============================================================%%
%% Prefix	-> \pfx{Dr}
%% GivenName	-> \fnm{Joergen W.}
%% Particle	-> \spfx{van der} -> surname prefix
%% FamilyName	-> \sur{Ploeg}
%% Suffix	-> \sfx{IV}
%% NatureName	-> \tanm{Poet Laureate} -> Title after name
%% Degrees	-> \dgr{MSc, PhD}
%% \author*[1,2]{\pfx{Dr} \fnm{Joergen W.} \spfx{van der} \sur{Ploeg} \sfx{IV} \tanm{Poet Laureate} 
%%                 \dgr{MSc, PhD}}\email{iauthor@gmail.com}
%%=============================================================%%

\author[1]{\fnm{Huaqing} \sur{Cheng}}
%\email{hqcheng@nao.cas.cn}\equalcont{These authors contributed equally to this work.}

\author*[1,2]{\fnm{Chen} \sur{Zhang}}
\email{chzhang@nao.cas.cn}
%\equalcont{These authors contributed equally to this work.}
\author[1,2]{\fnm{Zhixing} \sur{Ling}}
\author[3]{\fnm{Xiaojin} \sur{Sun}}
\author[3]{\fnm{Shengli} \sur{Sun}}
\author[1]{\fnm{Yuan} \sur{Liu}}
\author[1]{\fnm{Yanfeng} \sur{Dai}}
\author[1]{\fnm{Zhenqing} \sur{Jia}}
\author[1]{\fnm{Haiwu} \sur{Pan}}
\author[1]{\fnm{Wenxin} \sur{Wang}}
\author[1]{\fnm{Donghua} \sur{Zhao}}
\author[3]{\fnm{Yifan} \sur{Chen}}
\author[3]{\fnm{Zhiwei} \sur{Cheng}}
\author[3]{\fnm{Wei} \sur{Fu}}
\author[3]{\fnm{Yixiao} \sur{Han}}
\author[3]{\fnm{Junfei} \sur{Li}}
\author[3]{\fnm{Zhengda} \sur{Li}}
\author[3]{\fnm{Xiaohao} \sur{Ma}}
\author[3]{\fnm{Yulong} \sur{Xue}}
\author[3]{\fnm{Ailiang} \sur{Yan}}
\author[3]{\fnm{Qiang} \sur{Zhang}}
\author[4]{\fnm{Yusa} \sur{Wang}}
\author[4]{\fnm{Xiongtao} \sur{Yang}}
\author[4]{\fnm{Zijian} \sur{Zhao}}
\author[5]{\fnm{Longhui} \sur{Li}}
\author[5]{\fnm{Ge} \sur{Jin}}
\author[1,2]{\fnm{Weimin} \sur{Yuan}}

%\equalcont{These authors contributed equally to this work.}

\affil[1]{\orgdiv{Key Laboratory of Space Astronomy and Technology}, \orgname{National Astronomical Observatories, Chinese Academy of Sciences}, \orgaddress{\street{Datun Road 20A}, \city{Beijing}, \postcode{100101}, \country{China}}}

\affil[2]{\orgdiv{School of Astronomy and Space Science}, \orgname{University of Chinese Academy of Sciences}, \orgaddress{\street{Yuquan Road 19A}, \city{Beijing}, \postcode{100049}, \country{China}}}

\affil[3]{\orgdiv{Shanghai Institute of Technical Physics}, \orgname{Chinese Academy of Sciences}, \orgaddress{\street{Yutian Road 500}, \city{Shanghai}, \postcode{200083}, \country{China}}}

\affil[4]{\orgdiv{Institute of High Energy Physics}, \orgname{Chinese Academy of Sciences}, \orgaddress{\street{Yuquan Road 19B}, \city{Beijing}, \postcode{100049}, \country{China}}}

\affil[5]{\orgdiv{North Night Vision Technology Co., LTD},
\city{Nanjing}, \postcode{211106}, \country{China}}
%%==================================%%
%% sample for unstructured abstract %%
%%==================================%%

\abstract{We report on results of the on-ground X-ray calibration of the Wide-field X-ray Telescope (WXT), built from novel lobster-eye micro-pore optics, on board the \textit{Einstein Probe (EP)} satellite. 
To fully characterize the instrumental performance and properties of the WXT instrument, a series of tests and calibrations have been carried out at different levels of devices, assemblies and the complete module before the launch of \textit{EP}. 
In this paper, we present the calibration results of three flight model modules (FM1, FM5 and FM11) obtained during their end-to-end module calibration experiments carried out at the 100-m X-ray Test Facility (100XF) of the Institute of High Energy Physics (IHEP), Chinese Academy of Sciences (CAS). 
Measurements of the Point Spread Function (PSF), effective area, and energy response were performed for multiple incident directions and several characteristic X-ray emission line energies.
Specifically, the distributions of the PSF and effective areas are found to be roughly uniform across the FoV, in large agreement with the prediction of lobster-eye optics.
Their energy dependence behavior aligns well with theoretical predictions and Monte Carlo simulations.
At $\sim1.25$ keV, the full width at half maximum (FWHM) of the focal spot is in range of $3-7$ arcmin (with a median of $\sim4.2$ arcmin) and the effective area in range of $\sim2-3~{\rm cm^2}$.
Noticeably, the flight model instruments demonstrate a $\sim1.5$ arcmin spatial resolution improvement over the previously launched Lobster Eye Imager for Astronomy.
The properties of the complementary metal-oxide semiconductor (CMOS) sensors were also calibrated. The gain coefficients are in range of $6.4-6.9~{\rm eV/DN}$. 
The energy resolutions are in range of $\sim120$ to $140$ eV at $1.25$ keV, meeting design requirements.
These calibration results have been ingested into the first version of calibration database (CALDB) and applied to the analysis of the scientific data acquired by WXT after the launch of \textit{EP}.
}


\maketitle

\section{Introduction}\label{sec:intro}

Lobster eye micro-pore optics (MPO), inspired by the structure of lobster compound eyes, represent an innovative X-ray focusing technique well-suited for wide-field imaging applications \cite{1979ApJ...233..364A, 1992SPIE.1546...41F, 1998ExA.....8..281W, 2002SPIE.4497..115F, 2016SPIE.9905E..1YW}. 
The Wide-field X-ray Telescope (WXT) is one of the two main scientific payloads on board the \textit{Einstein Probe} satellite (EP) \citep{Yuan2015, Yuan2018, Yuan2022, Yuan2025}, a time-domain X-ray mission of the Chinese Academy of Sciences (CAS), in collaboration with the European Space Agency (ESA), the Max-Planck Institute for Extraterrestrial Physics (MPE) and the France Space Agency (CNES).
It is featured with the novel lobster-eye micro-pore optics, and designed to monitor the soft X-ray (0.5--4 keV) sky at high cadence (a coverage of the night sky every $\sim5$ hours), with a high sensitivity of $\sim 1$ mCrab ($\sim (2-3) \times 10^{-11}~{\rm erg~s^{-1}~cm^{-2}}$ in 0.5--4 keV) at an exposure of $\sim1000~{\rm s}$. 
This is several-tens times more sensitive than the other currently operating wide-field monitors.

The \textit{Einstein Probe} was launched on January 9, 2024.
The X-ray mission successfully passed the review of its in-orbit commissioning in July 2024 and went into the nominal scientific operational phase thereafter. \textit{EP} operates in a circular orbit of 593 km with an orbital period of 96.5 min and an inclination of 29\deg. 
Its first-light results have been released in April 2024, soon after its launch\footnote{\url{https://ep.bao.ac.cn/ep/cms/article/view?id=166}, \url{https://www.mpe.mpg.de/8007188/news20240427}}.
During the first year of operation, \textit{EP} has achieved a series of significant scientific discoveries across multiple high-energy astrophysical fields, including extragalactic fast X-ray transients \citep{Yin2024, LiuY2025, 2024HSun, Jiang2025}, tidal disruption events (TDEs) \citep{2025Jin}, X-ray binaries \citep{Marino2025, 2025YangBeXRB}, Active Galactic Nuclei (AGNs) \citep{LiuMJ_AGN} and other X-ray transient phenomena with unclear nature \citep{ZhangWD2025}.

\begin{figure}[!htbp]
    \centering
    \includegraphics[width=0.7\linewidth]{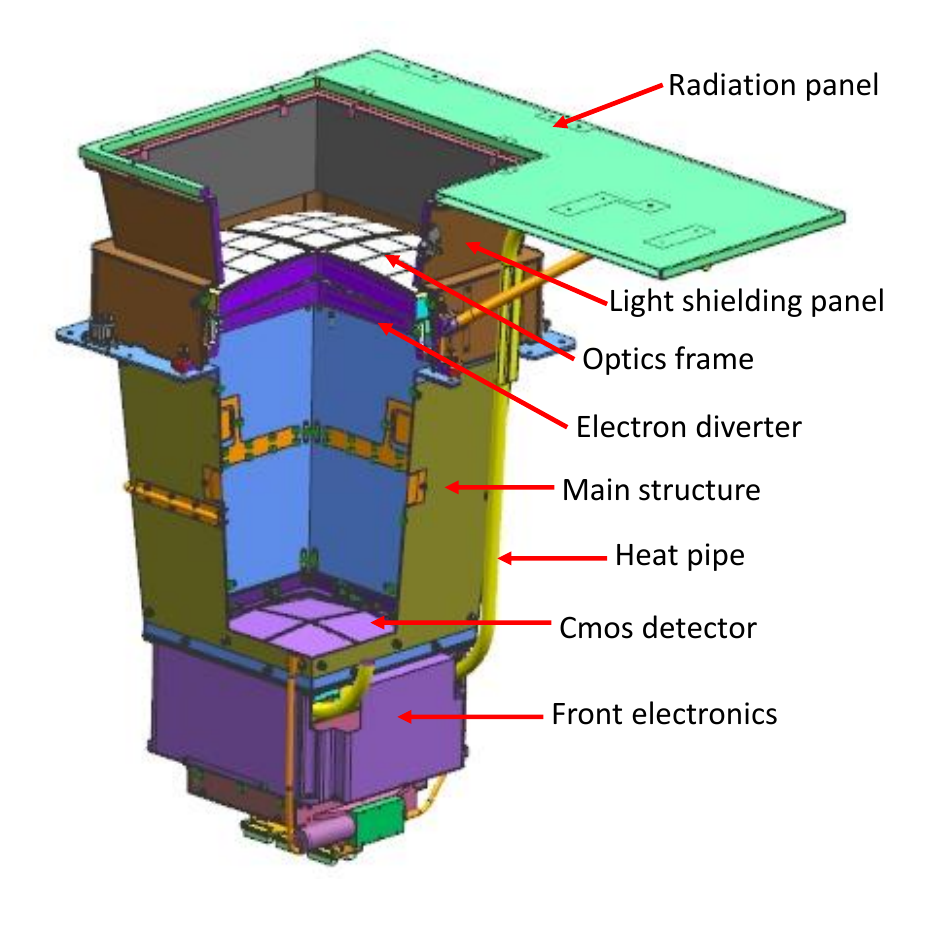}
    \caption{Schematic diagram showing the design layout of a single module of WXT instrument (cited from Figure 7 of \cite{Yuan2022}).}
    \label{fig:wxt_design}
\end{figure}

The WXT instrument onboard EP consists of twelve identical flight model (FM) modules,
with the optical design and layout of one representative module illustrated in 
Fig. \ref{fig:wxt_design}.
Each WXT module comprises primarily two key components: a lobster-eye focusing mirror assembly (MA) and a focal-sphere detector module. 
The MA is built from 36 micro-pore optics plates, forming the optical system, while the detector module, located at the bottom of the module, consists of four large-format, back-illuminated scientific CMOS sensors \citep{2022WangWXa,2022WangWXb,Wu2022,Wu2023a,Wu2023b,LiuMJ2023,ChenMX2024,LiuMJ2025} and an associated front-end electronics box. 
Since the focal surface of the lobster-eye optics is curved, the CMOS sensors are carefully tilted relative to one another to align optimally with this geometry \cite{2022ChenYifan}.
Each WXT module features a large field of view (FoV) of $18.6\deg\times18.6\deg$, leading to a total FoV of over 3600 square degrees subtending around $1/11$ of the whole sky.
Each CMOS sensor, which subtends one quadrant of a single module's FoV, provides a resolution of 4096 × 4096 pixels with a pixel size of 15 micro × 15 micro. To efficiently prevent contamination from optical and ultraviolet light, the sensors are coated with a 200 nanometer aluminum layer.
Detailed specifications of the WXT instrument are listed in Table 1 of \cite{Yuan2025}. 

Before the launch of \textit{EP}, extensive experiments were carried out to detailedly characterize the instrumental performance and calibrate the properties of each of the twelve WXT modules.
These tests were carried out on different levels: device, assembly and the complete instrument. 
Firstly, before mounting onto the MA, each of the MPO plates was individually measured in order to derive its parameters and to assess its optical performance.
Then the MA was calibrated for their focal lengths, PSFs and effective areas at the X-ray Imaging Beamline (XIB) at NAOC \cite{2012SPIE.8443E..3XZ}. 
These calibration results will be presented in a forthcoming paper (Zhang et al. in preparation). 
The MA aboard the fifth FM module (FM5) was also independently calibrated for its focal length, PSF and effective area at the Panter X-ray Test Facility of MPE, Germany\cite{2005ExA....20..405F, 2019SPIE11119E..16B}, with the calibration results been reported \cite{EPWXT_Panter2023}.
In the meantime, the performances and properties of the 48 CMOS detectors on board \textit{EP} were tested separately at NAOC, whose results will be presented elsewhere (Ling et al. in preparation).
Then the MA, the detectors and the thermal control unit were integrated together to build each of the twelve complete modules at the Shanghai Institute of Technical Physics (SITP).
Due to the tight schedule of the Phase D, only three complete modules (FM1, FM5 and FM11) were calibrated at the 100-m X-ray Test Facility (100XF) of the Institute of High Energy Physics (IHEP) \cite{2023ExA....55..427W} to fully characterize their performances at final state. FM1 was calibrated in August 2022, while FM5 and FM11 in May 2023. 

It is important to note that, prior to EP-WXT, the calibration procedures and methods for such kind of wide-field lobster-eye instruments have been successfully verified on its pathfinder module, the Lobster Eye Imager for Astronomy (\textit{LEIA}, \cite{2023LingZXRAA, 2022ZhangChenApJL}). 
The calibration campaign of \textit{LEIA}, carried out in November 2021, provides a standard template for the design and optimization of the calibration experiments of the WXT instrument, regarding the calibration targets, focal plane mapping pattern, exposure time for each test point, etc.
The detailed results of the \textit{LEIA} calibration have been presented in our earlier paper \cite{Cheng2024}. 
In this paper, we present the results of two comprehensive end-to-end calibration campaigns carried out for the three complete WXT flight model modules (FM1, FM5 and FM11) at the 100XF, with the main focus on the characterization of the properties of the PSF, effective area, and energy response of the detectors. 
This paper is organized as follows. The basic information including the experimental setup, procedure and data processing are described in Section \ref{sec:overview}. The calibration results are detailedly presented in Section \ref{sec:calib_results} and summarized in Section \ref{sec:summary}.

\section{The calibration campaigns}
\label{sec:overview}

The calibration of WXT flight model modules largely follows the approach used for the \textit{LEIA} instrument, including the test facility, experimental setup, calibration procedures, and data processing methods. For a detailed explanation one may refer to Section 2 of \cite{Cheng2024}. Here we focus on briefly summarizing the key experimental settings.

\begin{figure*}[!htbp]
    \centering    
    \includegraphics[width=0.5\textwidth]{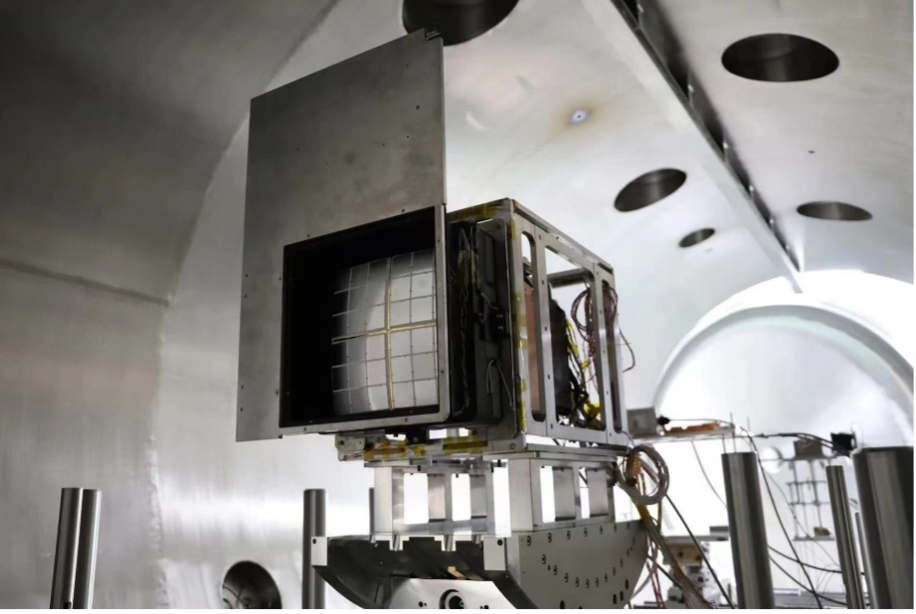}
    \includegraphics[width=0.5\textwidth]{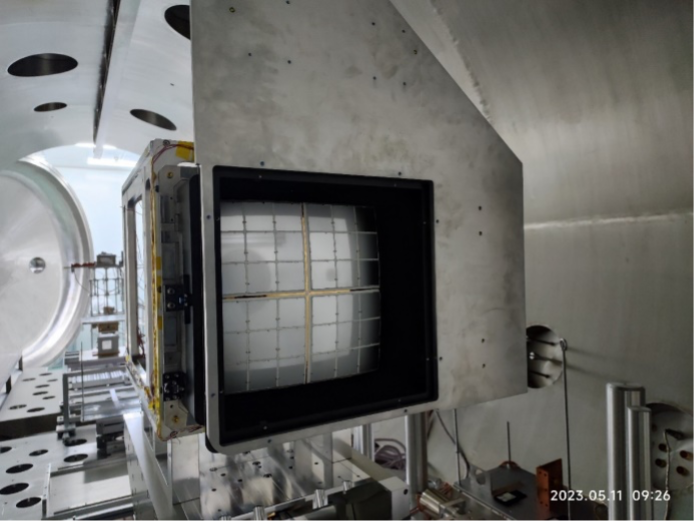}
    \includegraphics[width=0.5\textwidth]{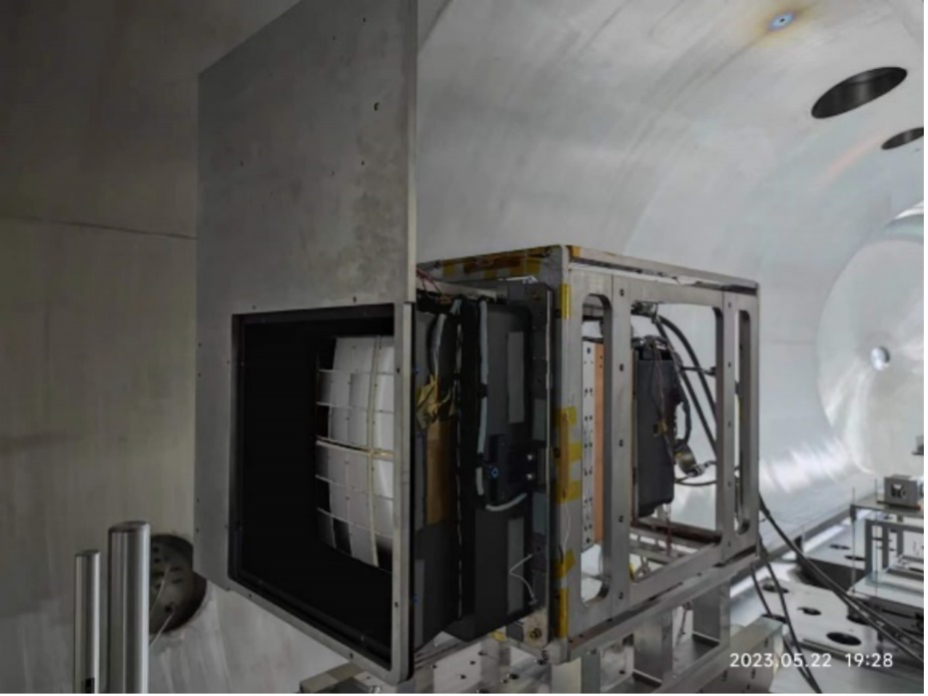}
    \caption{The three flight model modules FM1, FM5 and FM11 (from upper to bottom) placed in the large instrument chamber of the 100XF at IHEP.}
    \label{fig:fm1_5_11_at_ihep}
\end{figure*}

The calibration for the three complete modules, including FM1, FM5 and FM11, were carried out at the 100-m X-ray Test Facility (100XF), a dedicated facility built by IHEP, CAS, to calibrate X-ray telescopes (a detailed introduction to 100XF is presented in \cite{2023ExA....55..427W}). % 除气过程
The calibration for FM1 was carried out from August 20th to 26th, 2022. 
For the other two modules FM5 and FM11 the calibrations were conducted in sequence from May 13th to 26th, 2023. 
For each module, a degassing procedure aimed to eliminate the contaminants attached to the surface of the instrument, was performed as the first step, during when the instrument was placed in a small vacuum tank of the 100XF for $\sim3$ days and vacuumed to a degree of $\sim10^{-3}~{\rm Pa}$. 
Then the instrument was transferred to the large chamber (see Fig. \ref{fig:fm1_5_11_at_ihep}) and vacuumed to a higher degree of $\sim10^{-5}~{\rm Pa}$. 
The X-ray source locating at a distance of $\sim100$ meters from the large chamber generates a quasi-parallel X-ray beam with a size of 0.6 meters in diameter, which has a divergence angle of $<9$ arcmin when reaching the exit end of the chamber\cite{2023ExA....55..427W}. 

Photons with different energies of several characteristic X-ray emission lines were generated by the multi-target electron impact X-ray source using various targets, including magnesium (Mg), titanium (Ti), silicon dioxide (SiO$_2$), silicon (Si), silver (Ag) and copper (Cu). The X-ray emission lines in the subsequent analysis include the O K$\alpha$ (525 eV), Cu L (930 eV), Mg K$\alpha$ (1254 eV), Si K$\alpha$ (1740 eV), Ag L$\alpha$ (2980 eV), Ti K$\alpha$ (4511 eV), and Ti K$\beta$ (4932 eV). The absolute count rate of the X-ray beam is real-timely monitored by a standard silicon drift detector (SDD) of which the quantum efficiency (QE) has been well calibrated by a PNCCD. Due to the finite distance of the point source, the WXT module was $1.4$ millimeter out of focus during the whole experiment. The systematics introduced by the variations in the X-ray beam strength and de-focus effect are negligible (a detailed discussion can be found in Section 4.2 of \cite{Cheng2024}).

\begin{figure*}[!htbp]
    \centering
    \includegraphics[width=0.7\textwidth]{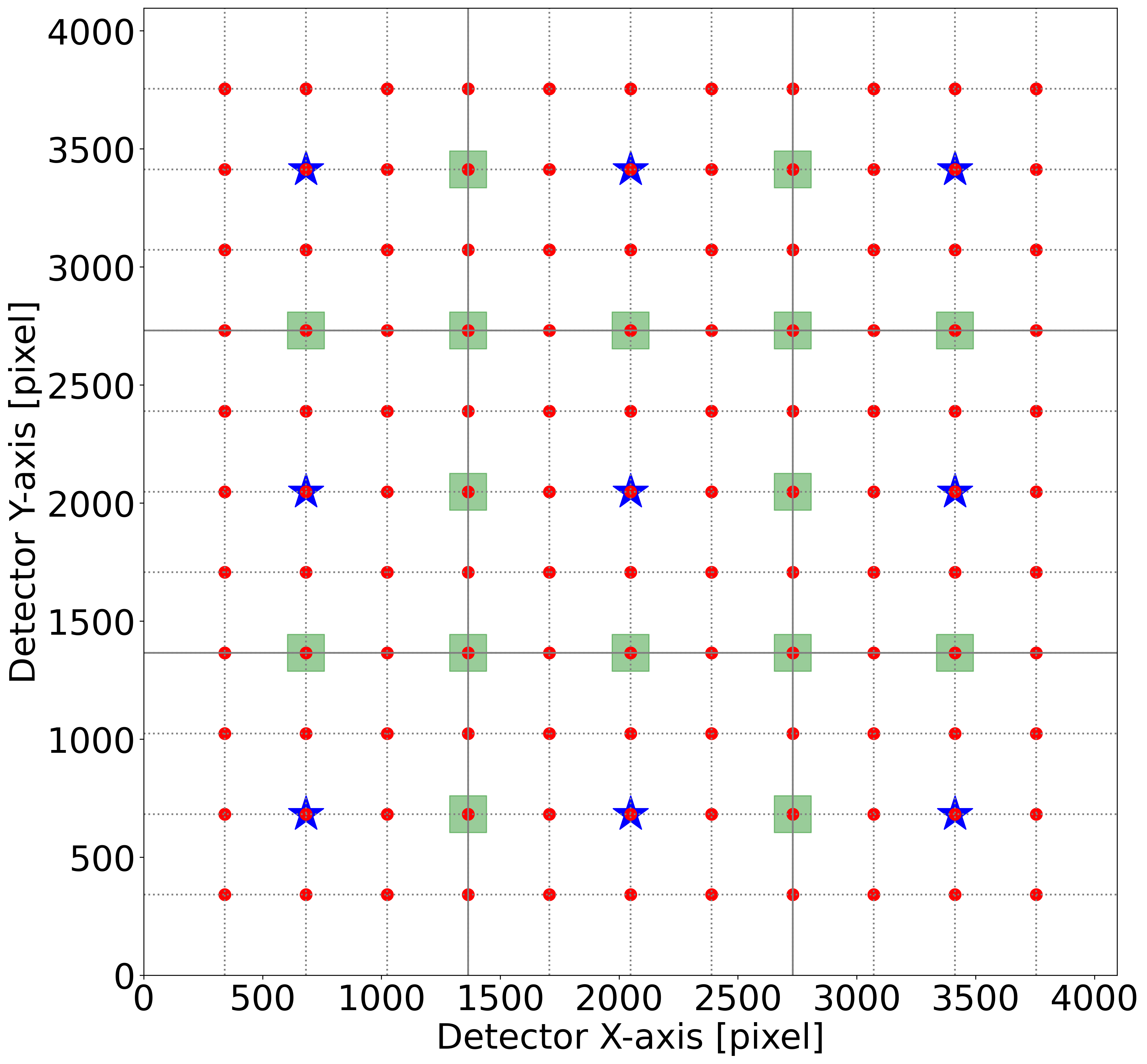}
    \caption{The diagram of the scan array for the mapping of the PSF and effective area across the FoV of a specific CMOS detector, including the `fine scan' array ($11\times11$ test points, red dots) adopted in the experiment using the Mg source, the `chip center scan' array ($3\times3$ test points, blue pentagrams) adopted in the experiment of SiO$_2$, Ti and Ag sources, and the`chip edge scan' array (16 test points, green rectangles) adopted in the thermal test of FM1 only. See text for more details.}
    \label{fig:scanarray}
\end{figure*}

To investigate the imaging quality (PSF) and the effective area across the FoV, the module was mounted on a movable stage in the large chamber (see Fig. 3 of \cite{Cheng2024}). Different combinations of the pitch and yaw angles correspond to different incident angles of the X-ray beam. 
The schematic diagram for the sampling array on one of the four CMOS detectors aboard one module (corresponding to a quadrant of the entire FoV of the module) is shown in Fig. \ref{fig:scanarray}.
Specifically, we define a grid consisting of 11 $\times$ 11 points corresponding to the theoretical positions of the central focal spots from 121 incident directions, termed `fine scan' array (the red dots). 
This array is utilized for the calibration at the energy of Mg K$\alpha$. 
We also define a sub-grid of 3$\times$3 points, termed `chip center' array (the blue pentagrams), and use it for the calibration at several other energies including O K$\alpha$, Si K$\alpha$, Ag L$\alpha$ and Ti K$\alpha$. 
For the calibration at the energy of Cu L, only the direction passing through the CMOS chip center is sampled.
A supplementary 16-point `chip edge' array (green rectangles) is further defined at boundary locations between adjacent MPO plates. 
By combining the `chip edge' array with the `chip center' array, we generate a hybrid $5\times5$ grid and adopt it in the FM1's thermal test.

Both FM1 and FM5 underwent a comprehensive calibration at several photon energies, while FM11's calibration was restricted to the energy of Mg K$\alpha$ due to schedule constraints.
The log of the calibration experiment for these three modules is summarized in Table \ref{tab:epwxt_schedule}. 
During the calibration campaign, dark images were taken as a daily basis to continuously monitor the status of the CMOS detectors. A small number of bad pixels were identified during the experiment, which were then disabled onboard and documented in the calibration database. 
X-ray photon events were extracted from the CMOS detectors (see \cite{Wu2022} for more details of event extraction) using an adjustable low-energy threshold ($\sim350$ eV in nominal working mode). Each event was assigned by a pulse-height amplitude (PHA) and a grade of the event split pattern. Single-, double-, triple-, and quadruple- events were screened to generate scientific products (images, spectra, light curves) for subsequent analysis.

\begin{table}[htbp]
\caption{Log of the calibration experiment.}
\begin{tabular}{ccccccc}
\toprule%
\centering
FM & CMOS ID & Date & Target & Line energies & Points / CMOS & Exposure (s) \\ \hline
1 & 1-4 & 2022.8.20 & Cu & 930 eV & $1$ & $1800$ \\
 & & 2022.8.22 & SiO2 & 525 eV, 1740 eV & $3\times3$ & 600 \\
 & & 2022.8.22-8.23 & Ag & 2980 eV & $3\times3$ & 500 \\
 & & 2022.8.23 & Ti & 4511 eV, 4932 eV & $3\times3$ & 500 \\
 & & 2022.8.24-8.25 & Mg & 1254 eV &$11\times11$ & 80 \\
 & & 2022.8.26 & Mg$^\dagger$ & 1254 eV &$5\times5^\star$ & 80 \\ 
5 & 17-20 & 2023.5.13 & Cu & 930 eV & $1$ & $1200$ \\
  & & 2023.5.15 & Si & 1740 eV & $3\times3$ & $600$ \\
  & & 2023.5.15-5.16 & Ag & 2980 eV & $3\times3$ & $500$ \\
  & & 2023.5.16 & Ti & 4511 eV, 4932 eV & $3\times3$ & $500$ \\
  & & 2023.5.17-5.18 & Mg & 1254 eV &$11\times11$ & 80 \\
  & & 2023.5.20 & SiO2 & 525 eV & $3\times3$ & $500$ \\
11 & 41-44 & 2023.5.24-5.26 & Mg & 1254 eV &$11\times11$ & 80 \\
\botrule
\end{tabular}
{{\rm Notes:} A summary of the end-to-end calibration procedures for the three FM modules, with details including the date, target, characteristic X-ray line energies, sampling points on each CMOS detector, and the exposure time for each test point.

$\dagger$: the thermal test with the MPO plates heating up from $\sim16\deg$C to $\sim28\deg$C.

$\star$: `chip center' array (9 points) + `chip edge' array (16 points).

}
\label{tab:epwxt_schedule}
\end{table}

\section{Calibration Results}
\label{sec:calib_results}

In this section, we present the calibration results of the properties including the PSF, effective area as well as their variations across the whole field of view (vignetting), and the energy scale and spectral resolution of the CMOS detectors. 
The result of the complementary thermal test carried out for FM1 is also presented.

\subsection{Point Spread Function}
\label{sec:psfscan}

The lobster eye optics show distinct PSF properties compared with traditional Wolter-I telescopes.
The PSF consists of a central focal spot surrounded by cruciform arm structures. Photons undergoing even numbers of reflections (predominantly two times) within orthogonal micro-pore channel walls concentrate to form the focal spot, while those experiencing odd reflections contribute to the cruciform arms.
For a given photon energy, in theory there is a good uniformity for different incident directions within the FoV showing slight variations in the shapes and dispersions of the PSF \citep{1979ApJ...233..364A}.
The homogeneity of the PSF thus represents an important indicator for the imaging quality. 
For each FoV quadrant of FM1 and FM5, the PSF was measured at a grid of $11\times11$ different incident angles at the energy of Mg K$\alpha$, a sub-grid of $3\times3$ different incident angles at the energies of O K$\alpha$, Ti K$\alpha$, Si K$\alpha$, Ag L and Ti K$\alpha$, and the central direction at the energy of Cu L.
For FM11, the PSF was tested at the energy of Mg K$\alpha$ line only.

\subsubsection{Dependence on the incident direction}
\label{sec:psf_dif_direction}

\begin{figure*}[!htbp]
\centering
\includegraphics[width=0.48\textwidth]{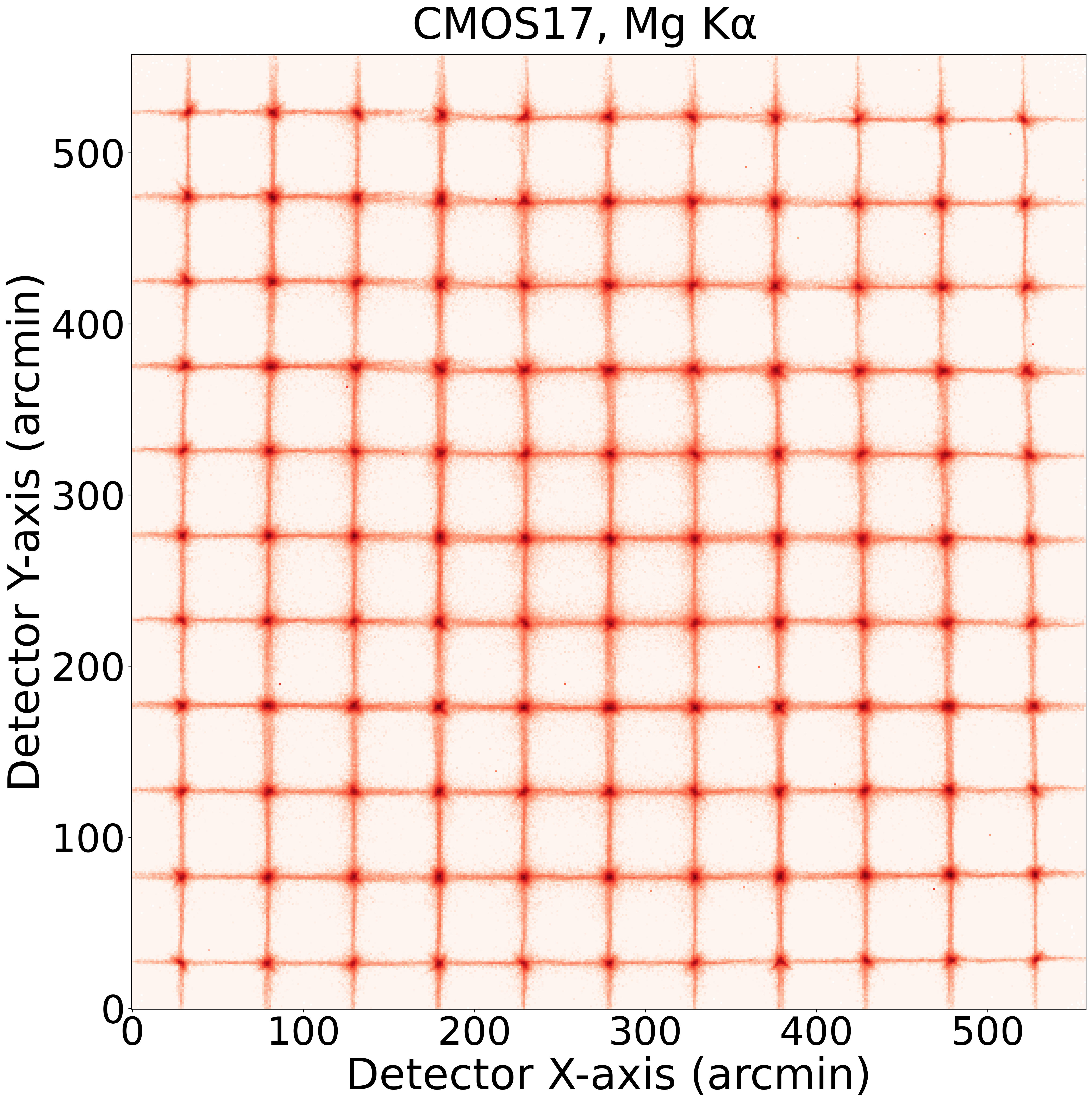}
\includegraphics[width=0.48\textwidth]{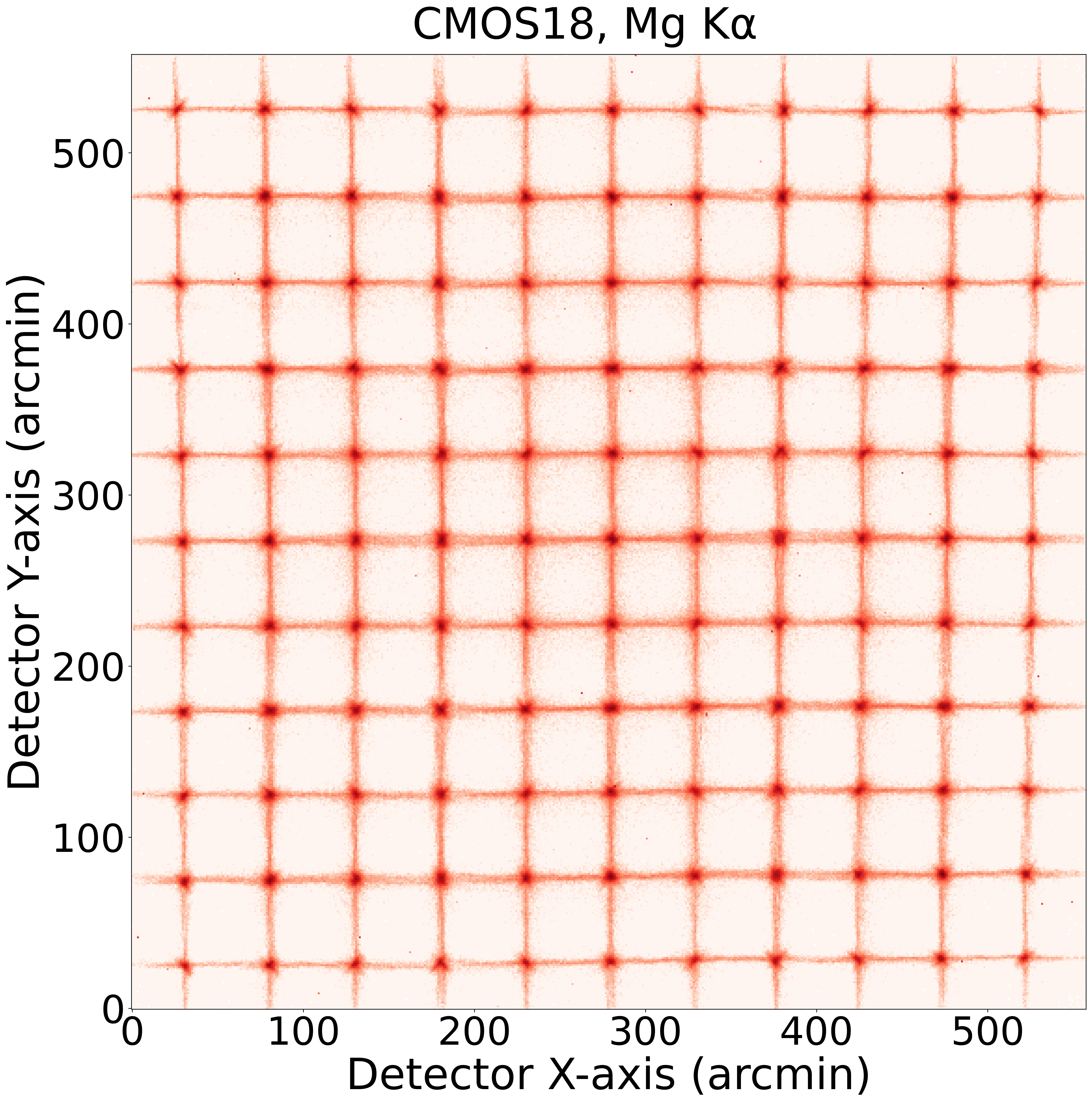}
\includegraphics[width=0.48\textwidth]{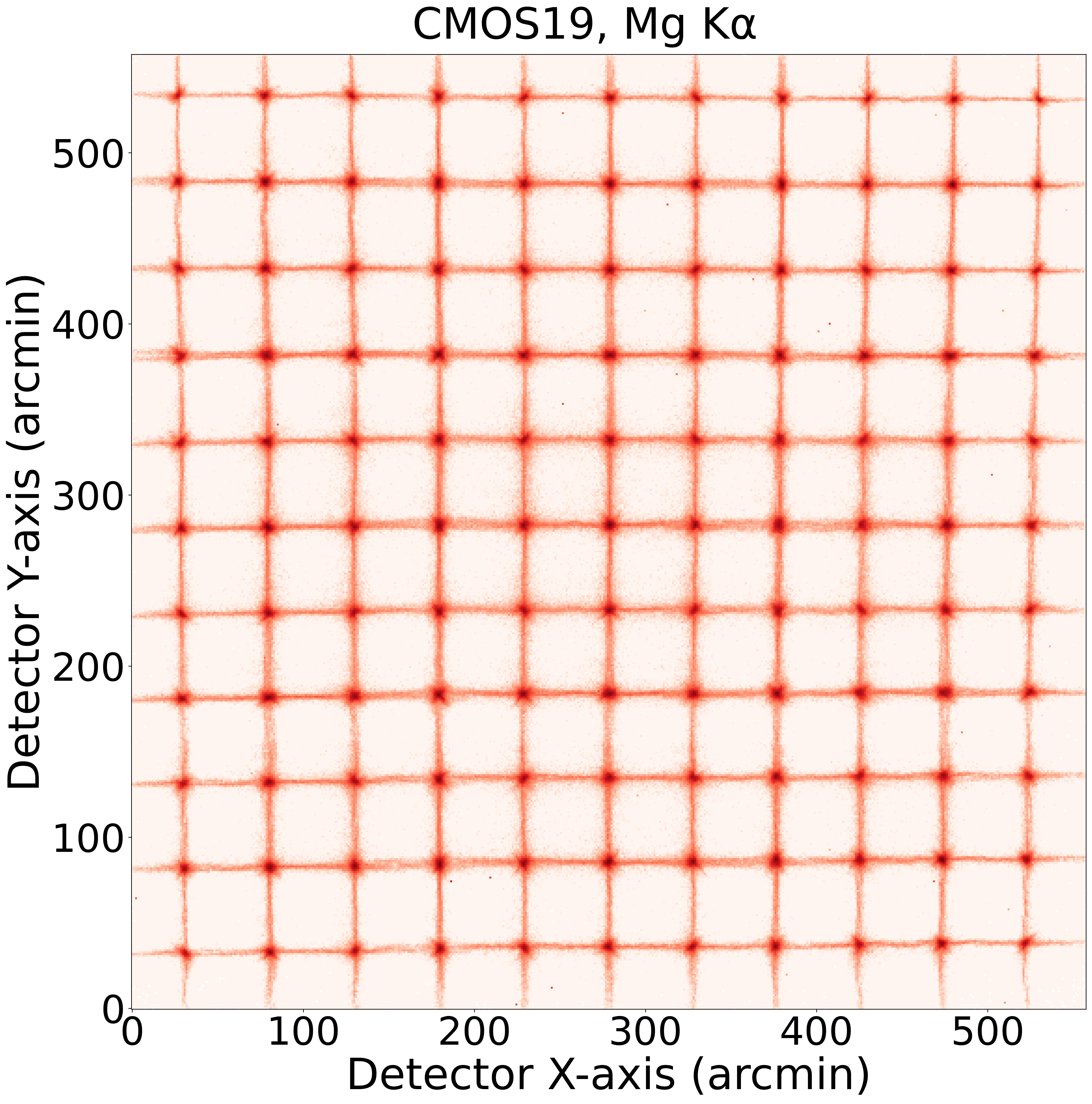}
\includegraphics[width=0.48\textwidth]{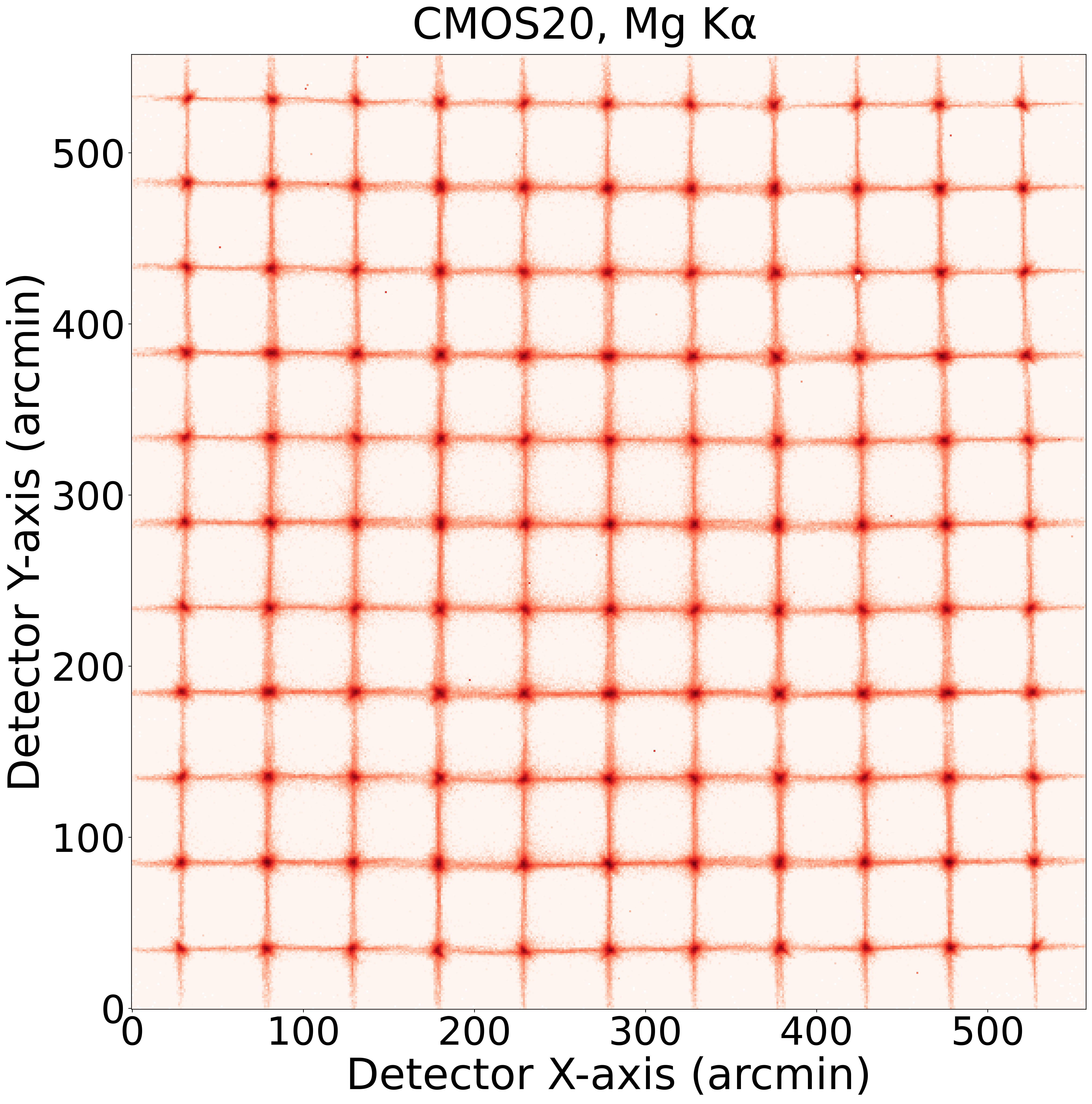}
\caption{\label{fig:psfscan_mgk}A mosaic of images of the PSF obtained at 1.25 keV (Mg K$\alpha$ line) across the entire FoV of FM5 (subtended by CMOS 17-20).}
\end{figure*}

\begin{figure*}[!htbp]
\centering
\includegraphics[width=0.48\textwidth]{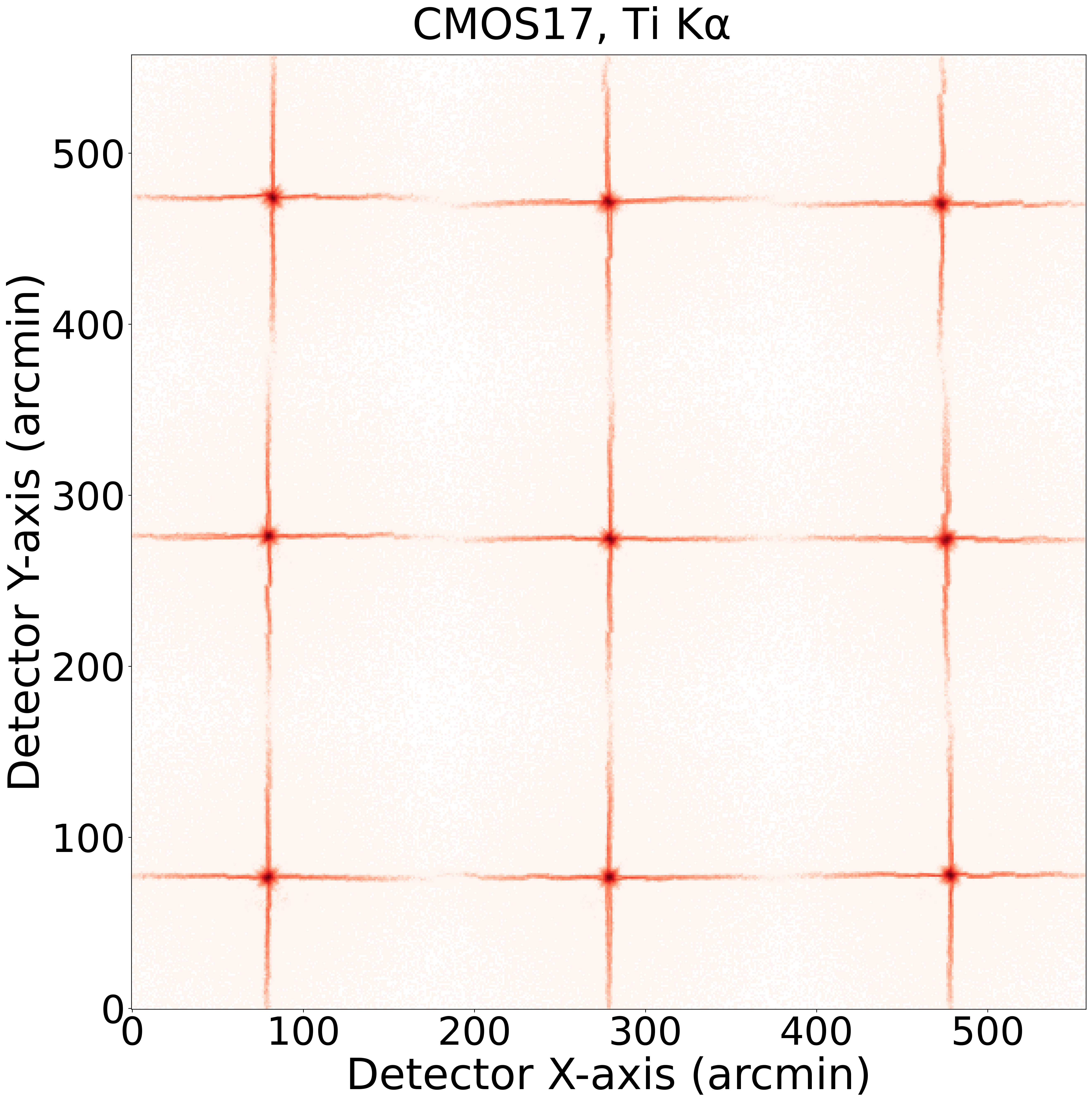}
\includegraphics[width=0.48\textwidth]{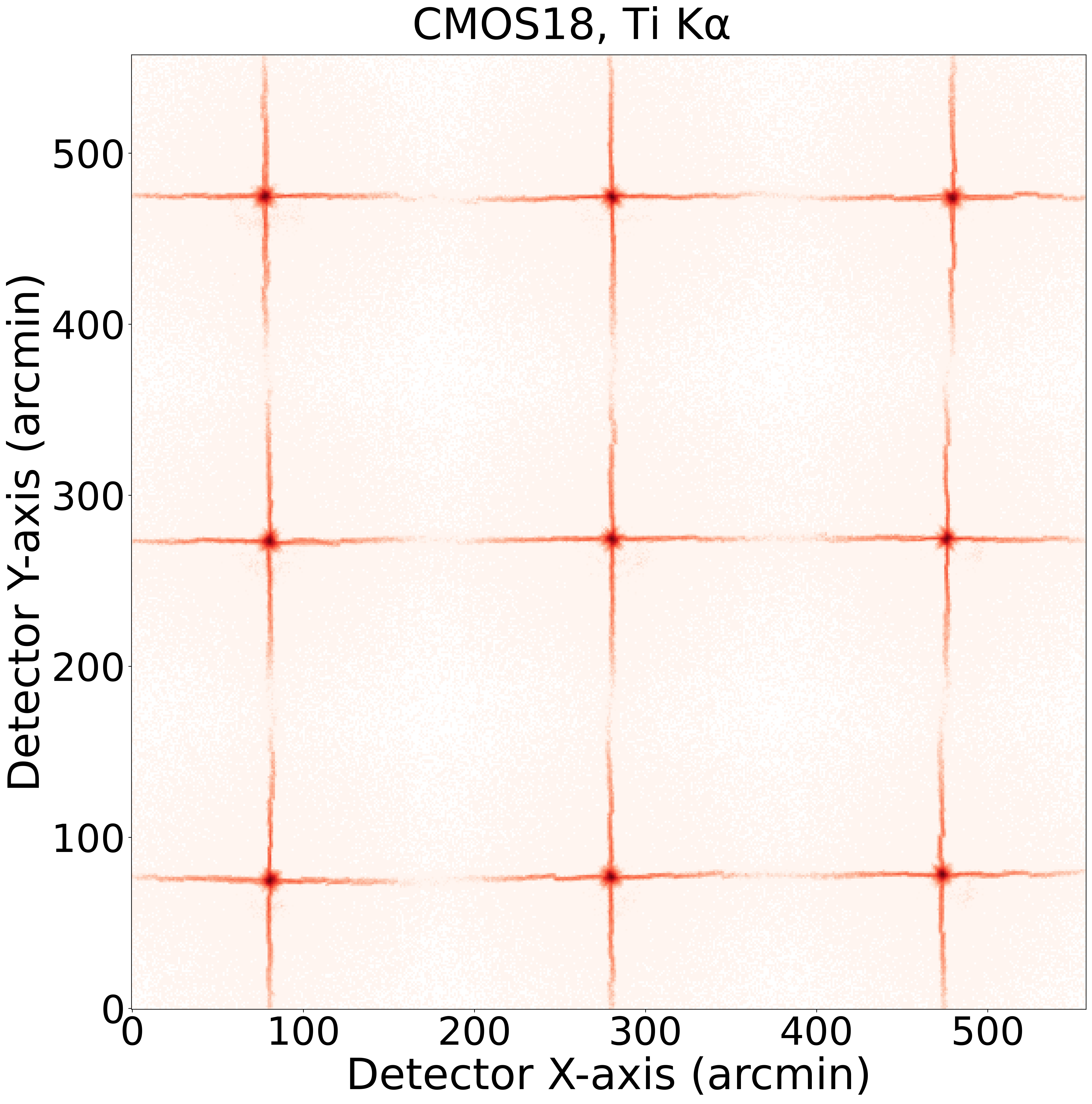}
\includegraphics[width=0.48\textwidth]{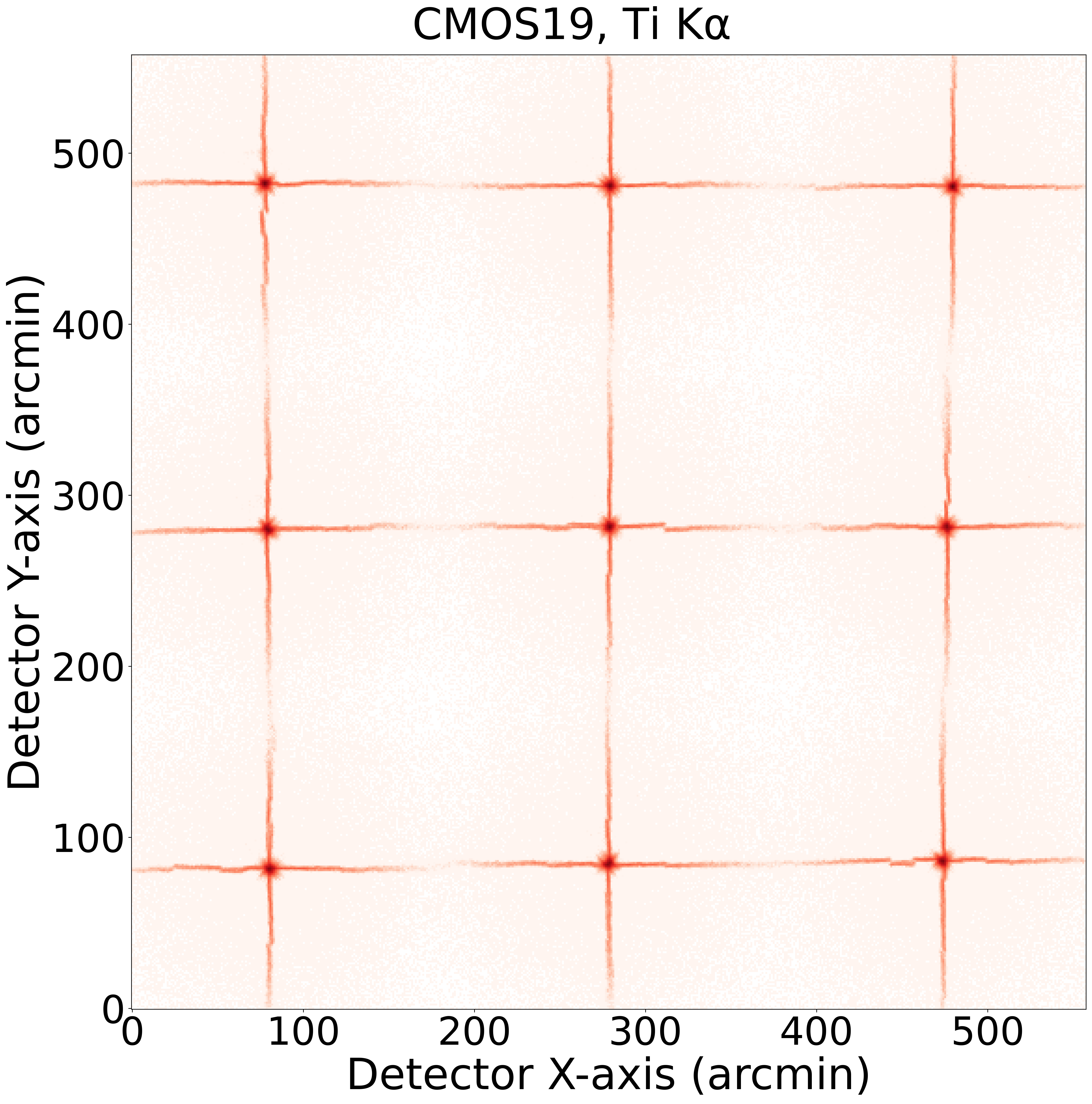}
\includegraphics[width=0.48\textwidth]{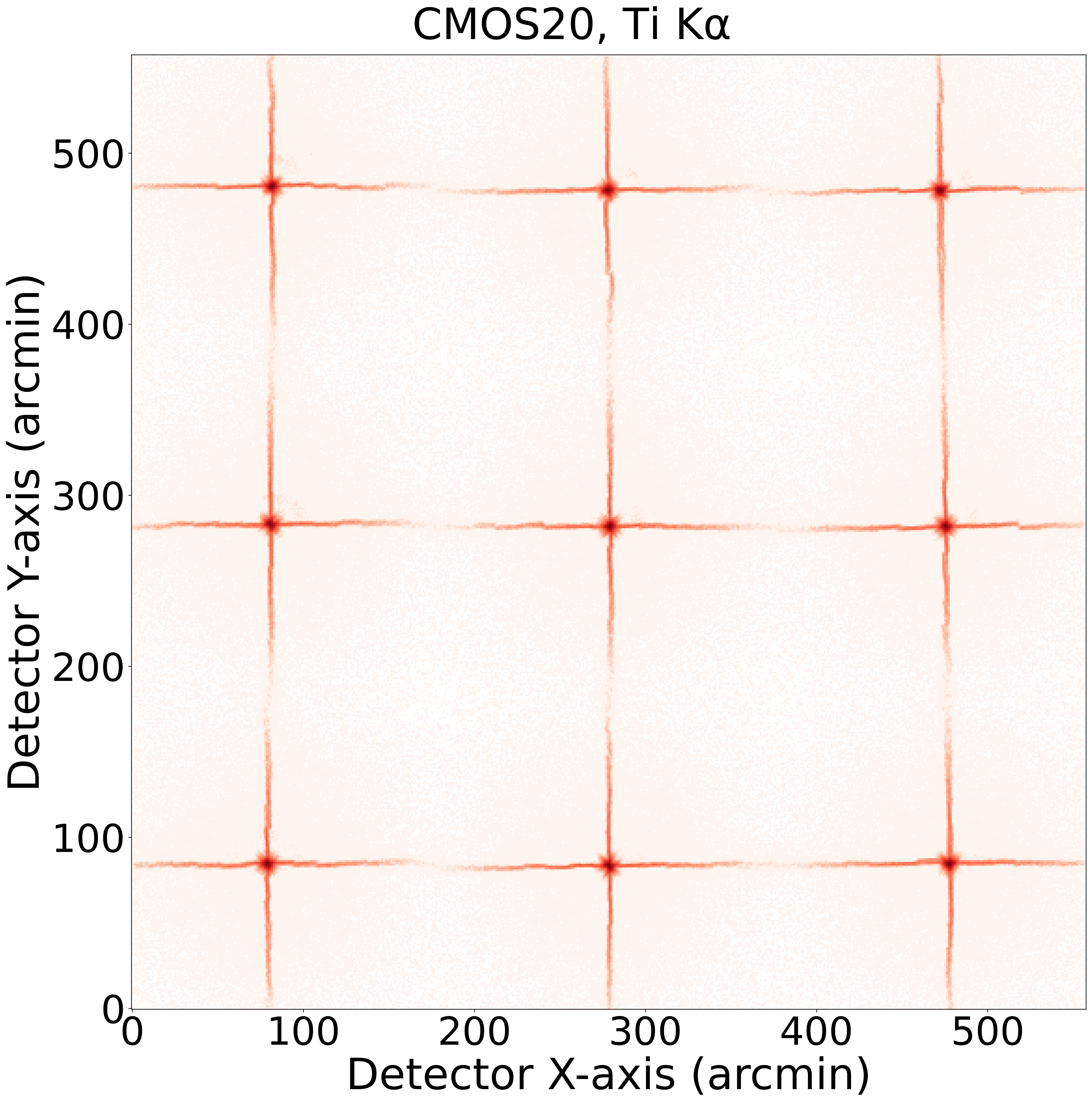}
\caption{\label{fig:psfscan_tika}A mosaic of images of the PSF obtained at 4.51 keV (Ti K$\alpha$ line) across the entire FoV of FM5 (subtended by CMOS 17-20).}
\end{figure*}

\begin{figure*}[!htbp]
    \centering
    \includegraphics[width=\textwidth]{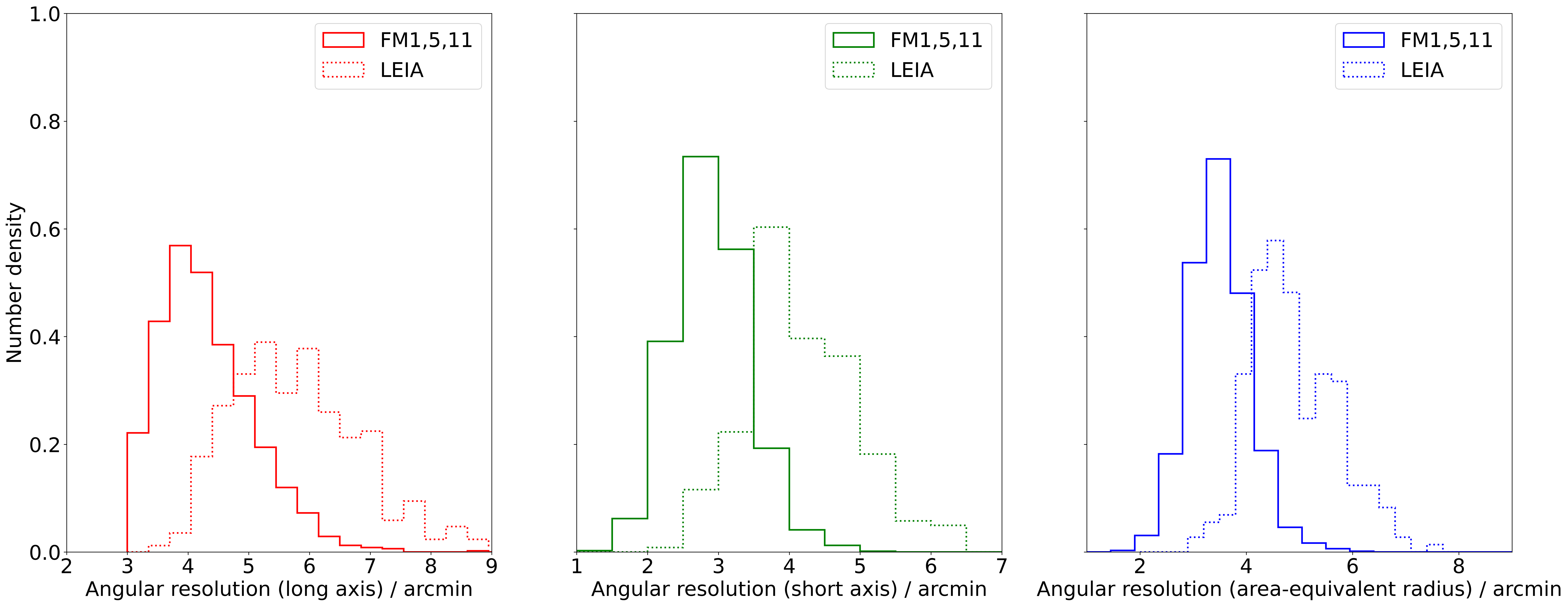}
    \caption{
    The distributions of all three measures of the FWHM (long axis, short axis and area-equivalent radius of the best-fit ellipse) for all the sampled PSFs of the three calibrated modules, with data taken at the energy of Mg K$\alpha$. The result obtained from LEIA is plotted as dashed lines for comparison.
    }
    \label{fig:psf_fitting_cdf}
\end{figure*}

As an example, the mosaics of the X-ray images of the PSF measured at the energy of Mg K$\alpha$ across the entire FoV of FM5 subtended by CMOS 17-20 is shown in Fig. \ref{fig:psfscan_mgk}. 
The result obtained at the energy of Ti K$\alpha$ is shown in Fig. \ref{fig:psfscan_tika}.
It is found that the characteristic cruciform shapes of the PSFs are largely consistent among most of the sampled directions for both energies.
We note that the results obtained for other two modules, as well as other energies are similar.
The measured spatial uniformity of PSF basically validates the inherent performance homogeneity predicted by lobster-eye X-ray optics theory \cite{1979ApJ...233..364A, 2014SPIE.9144E..4EZ, 2016SPIE.9905E..1YW}.
Also, the central focal spots and cruciform arms are very well aligned within the FoV, without discernible PSF misalignments as have been found in \textit{LEIA} (see Fig. 5 and Section 4.3 of \cite{Cheng2024} for details).
This indicates that the MPO plates are mounted precisely with respect to the focal plane detectors. 
In fact, no noticeable PSF misalignments are observed for all twelve WXT modules (Zhang et al. in preparation).

We follow our standard PSF analysis procedures as applied for \textit{LEIA} data (see Section 3.1 and Figure 6 in \cite{Cheng2024}).
Specifically, the barycenter and contours are firstly determined for the focal spot region of the PSF. Then the half-height contour is fitted with an elliptical function (which is found to be a more realistic description of the focal spot morphology due to inherent distortions of the micro-pore channels \cite{2021OptCo.48326656L, 2022PASP..134k5002L}) for derivations of the Full Width Half Maximum (FWHM) of the PSF, including the lengths of the long and short axes of the best-fit ellipse, as well as their product root on behalf of the focal spot area.
In Fig. \ref{fig:psf_fitting_cdf}, we present the distribution of all three measures of the FWHM for the sampled PSFs of these three modules obtained at the energy of Mg K$\alpha$, with the results of \textit{LEIA} over-plotted for a comparision.
It is found that the spatial resolutions (defined as the length of the long axes) range from $\sim3$ arcmin to $\sim7$ arcmin, with a median of $\sim4.2$ arcmin.
The slightly broad distribution in the PSF size suggests the presence of a mild non-uniformity across the FoV, due primarily to the inherent imperfections introduced in the MPO fabrication and mounting. The slight mismatch between the flat detector plane and spherical focal surface may also cause tiny distortions of the PSF particularly near the edge of the detector.
Notably, the flight model modules demonstrate a PSF quality improvement of $\sim1.5$ arcmin compared to \textit{LEIA}, highlighting their superior imaging capabilities. This enhancement primarily results from refined fabrication techniques for the MPO plates.

We note that while the PSF FWHM measurements are inevitably affected by the de-focus effect arising from the finite X-ray source distance ($\sim100$ m), the results obtained at 100XF remain representative of in-flight performance as well. 
This originates from intrinsic curvature variations inherent to MPO plate fabrication. Specifically, each MPO plate is fabricated with an individual curvature center. To optimize the overall imaging performance, that is, to maintain uniform PSF morphology and width across different incident angles, an `effective' curvature center is defined through plate alignment optimization for each module's FoV quadrant (comprising a $3\times3$ MPO plate array). 
Consequently, some of the MPO plates are strategically mounted with controlled de-focus offsets ($\sim1–2$ mm). In this sense, the calibration-induced out-of-focus effect instead compensates for plate-level curvature variations, improving the imaging quality for these plates. The global PSF property distribution remains statistically preserved under this compensatory interplay.

\subsubsection{Dependence on photon energy}
\label{sec:psf_on_energy}

To characterize the energy dependence of the PSF properties, for FM1 and FM5 the PSFs were also measured at several other energies (O K$\alpha$, Si K$\alpha$, Ag L$\alpha$ and Ti K$\alpha$) using a sub-grid of $3\times3$ points (the `chip center' array) for each CMOS detector. Due to time constraints, the Cu L measurements were confined to the central direction of the CMOS detector.

\begin{figure*}[!htbp]
    \centering
    \includegraphics[width=0.3\textwidth]{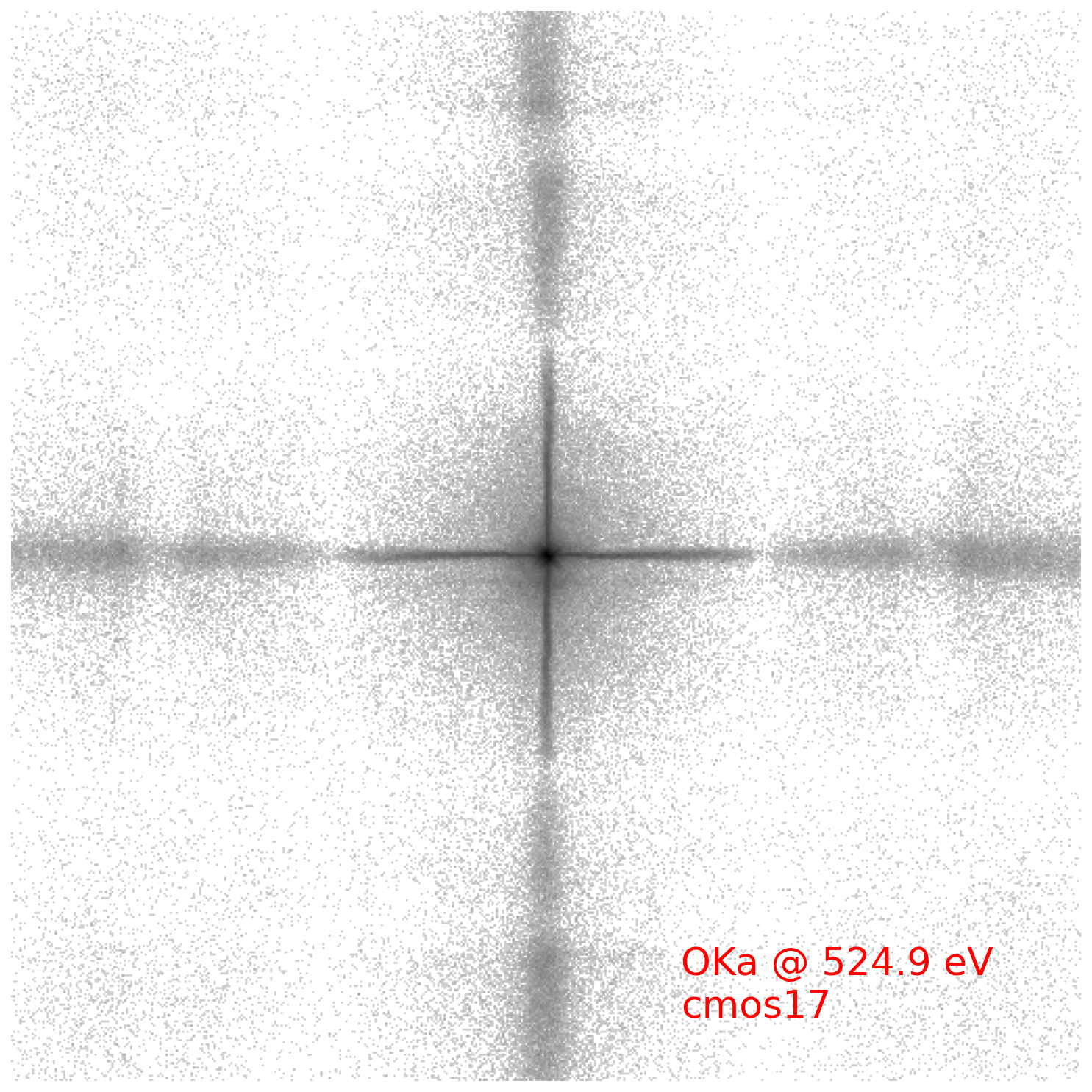}
    \includegraphics[width=0.3\textwidth]{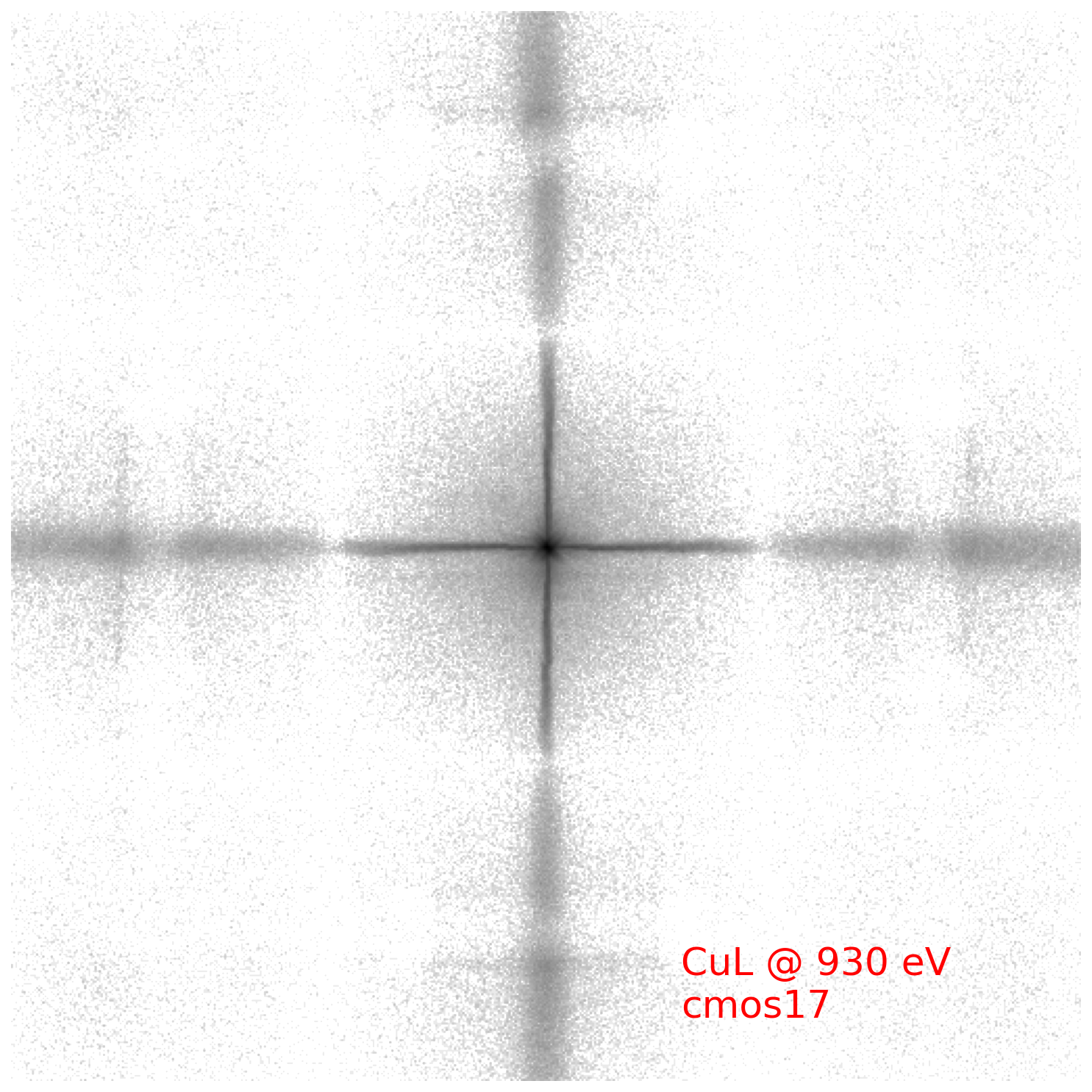}
    \includegraphics[width=0.3\textwidth]{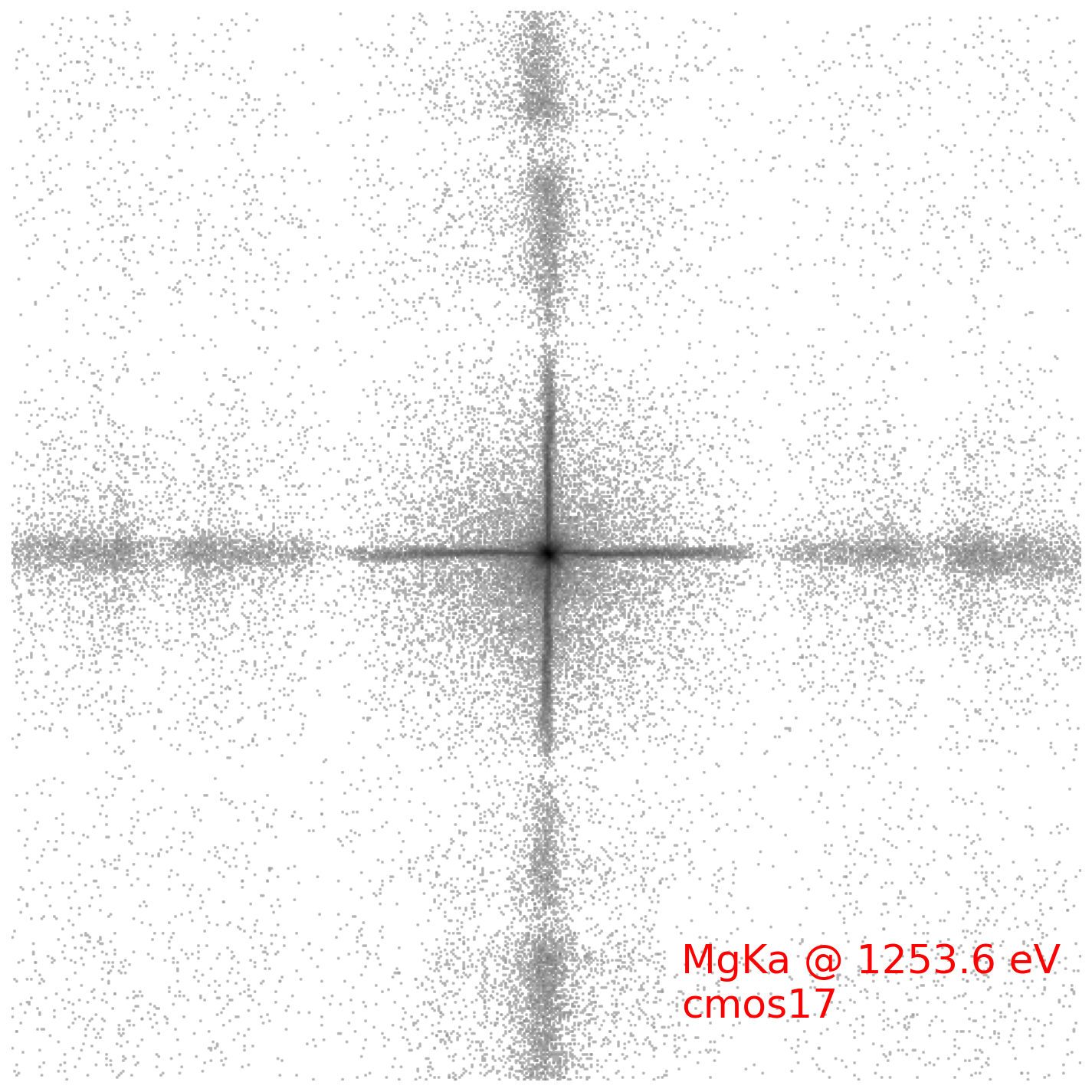}
    \includegraphics[width=0.3\textwidth]{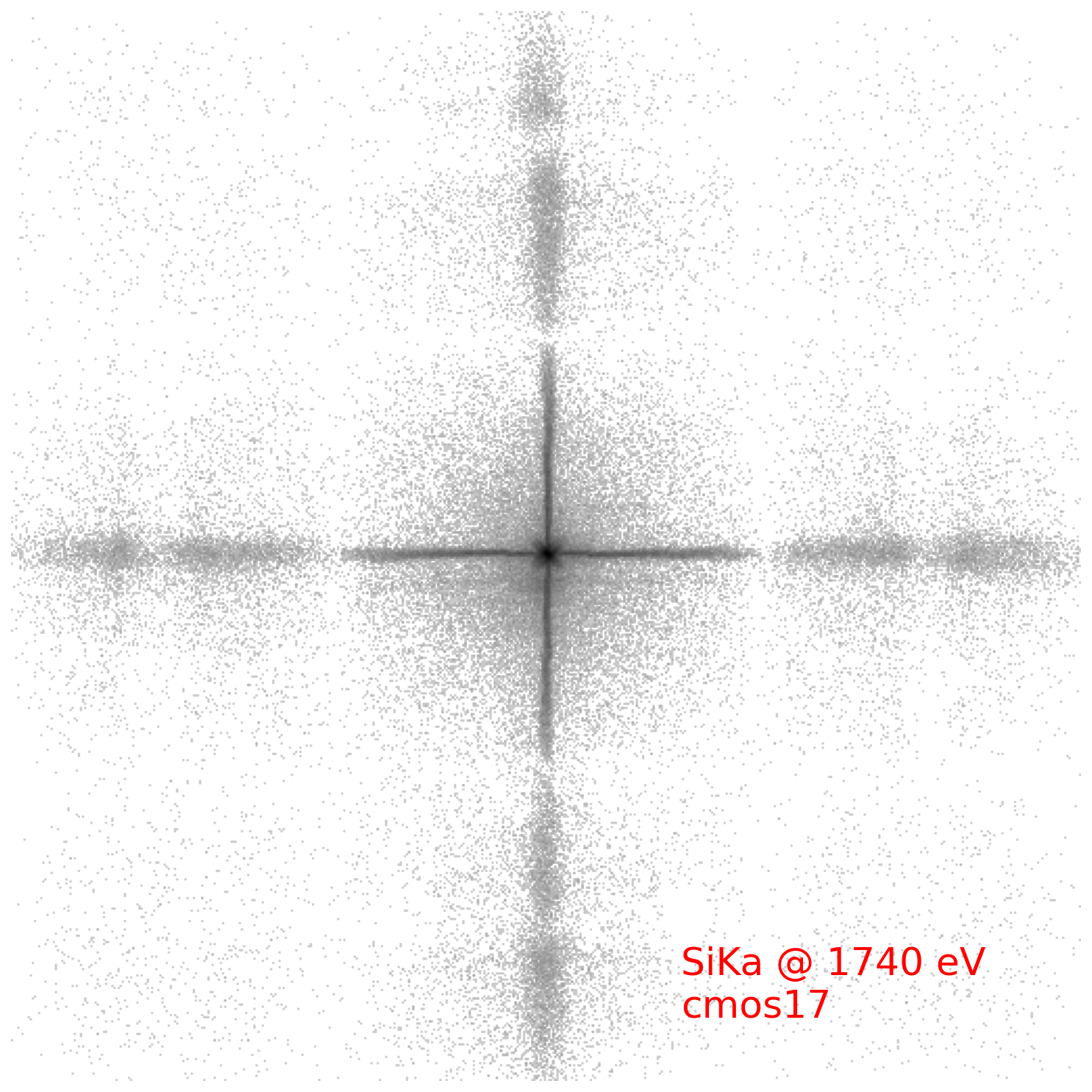}
    \includegraphics[width=0.3\textwidth]{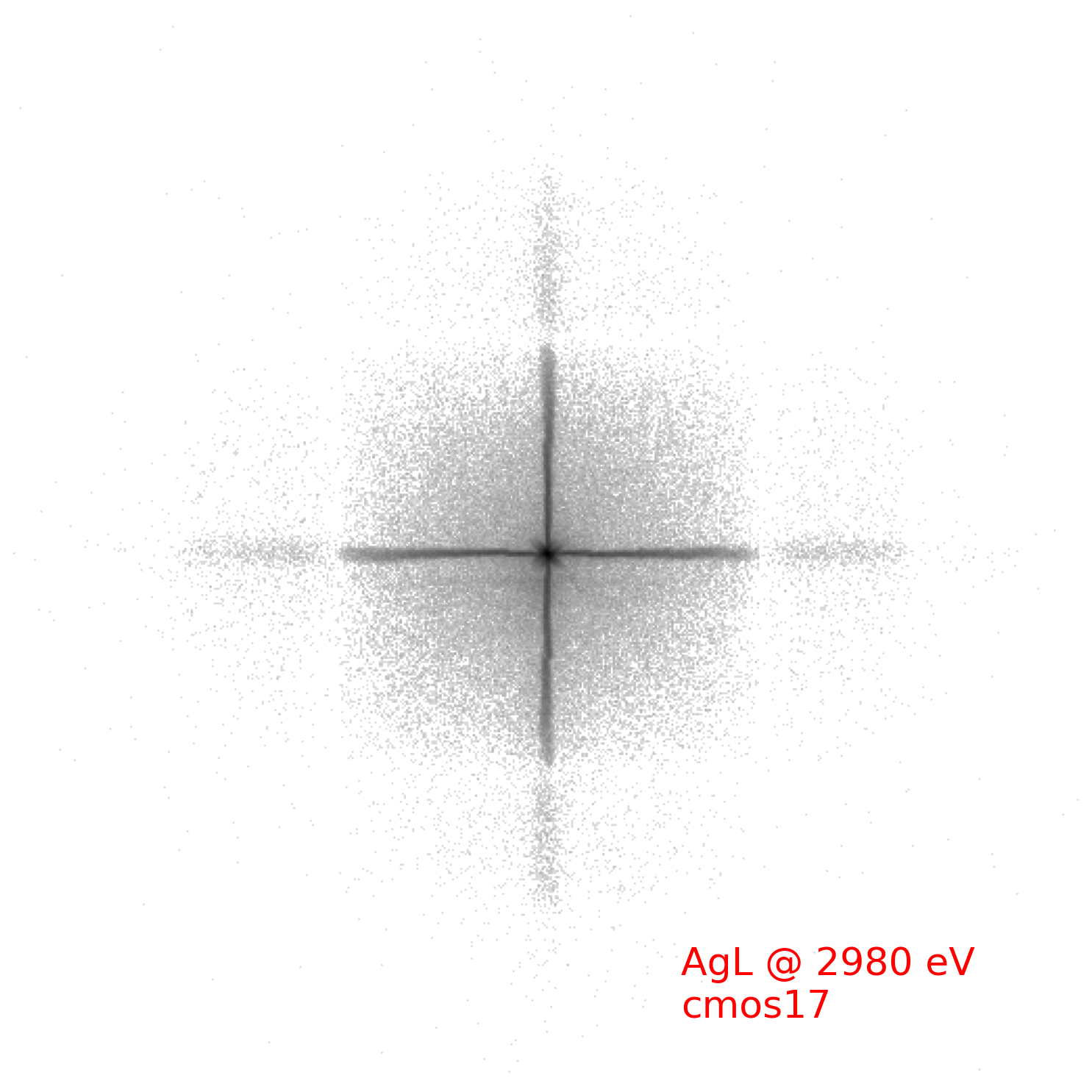}
    \includegraphics[width=0.3\textwidth]{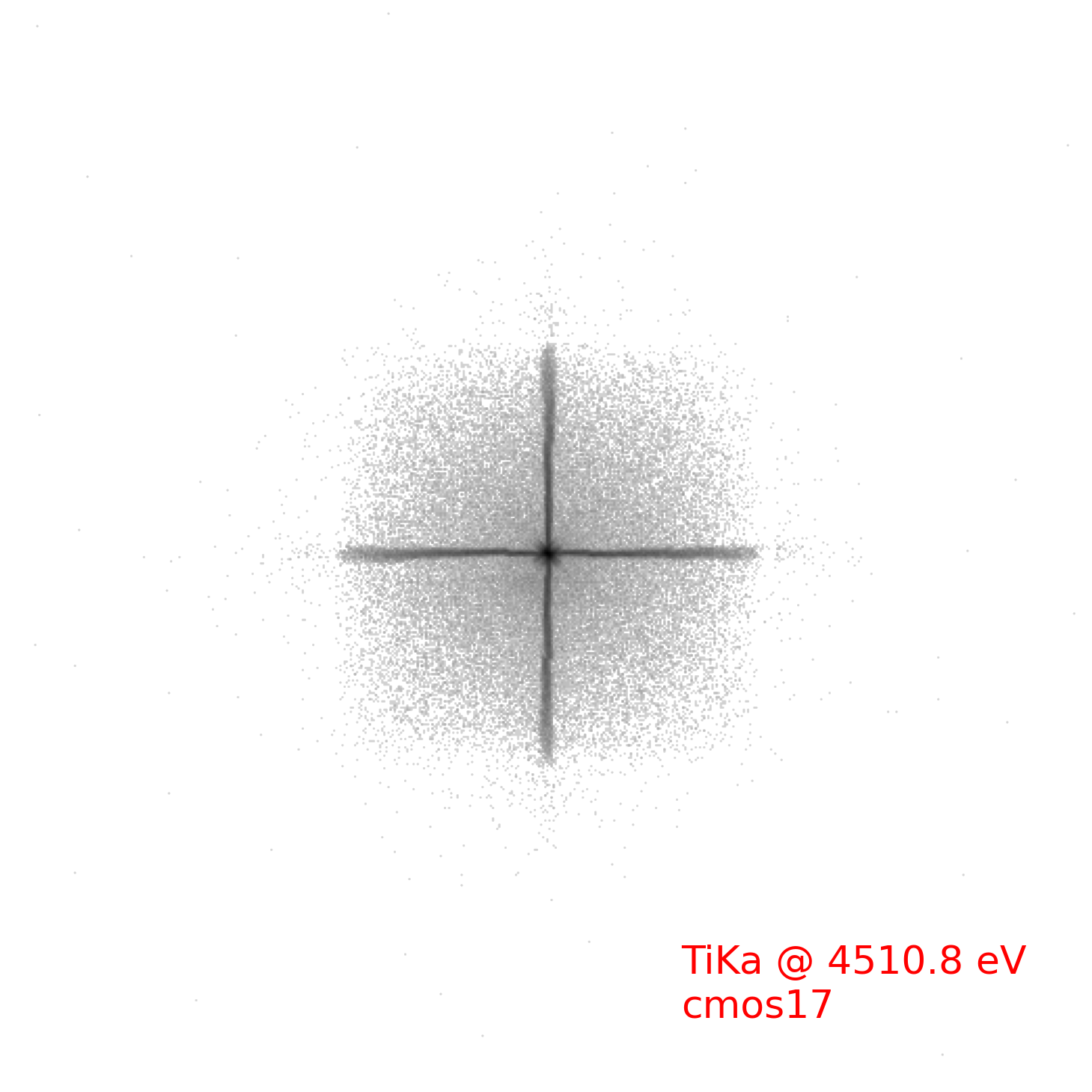}
    \caption{Images of the measured PSFs in the direction along the center of CMOS 17, with data taken at the line energies of O K$\alpha$, Cu L, Mg K$\alpha$, Si K$\alpha$, Ag L$\alpha$ and Ti K$\alpha$.}
    \label{fig:psf_dif_en}
\end{figure*}

\begin{figure}[!htbp]
    \includegraphics[width=0.8\textwidth]{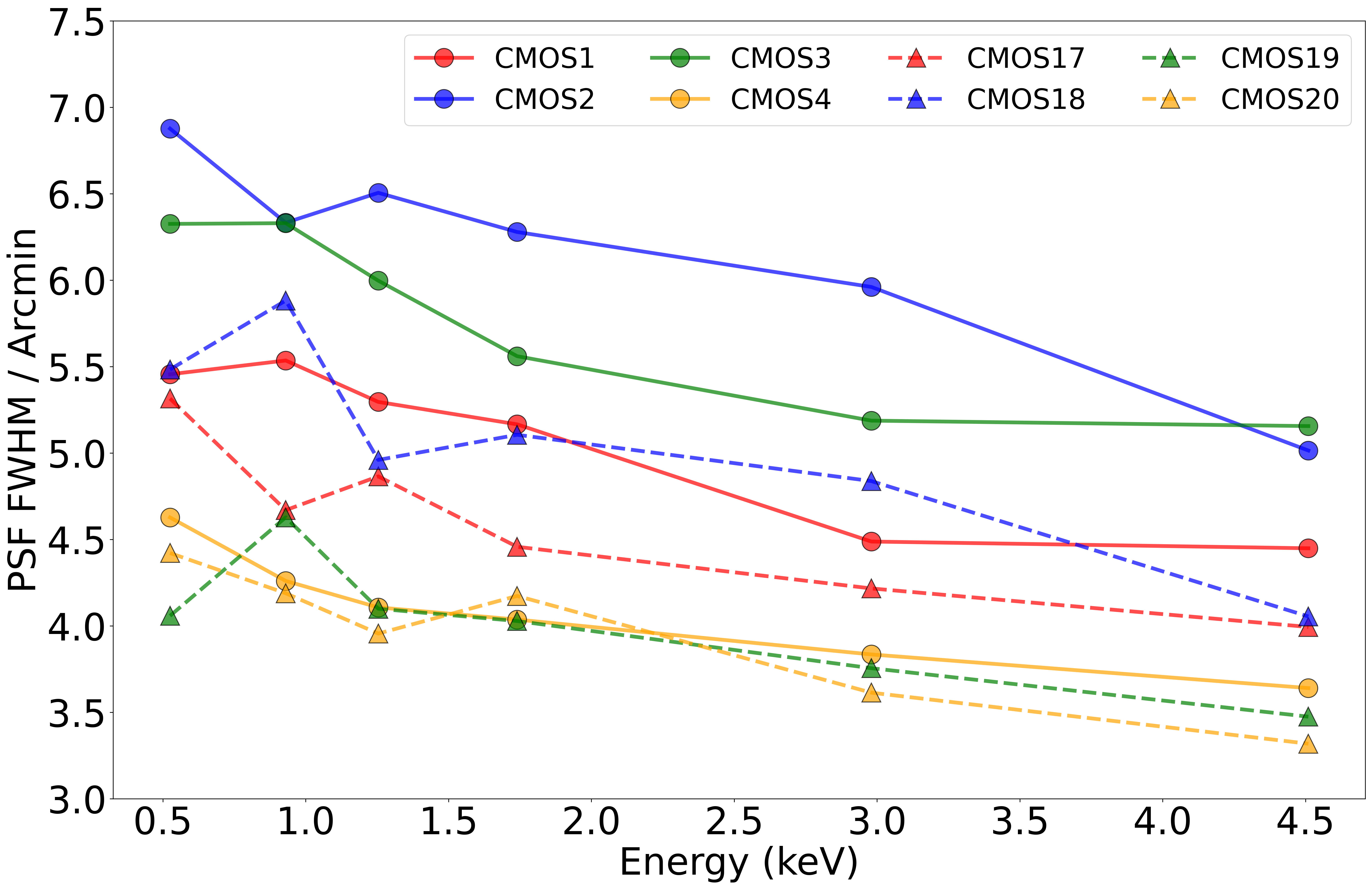}
    \caption{The dependence of the FWHM (the long axis) of the PSF on the photon energy, with the data taken along the central direction of CMOS 1-4 and 17-20.}
    \label{fig:psf_fwhm_vs_energy}
\end{figure}

As an example, Fig. \ref{fig:psf_dif_en} shows the mosaics of the X-ray images measured along the center of CMOS 17 aboard FM5, at various line energies. It is found that the PSF morphology is strongly dependent on the photon energy. 
On the other hand, the relation between $FWHM_{\rm PSF}$ and photon energy is shown in Fig. \ref{fig:psf_fwhm_vs_energy}; only those in the directions passing through the respective centers of the eight sensors are shown, as examples. It is found that the width of the focal spot exhibit a decreasing trend when the photon energy increases from $\sim0.53$ eV to $\sim4.51$ eV. 

The variations of the PSF morphology and FWHM with the changing photon energy arises intrinsically from the inverse relation between the photon energy and critical angle of grazing incidence. On one hand, the chances for multiple reflections (corresponding to large incident angles) are significantly reduced for higher energy photons, yielding more compact cruciform structures with attenuated higher-order components. 
On the other hand, the dispersion of the PSF focal spot is mainly determined by the integrated pointing deviations of the micro-pore channels (that reflect focal spot photons) from the respective curvature center \cite{2017ZhaoDH_simulation}.
At lower energies, the increased number of contributing channels amplifies the pointing non-uniformity, thereby broadening the FWHM as observed.
These results are consistent with the results of the independent calibration of the MA aboard FM5 carried out at the Panter facility \cite{EPWXT_Panter2023}, as well as that obtained by \textit{LEIA} \cite{Cheng2024}. 
The energy dependence of the PSF can be well reproduced by our Monte Carlo simulations \cite{2014SPIE.9144E..4EZ,2017ZhaoDH_simulation}.

\subsection{Effective Area}
\label{sec:effarea}

\begin{figure*}[!htbp]
    \centering
    \includegraphics[width=\textwidth]{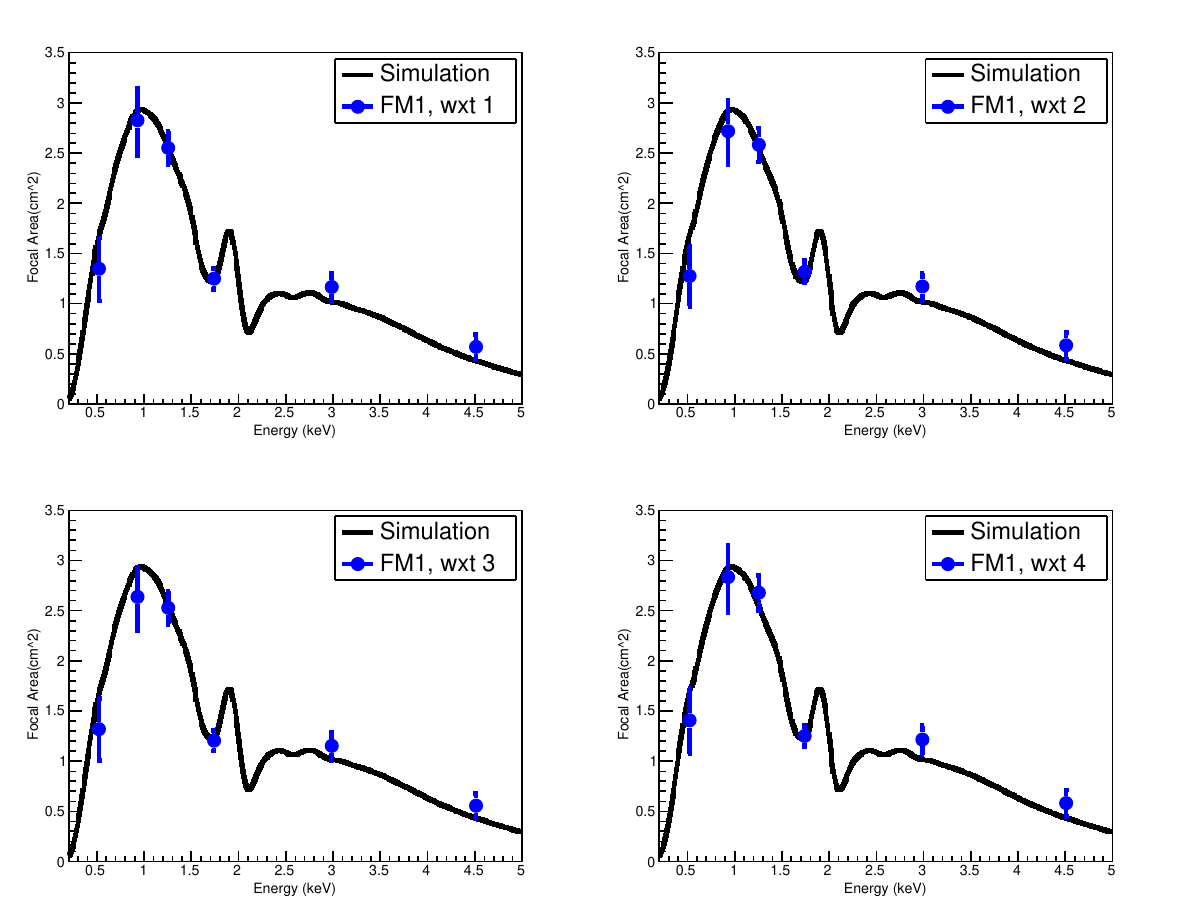}
    \caption{The effective area as a function of the energy of incident photons, measured in the direction along the center of the four CMOS detectors aboard FM1 (CMOS 1-4). The blue dots are the measurements at different energies and the black solid line denotes the simulated model over-plotted for comparison.}
    \label{fig:effarea_energy_curve_fm1}
\end{figure*}

\begin{figure}[!htbp]
    \centering
    \includegraphics[width=\textwidth]{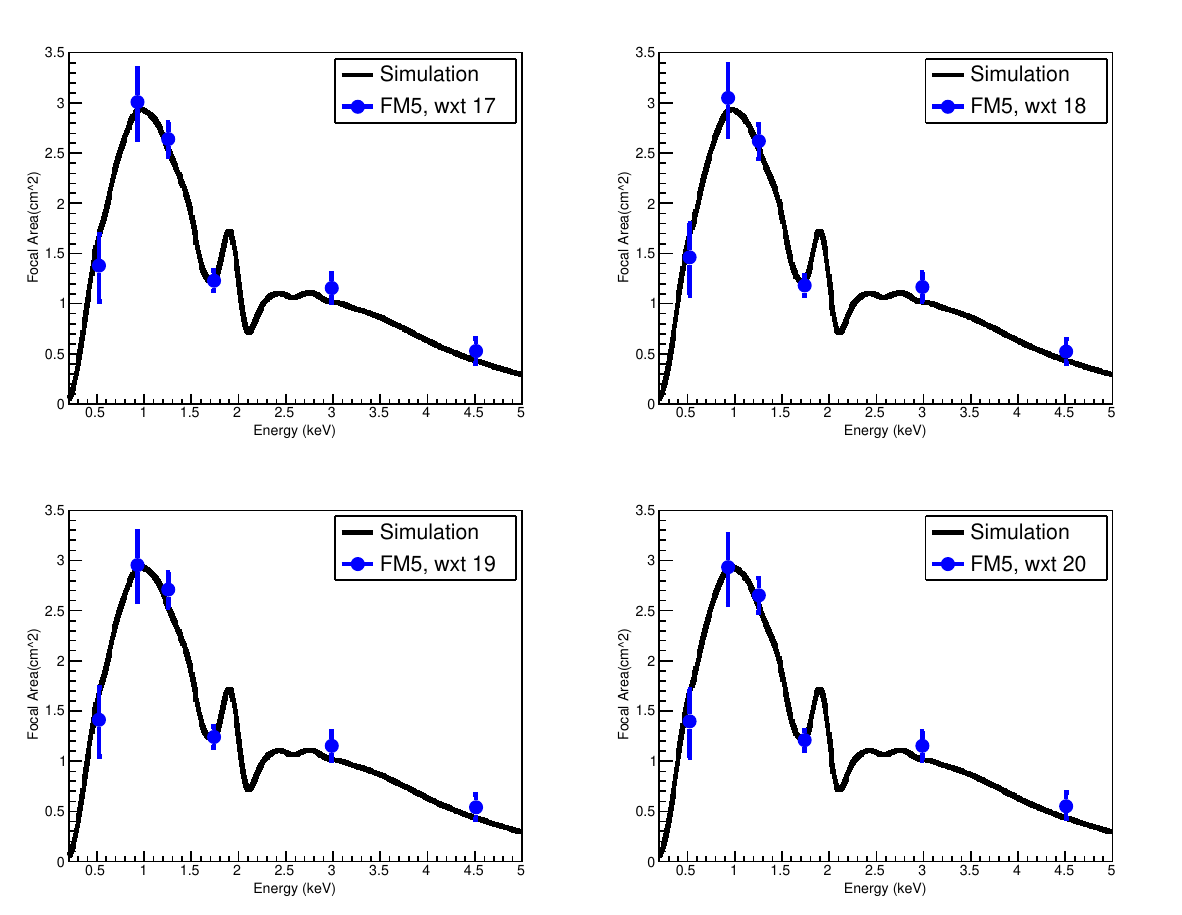}
    \caption{Same as Figure \ref{fig:effarea_energy_curve_fm1}, but for FM5 (CMOS 17-20).}
    \label{fig:effarea_energy_curve_fm5}
\end{figure}

\begin{figure}[!htbp]
    \centering
    \includegraphics[width=\textwidth]{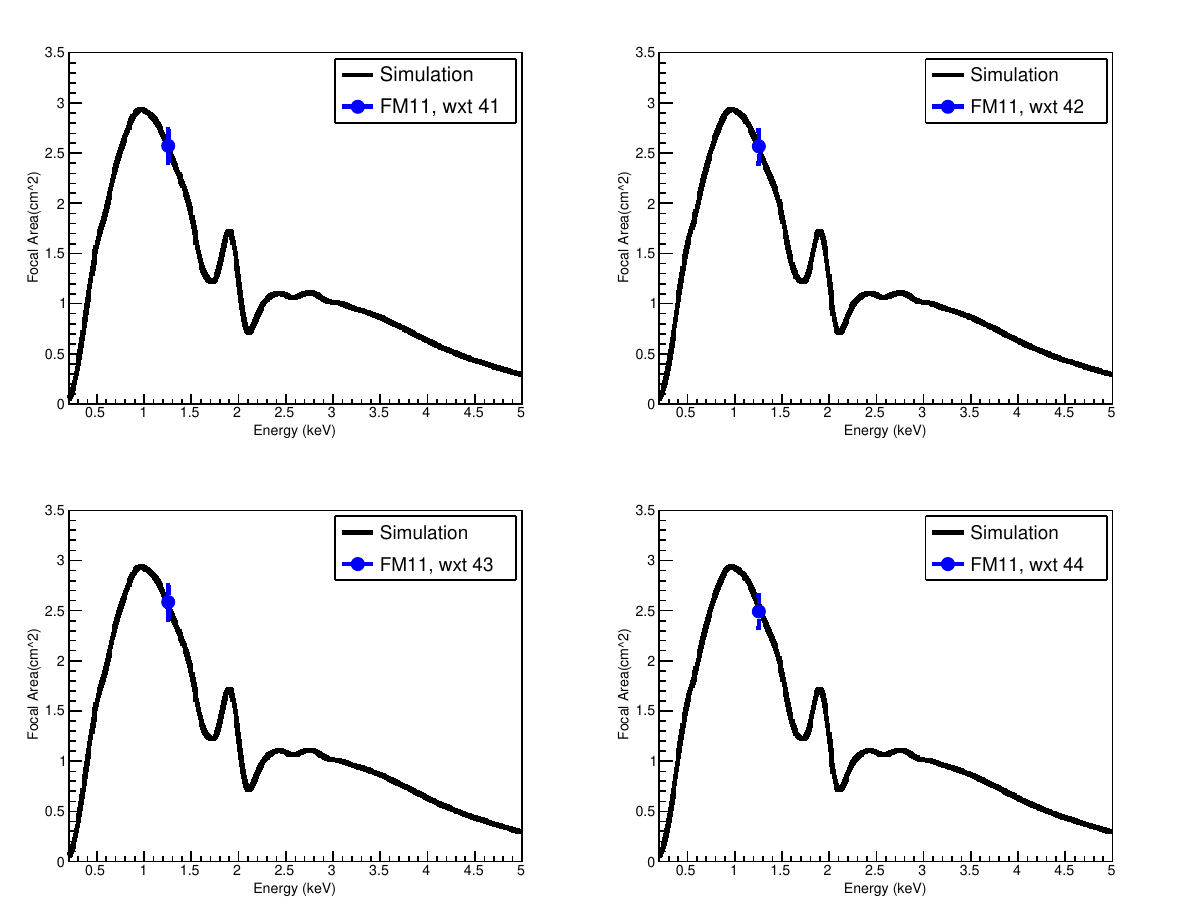}
    \caption{Same as Figure \ref{fig:effarea_energy_curve_fm1}, but for FM11 (CMOS 41-44).}
    \label{fig:effarea_energy_curve_fm11}
\end{figure}

\begin{figure*}[!htbp]
    \centering
    \includegraphics[width=0.48\textwidth]{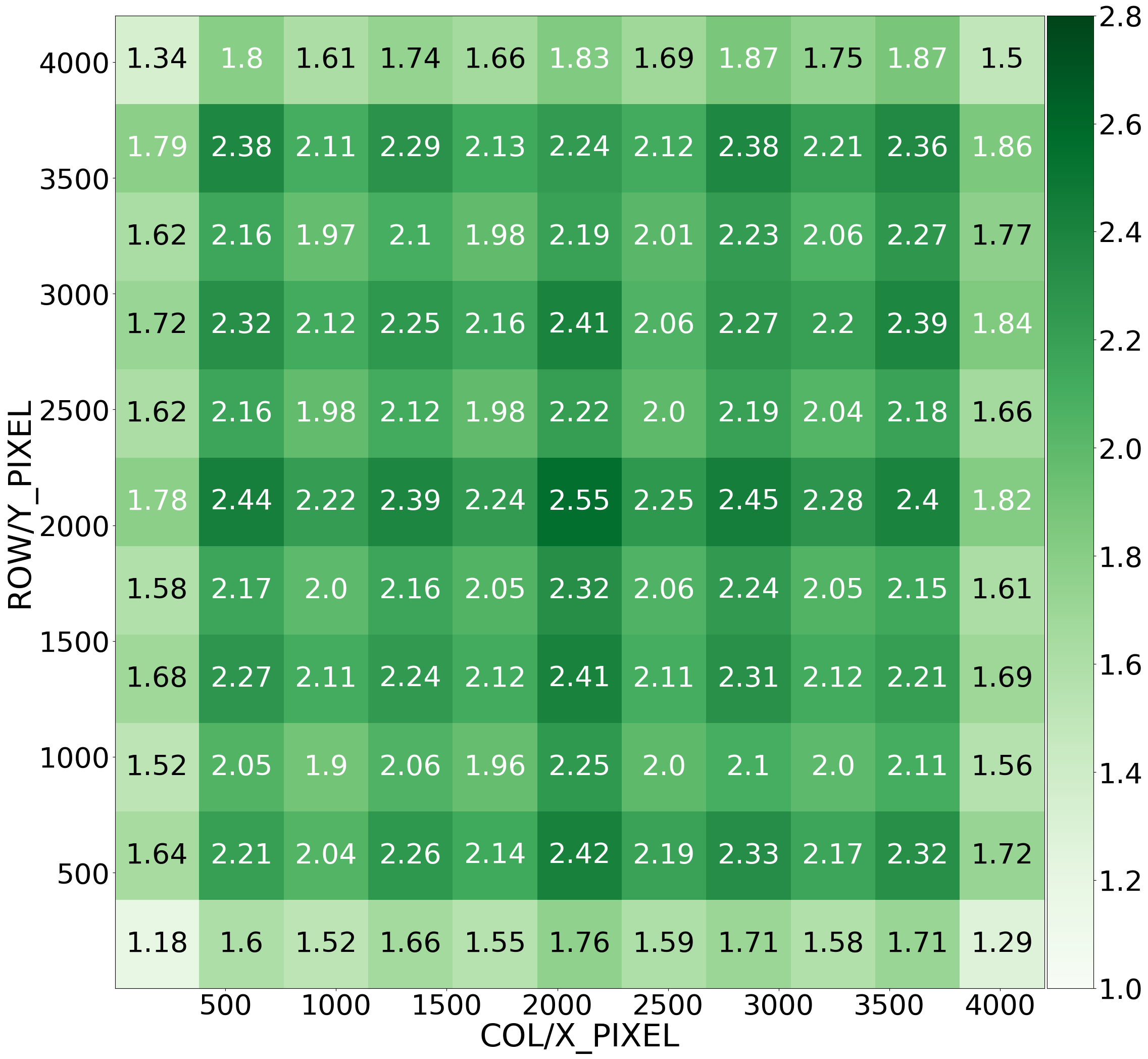}
    \includegraphics[width=0.48\textwidth]{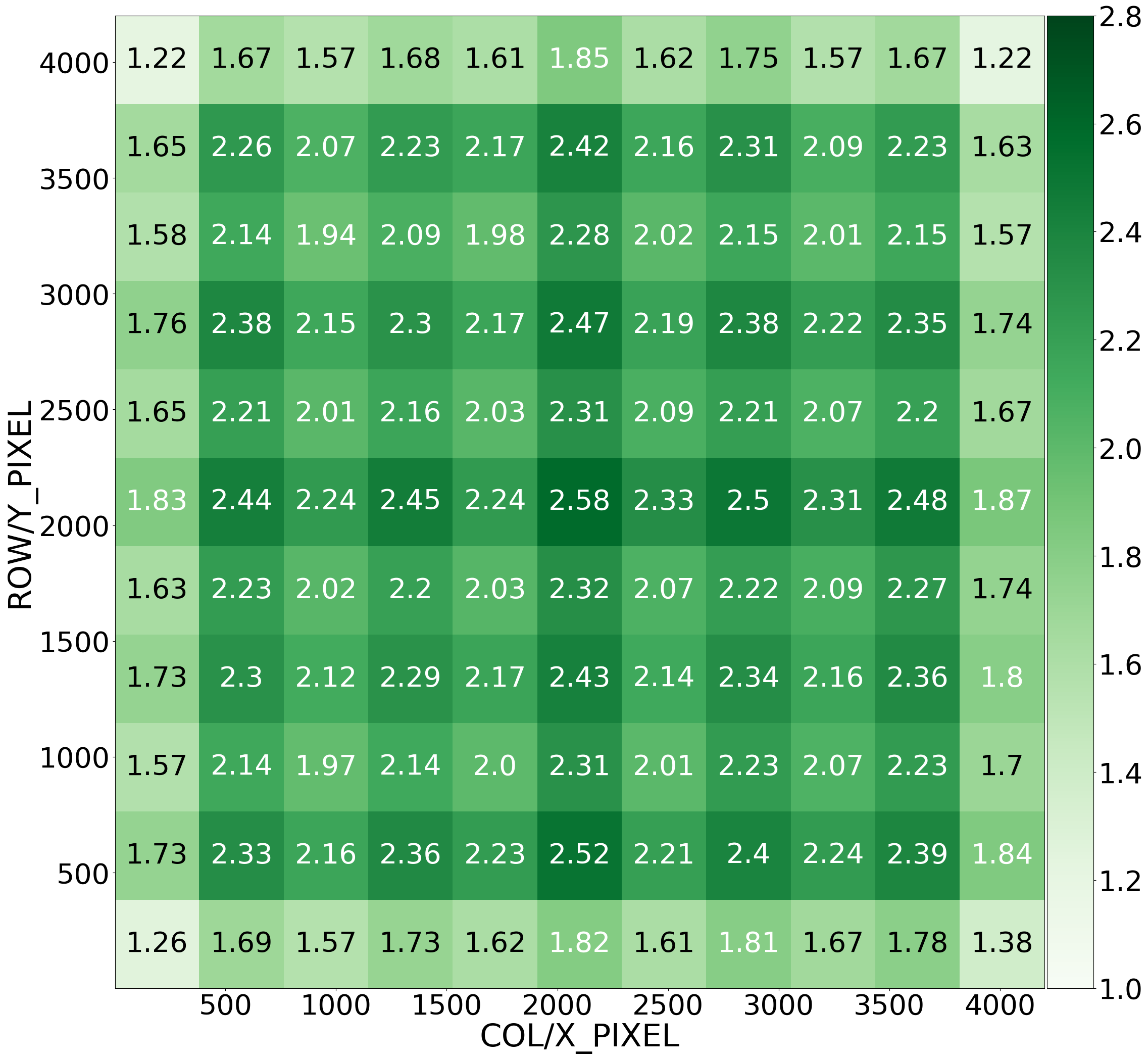}
    \includegraphics[width=0.48\textwidth]{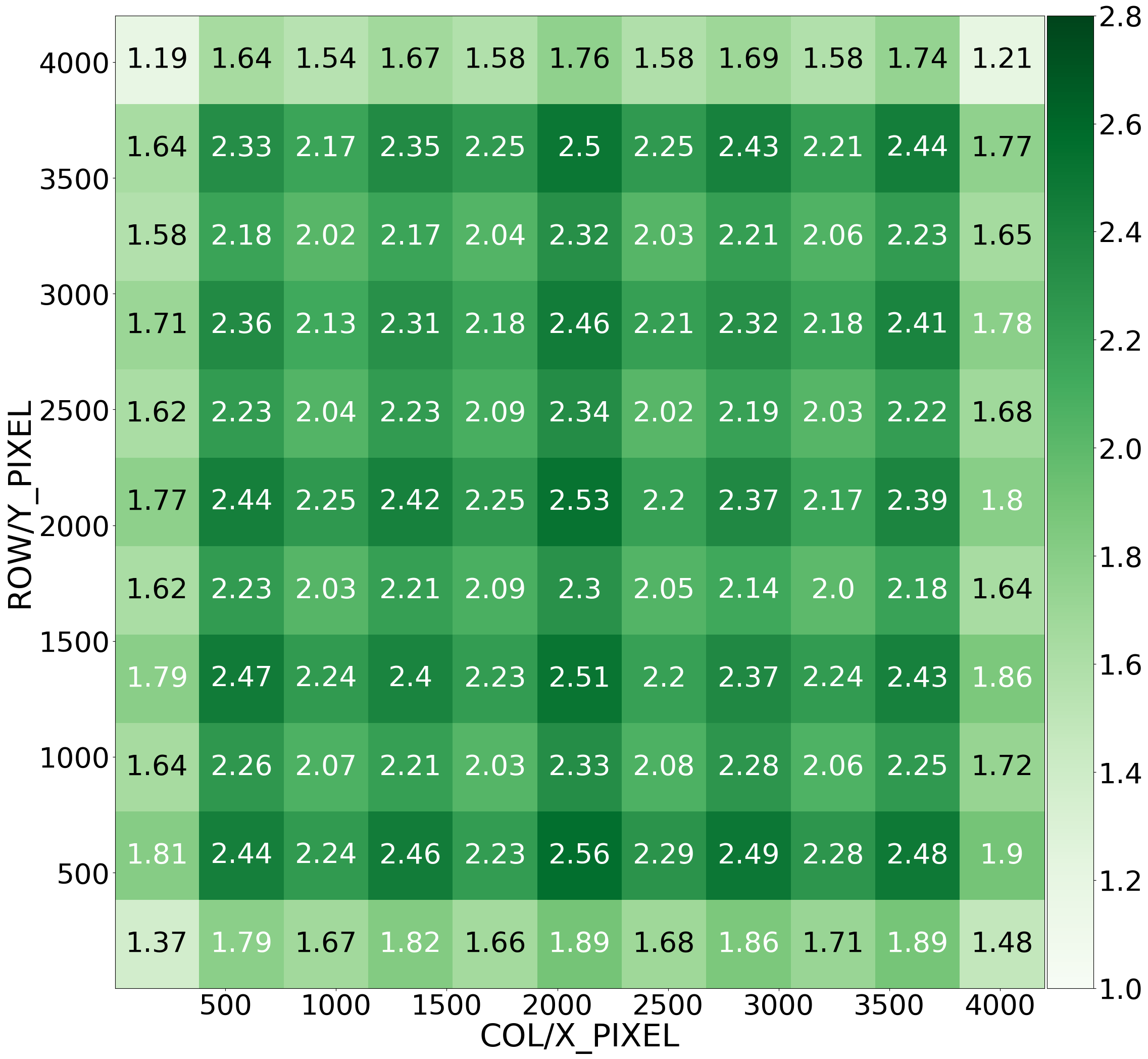}
    \includegraphics[width=0.48\textwidth]{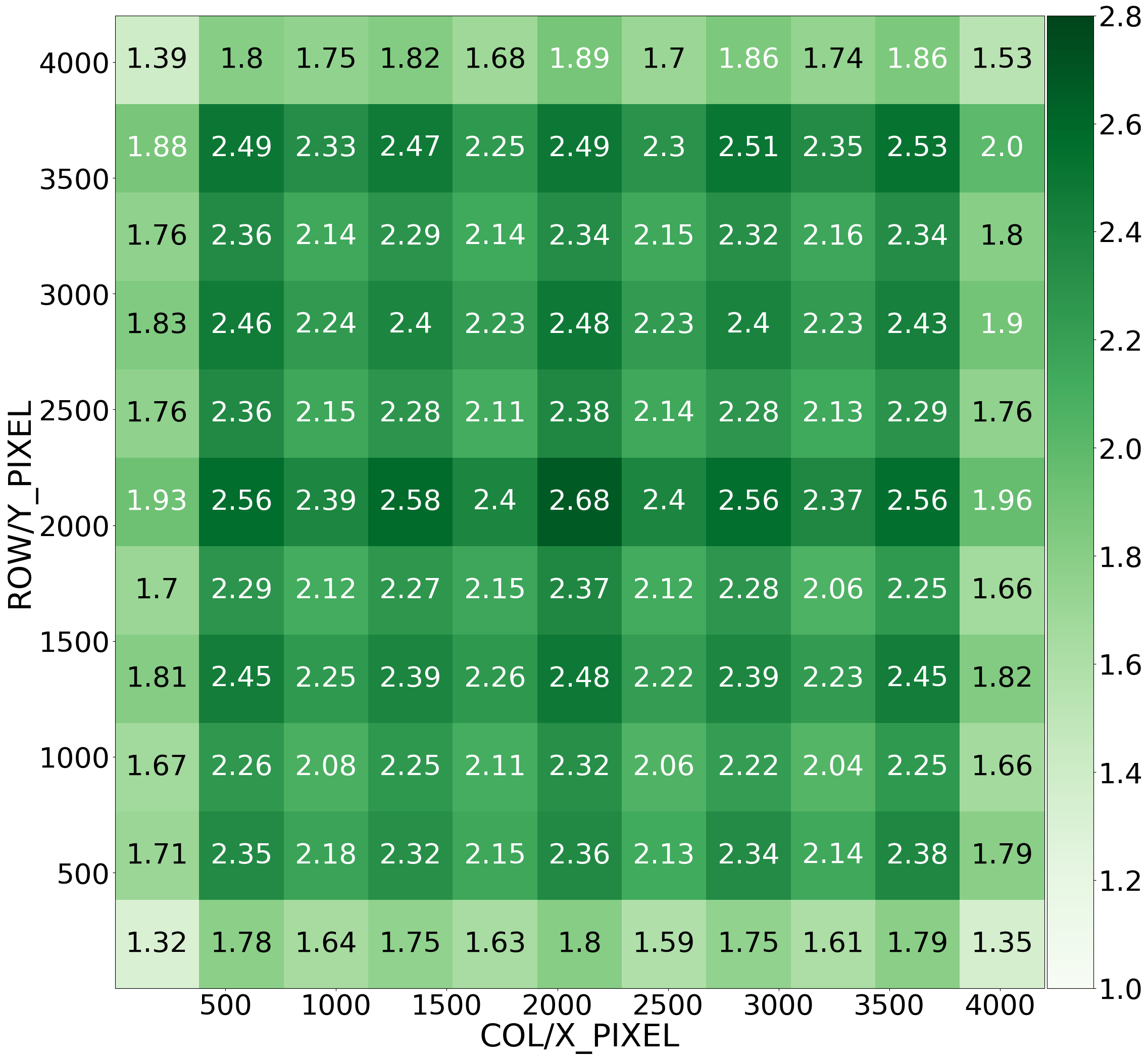}
    \caption{The distribution of the effective area of the focal spot for the CMOS 1--4, at the energy of $1.25$ keV.}
    \label{fig:fm1_effarea_distribution}
\end{figure*}

\begin{figure}[!htbp]
    \centering
    \includegraphics[width=0.48\textwidth]{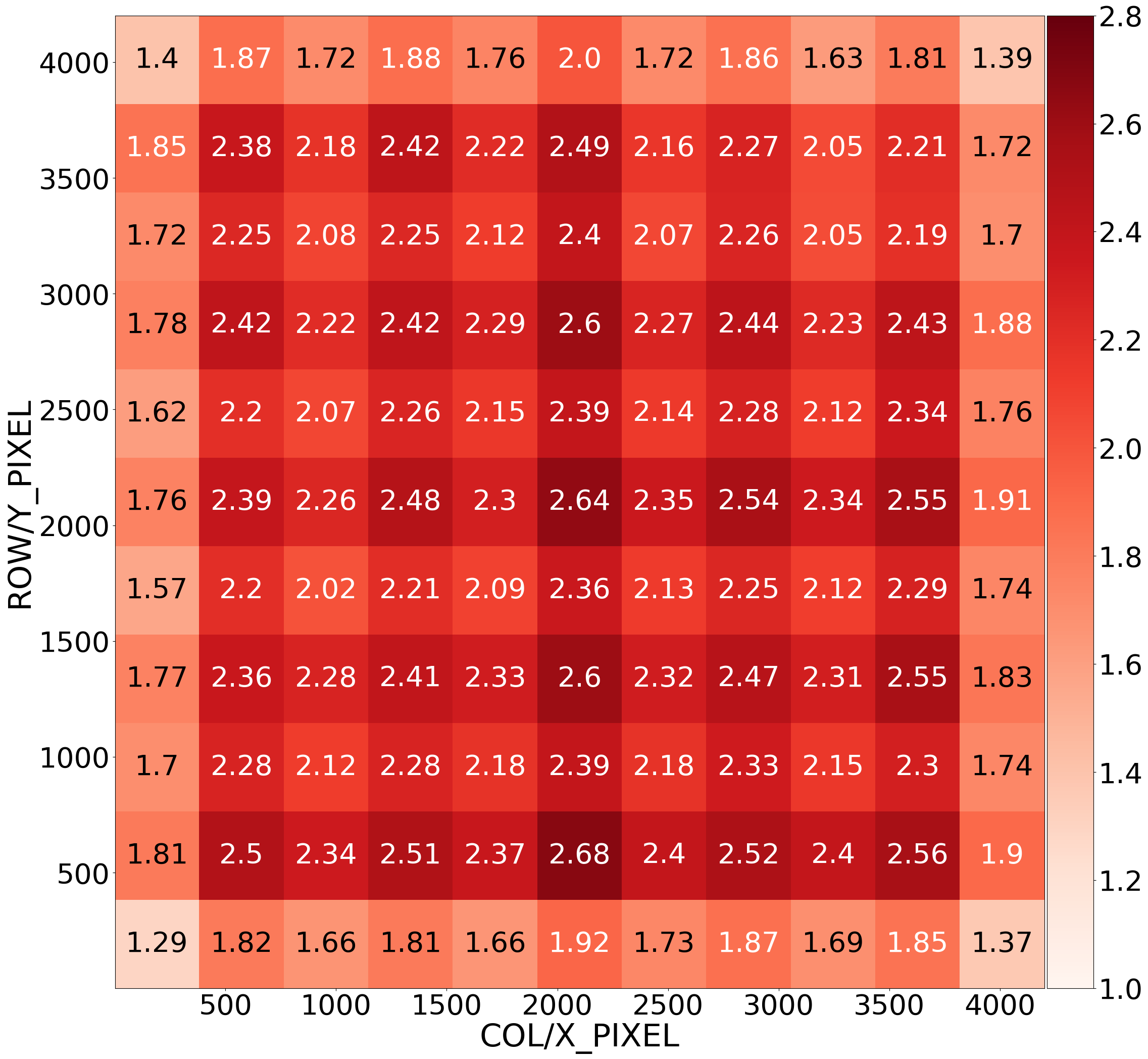}
    \includegraphics[width=0.48\textwidth]{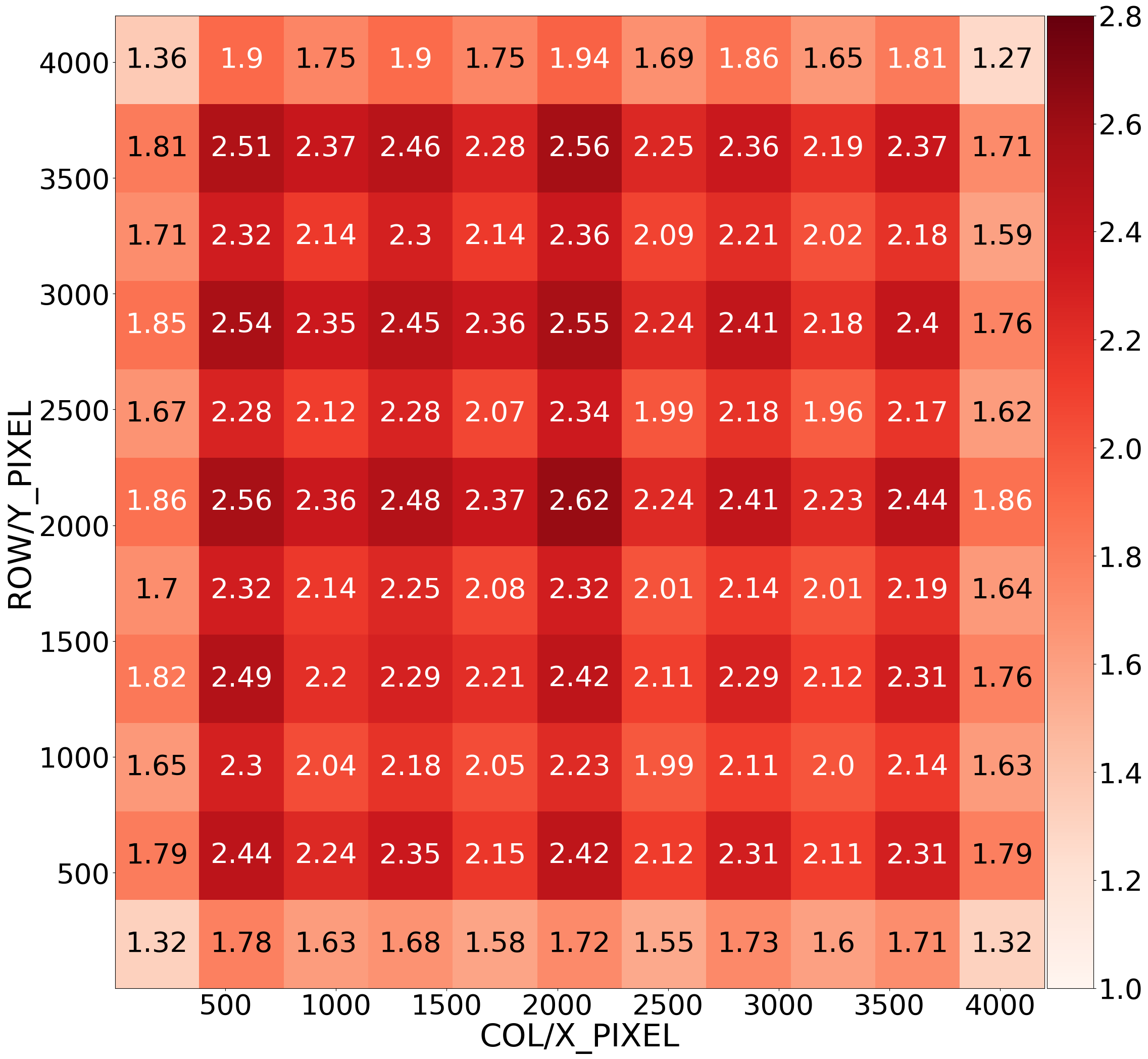}
    \includegraphics[width=0.48\textwidth]{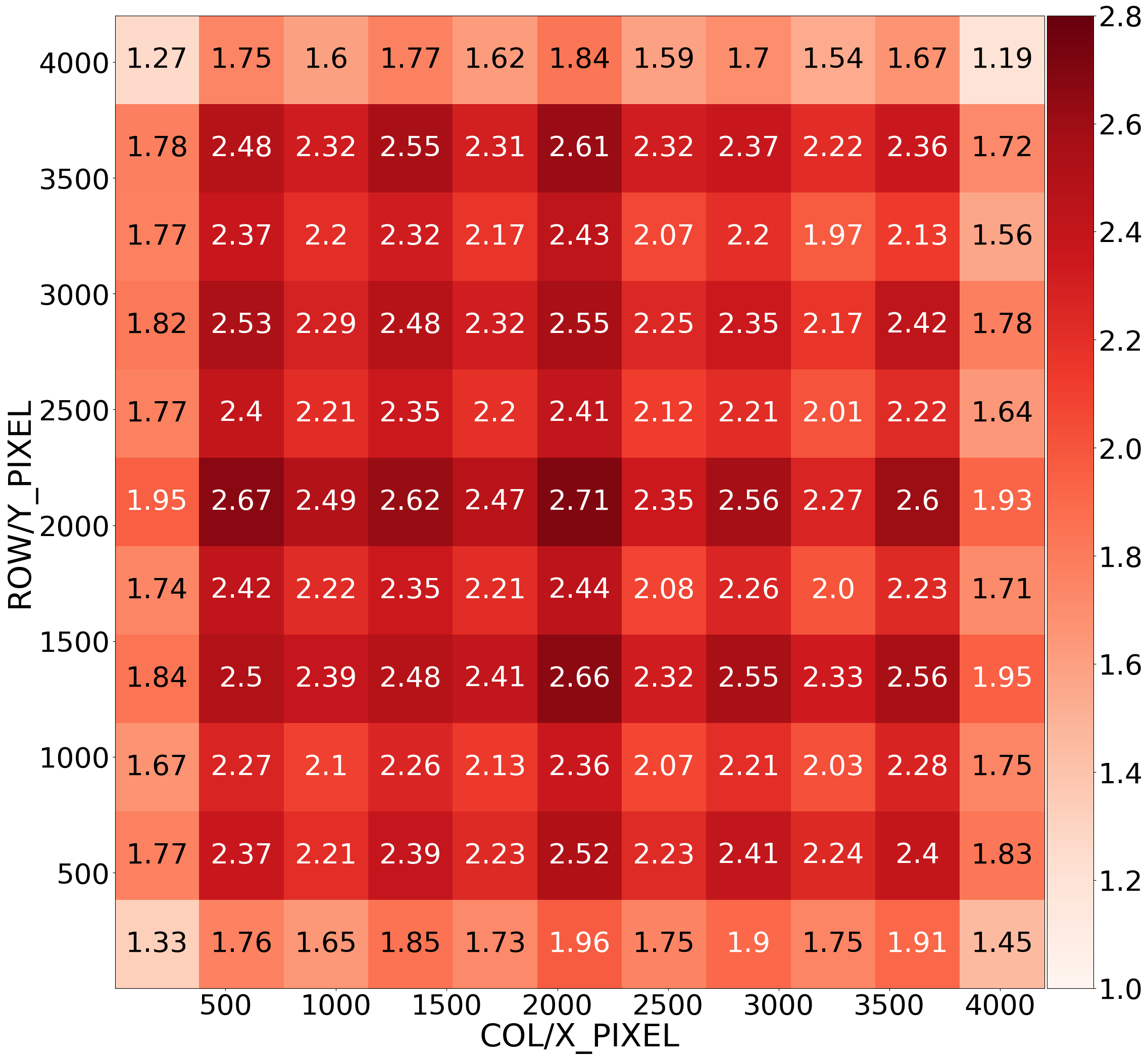}
    \includegraphics[width=0.48\textwidth]{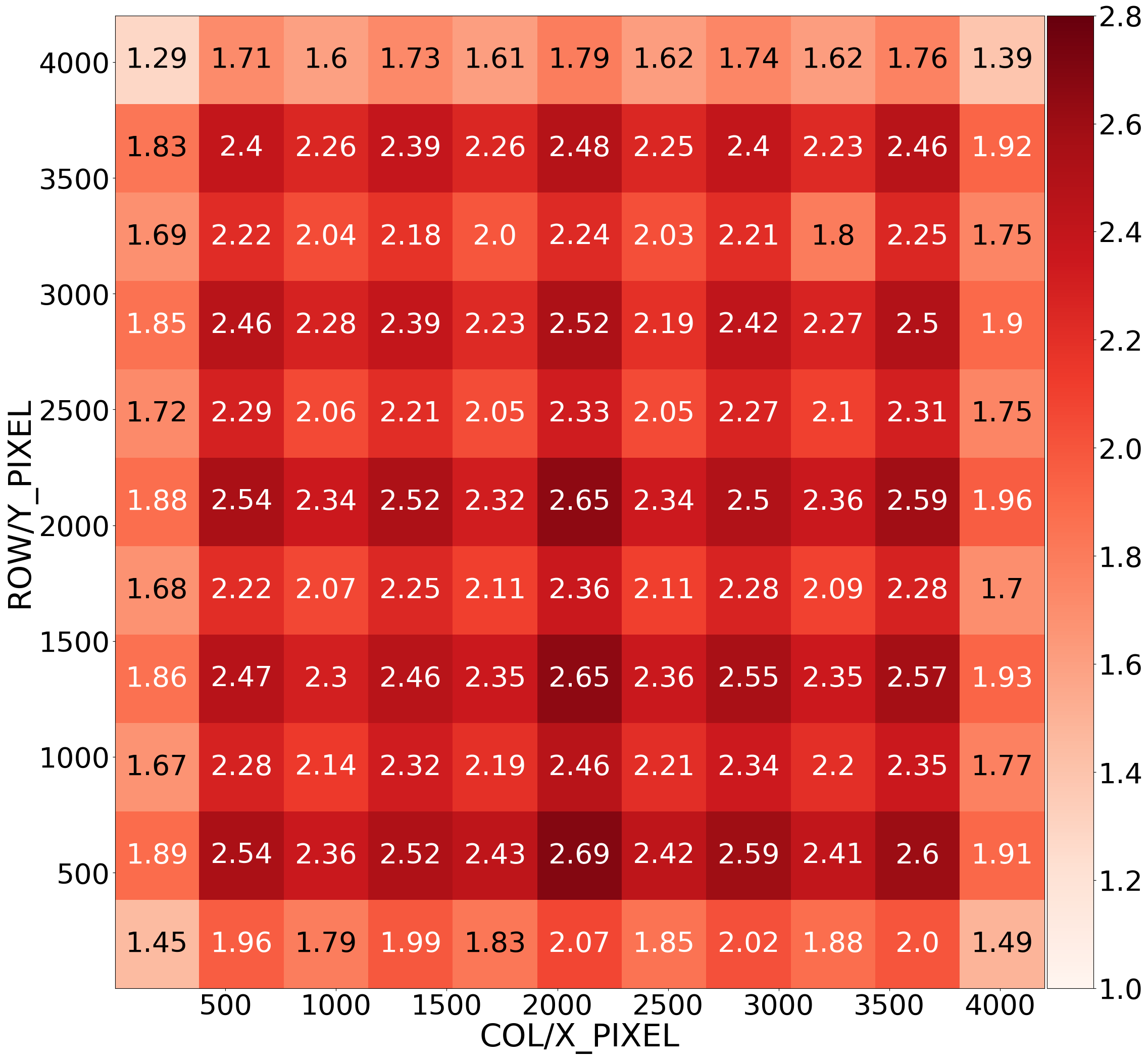}
    \caption{Same as Figure \ref{fig:fm1_effarea_distribution}, but for CMOS 17--20.}
    \label{fig:fm5_effarea_distribution}
\end{figure}

\begin{figure*}[!htbp]
    \centering
    \includegraphics[width=0.48\textwidth]{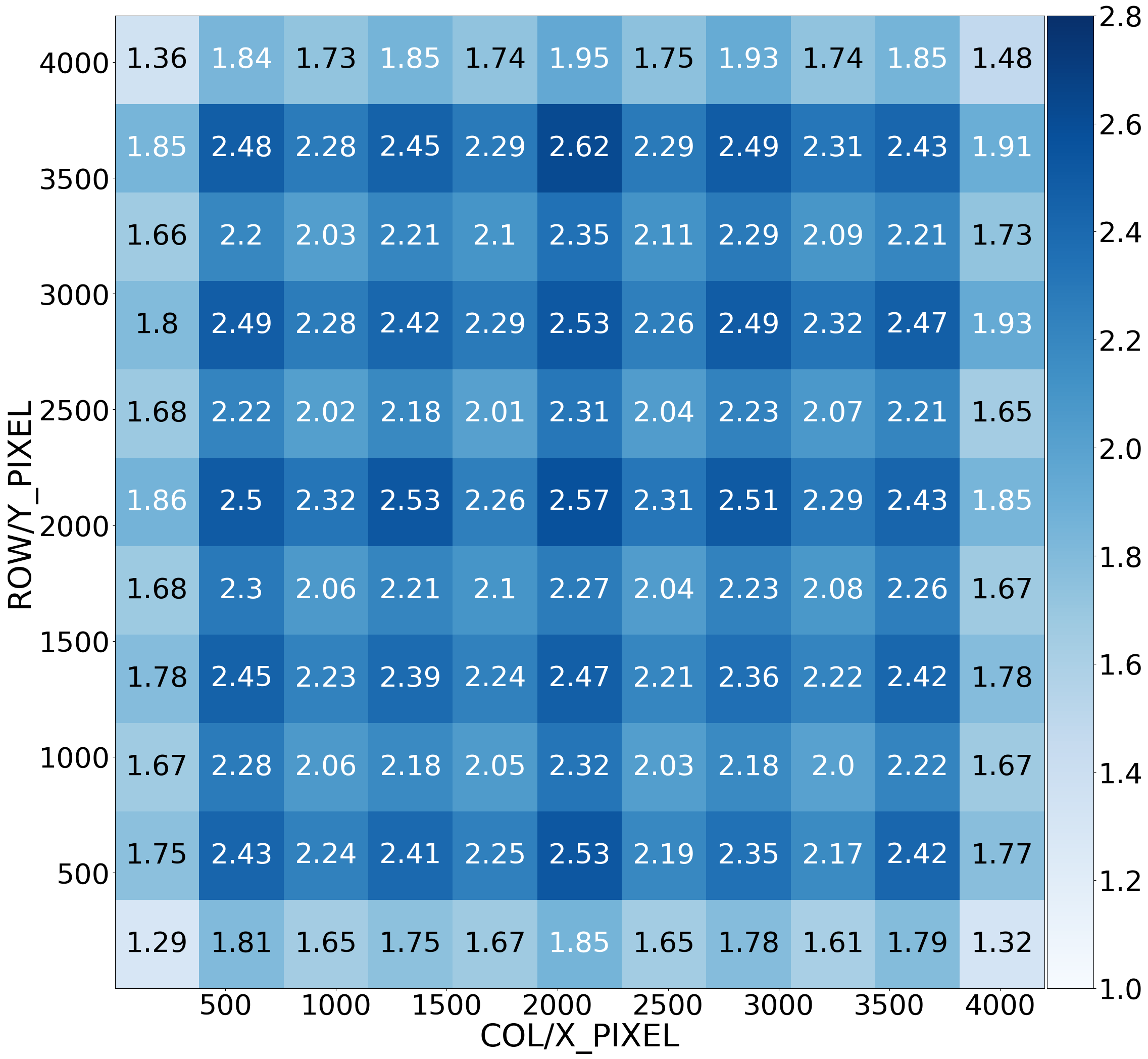}
    \includegraphics[width=0.48\textwidth]{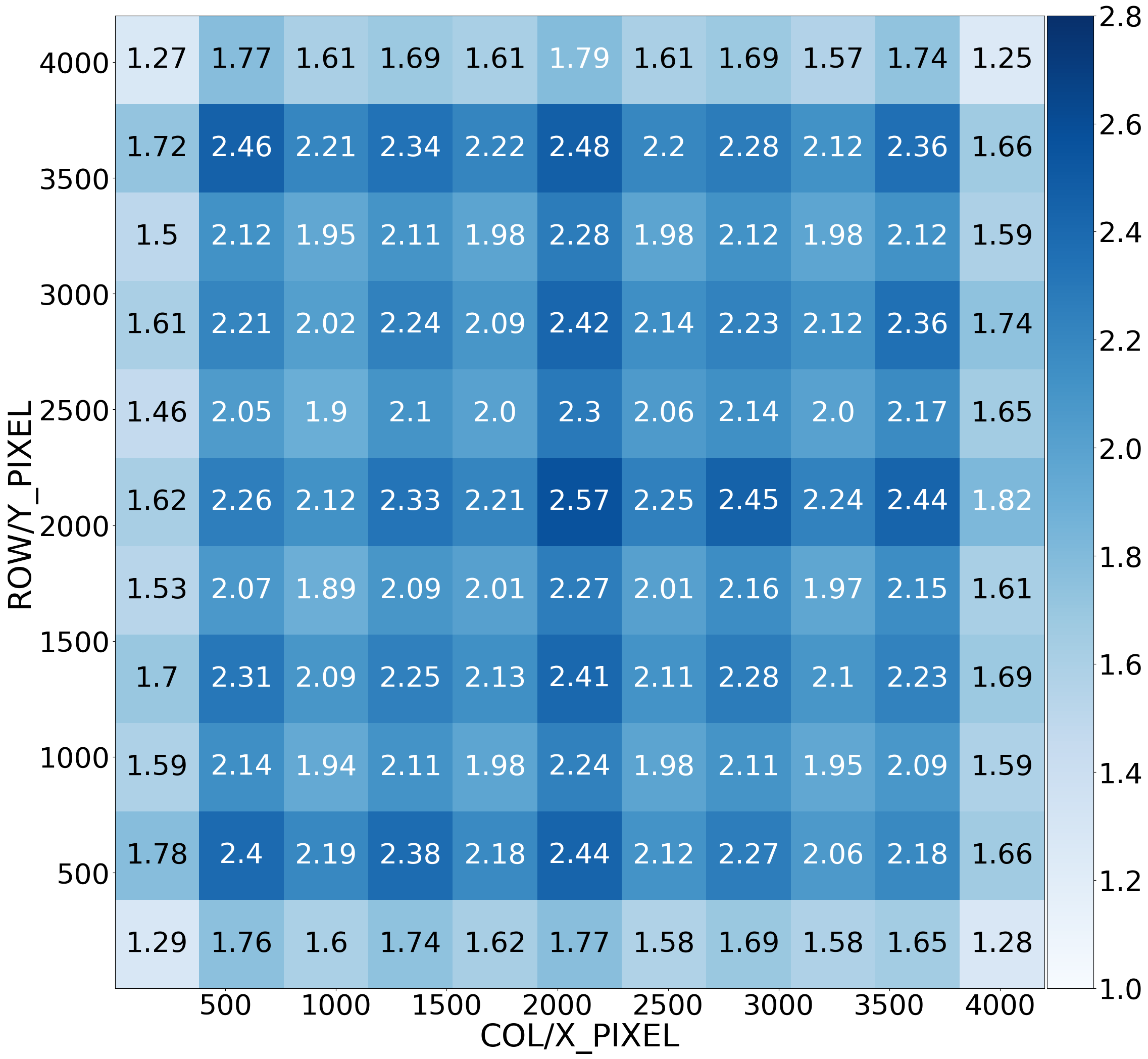}
    \includegraphics[width=0.48\textwidth]{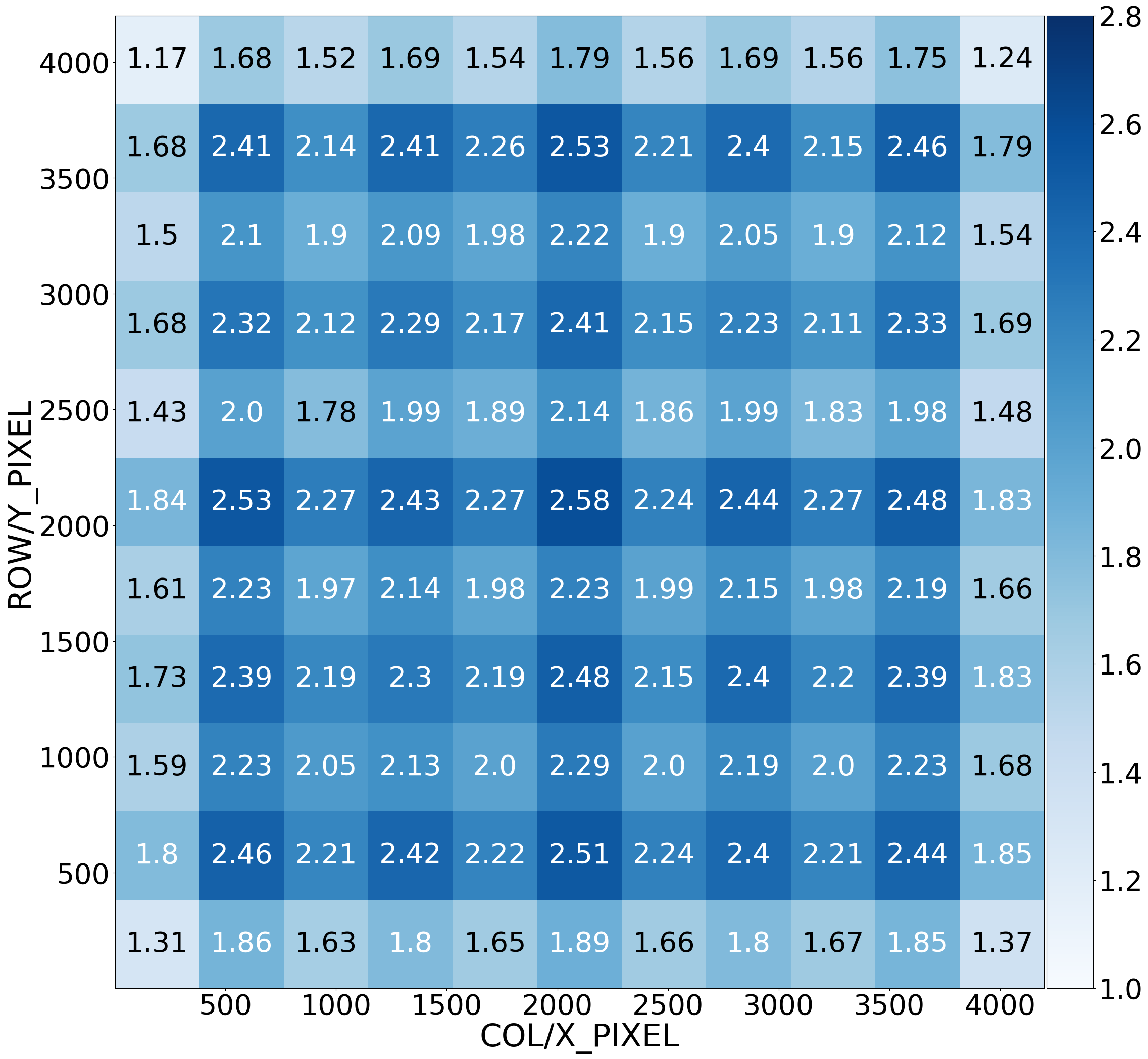}
    \includegraphics[width=0.48\textwidth]{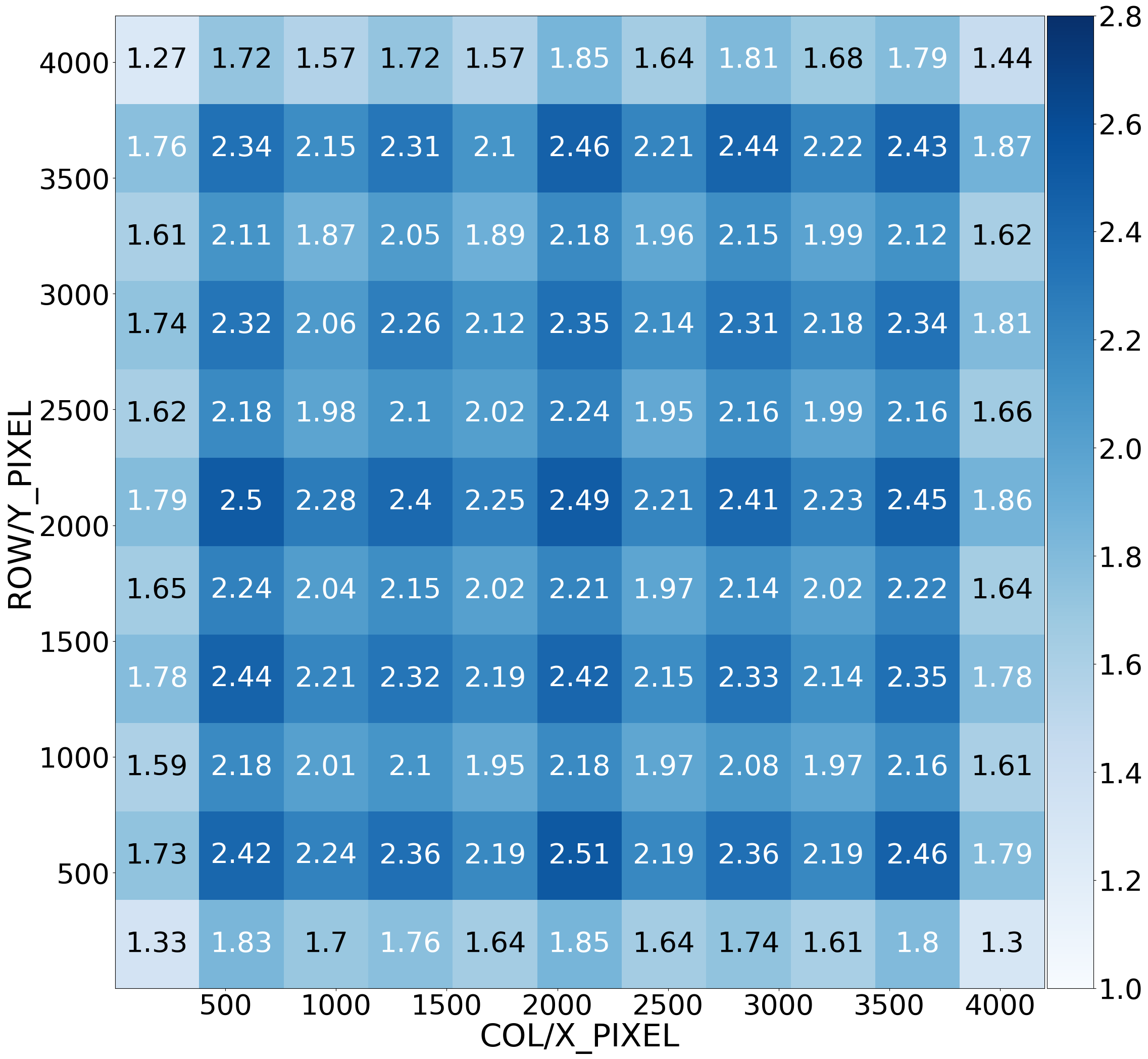}
    \caption{Same as Figure \ref{fig:fm1_effarea_distribution}, but for CMOS 41--44.}
    \label{fig:fm11_effarea_distribution}
\end{figure*}

The effective areas were measured at a wide range of incident directions and energies of various X-ray emission lines. 
Noted that the effective area here represents the total effective area of the module ($A_{\rm eff}$), i.e. a product of the effective area of the MPO optics and quantum efficiency of the CMOS detector ($A_{\rm eff} \equiv A_{\rm MA}\times~QE_{\rm CMOS}$).
During the calibration experiment, the beam flux --serving as the absolute flux reference-- was monitored in real time using a standard SDD of which the quantum efficiency has been calibrated.
At a given energy, the effective area ($A_{\rm eff}$) is defined as the ratio of the measured measured count rate ($\rm cts~s^{-1}$) by the detectors ($C_{\rm det}$) to the photon flux ($\rm cts~s^{-1}~{cm}^{-2}$) of the incident X-ray beam reaching the optics ($F_{\rm MA}$),
\begin{equation}
\label{eq:effarea}
A_{\rm eff} = \frac{C_{\rm det}}{F_{\rm MA}} 
\end{equation},
where $F_{\rm MA}$ can be derived from the count rate ($\rm cts~s^{-1}$) of the reference SDD detector ($C_{\rm SDD}$) which is placed in front of the MA, as
\begin{equation}
\label{eq:fmpo}
    F_{\rm MA}=\frac{C_{\rm SDD}}{A_{\rm SDD}\times QE_{\rm SDD}}\times G
\end{equation}
where $A_{\rm SDD}=~0.17~{\rm cm^2}$ is the collecting area and $QE_{\rm SDD}$ the quantum efficiency of the SDD. 
For the divergent X-ray beam, the geometrical correction factor $G$ is defined as 
\begin{equation}
\label{eq:geom_coeff}
    G=(d_{\rm source-SDD}/{d_{\rm source-MA}})^2
\end{equation}
where $d_{\rm source-SDD}$ denotes the distance from source to SDD and $d_{\rm source-MA}$ represents the distance from source to MA. During the whole experiment, $G=0.943$. 
The uncertainty in $A_{\rm eff}$ incorporates both the statistical fluctuations in the count rates and the systematic uncertainty in $QE_{\rm SDD}$, with the latter being dominant. To analyze the effective area, the photon extraction region for the focal spot is determined by a circular aperture centered on the PSF's barycenter, with a radius of 1.25 mm (equivalent to 11.5 arcmin). This aperture is chosen to ensure that the majority of photons originate from the focal spot, with little contamination from the cruciform arms. 

As an example, in Figs \ref{fig:effarea_energy_curve_fm1}, \ref{fig:effarea_energy_curve_fm5} and \ref{fig:effarea_energy_curve_fm11}, we present the effective areas for the central focal spot region, measured along the central direction of the twelve CMOS detectors. 
Specifically, the effective area is $\lesssim1.5~{\rm cm^2}$ at $0.53$ keV, $\sim3~{\rm cm^2}$ (the peak value) at $0.93$ keV, $\sim2.5~{\rm cm^2}$ at $1.25$ keV, $\sim1.2~{\rm cm^2}$ at $1.74$ keV, $\sim1~{\rm cm^2}$ at $2.98$ keV and $\sim0.5~{\rm cm^2}$ at $4.5$ keV, with some slight variations among different modules and detectors.
The modeled effective area based on Monte Carlo simulations \cite{2014SPIE.9144E..4EZ, 2017ZhaoDH_simulation} is also plotted for comparison. 
Overall, the measured effective areas at various energies align well with the simulated theoretical values (we note that this consistency is also maintained at other sampled directions), demonstrating the validity of applying the theoretical effective area curve to observational data analysis.

In theory, the effective areas for an ideal lobster-eye telescope should be the same for essentially all directions (except for edges where the focusing efficiency decreases due to the incompleteness of the optics), resulting in a negligible vignetting effect within the FoV \cite{1979ApJ...233..364A}.
We investigate the uniformity of the effective area with the experiment using Mg target. In Figs. \ref{fig:fm1_effarea_distribution}, \ref{fig:fm5_effarea_distribution} and \ref{fig:fm11_effarea_distribution}, we present the distribution of the effective area within the FoV of each module at $1.25$ keV. The main results include: 
(1) For all the CMOS sensors, the effective area is basically uniform across the FoV ranging from $2-3~{\rm cm^2}$, except for the edges where it decreases to smaller value of $\sim1-2~{\rm cm^2}$. This is generally consistent with theoretical predictions.
(2) Aside from the overall uniformity, the effective area also shows a pattern of mild variations. Specifically, in some of the sampled directions, the effective area decreases to below $\sim2~{\rm cm^2}$, due primarily to the blocking of the incident X-ray photons by the mounting frame between the individual MPO plates. 
In practice, the directions featured with relatively larger effective area values belong mostly to the `chip center' array and `chip edge' array.
In the former case, the direction along the center of each MPO chip is almost free from obscuring effect of the MPO mounting frame. 
As for the `chip edge’ locations, the focusing efficiency is also enhanced with contributions from the two/four MPO plates around the mounting frame.
(3) While the distributing patterns show a large degree of similarity among different modules, distinctions are also present regarding the absolute values of the effective area. This is mainly due to the intrinsic disparities in the imaging characteristics of discrete optical assemblies.

We note that the measured effective area of WXT instrument is very close to that of \textit{LEIA} (see Figures 9 and 10 in \cite{Cheng2024}). 
This alignment arises primarily from the fact the effective area is determined mainly by the fraction of photons enclosed within the aperture chosen for photon extraction. 
Despite of WXT's superior photon concentration (as indicated by smaller values of FWHM, see Fig. \ref{fig:psf_fitting_cdf} in Section \ref{sec:psf_dif_direction}), the physical extent of the focal spot is comparable for both instruments. 
The usage of the 1.25 mm aperture ensures that nearly all photons of the focal spot are included, resulting in similar measures of the effective area at all energies.

In summary, the effective area is found to be strongly energy-dependent, with a peak value of $\sim3~{\rm cm^2}$ at $\sim1$ keV. On the other hand, the effective area is generally direction-independent as theoretically predicted, though mild variations also exist as a result of the blocking effect of the mounting frame. 
These results are very well consistent with those of the \textit{LEIA} instrument, and can be well reproduced by Monte Carlo simulations.

\subsection{Performance of the CMOS detectors}
\label{sec:cmos_sensor}

\begin{figure*}[htbp]
\centering
\includegraphics[width=0.48\textwidth]{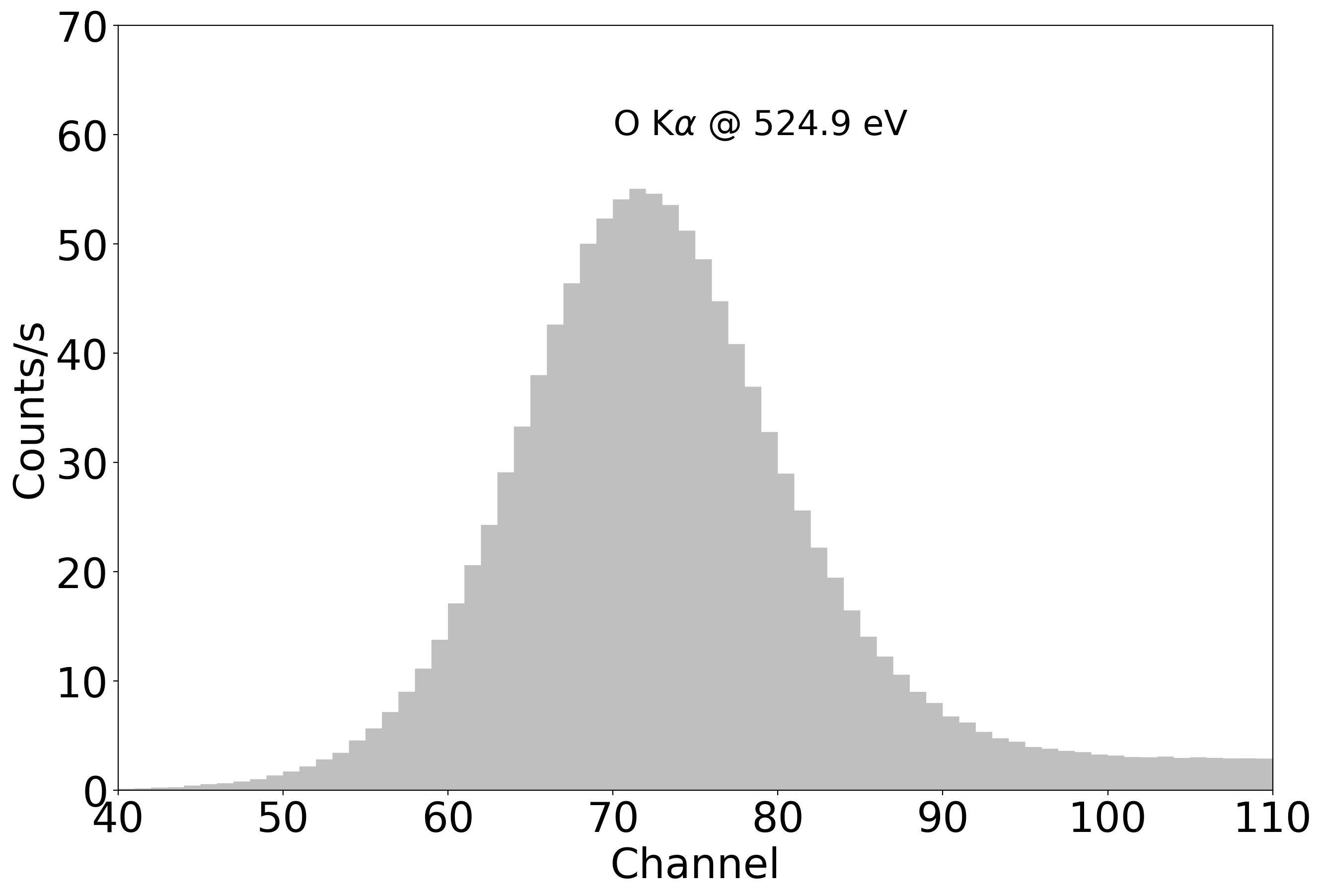}
\includegraphics[width=0.48\textwidth]{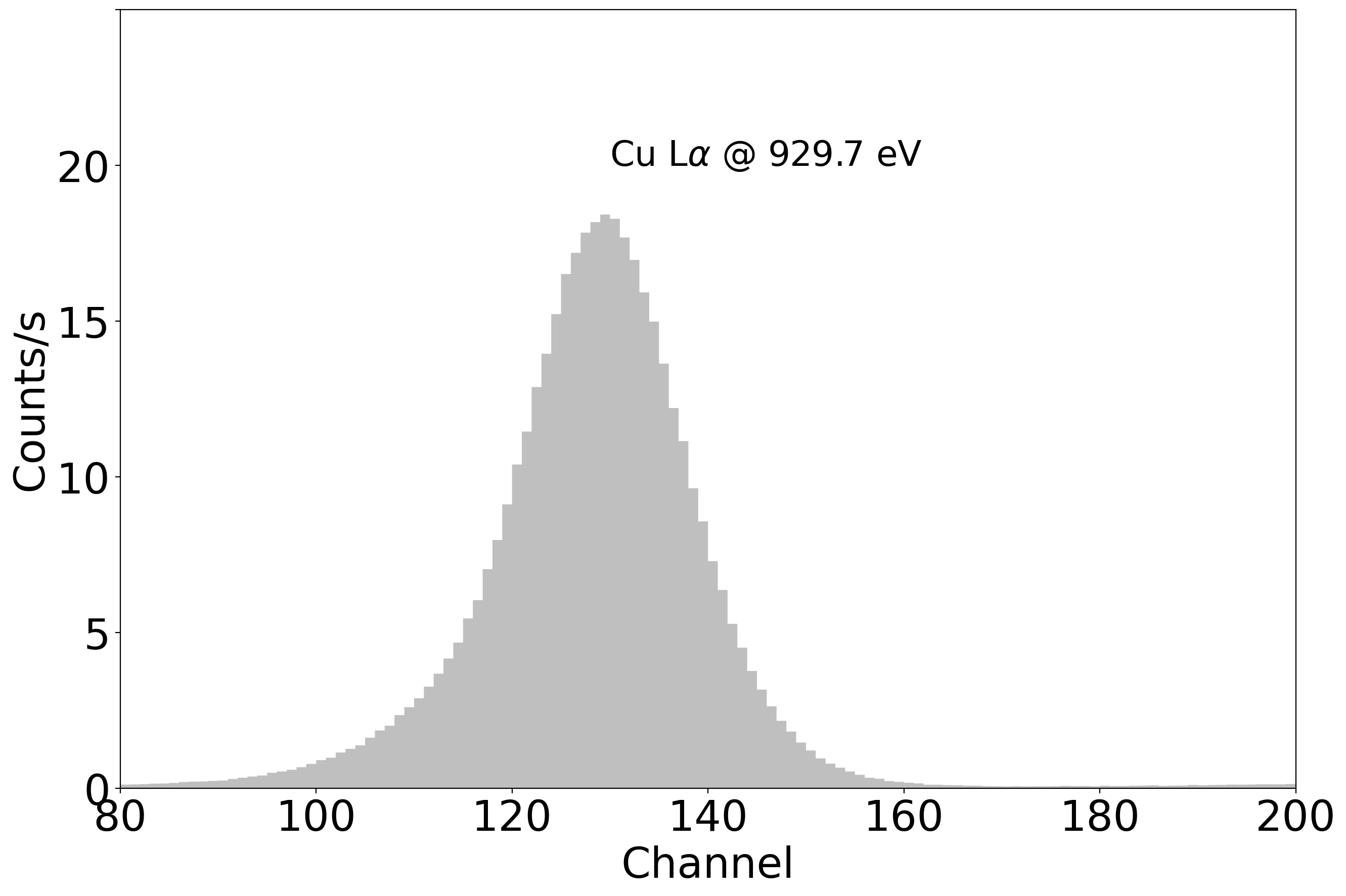}
\includegraphics[width=0.48\textwidth]{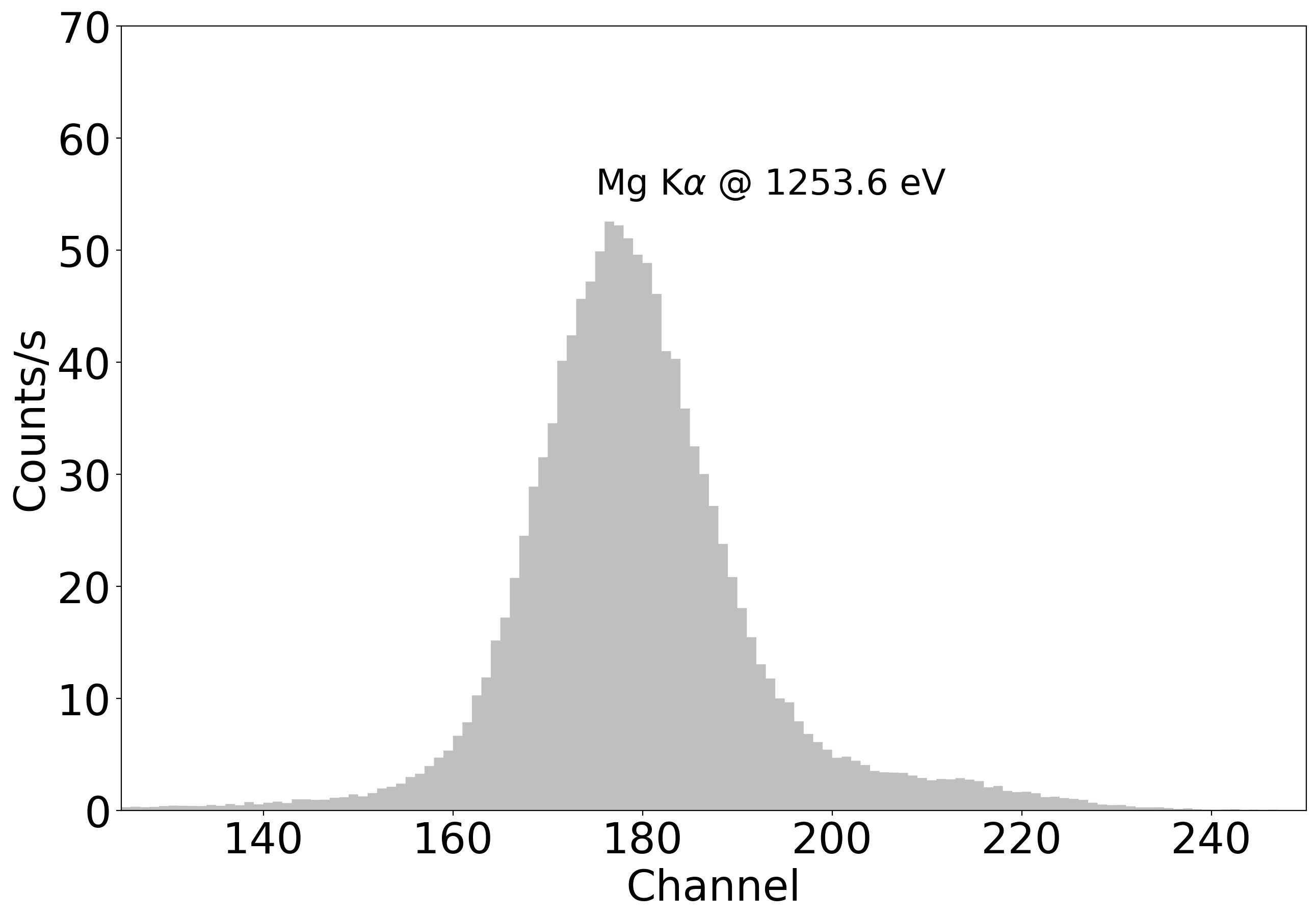}
\includegraphics[width=0.48\textwidth]{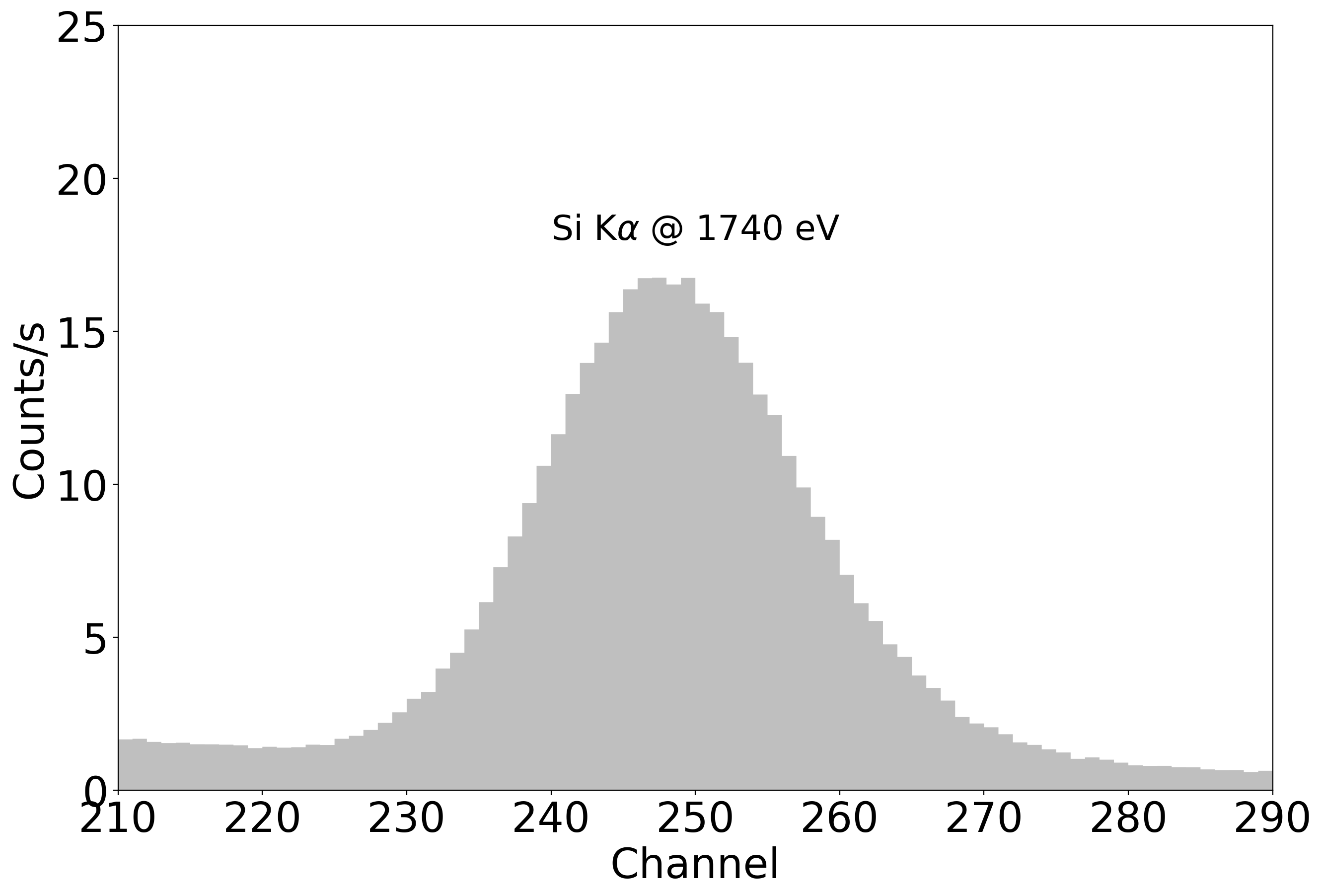}
\includegraphics[width=0.48\textwidth]{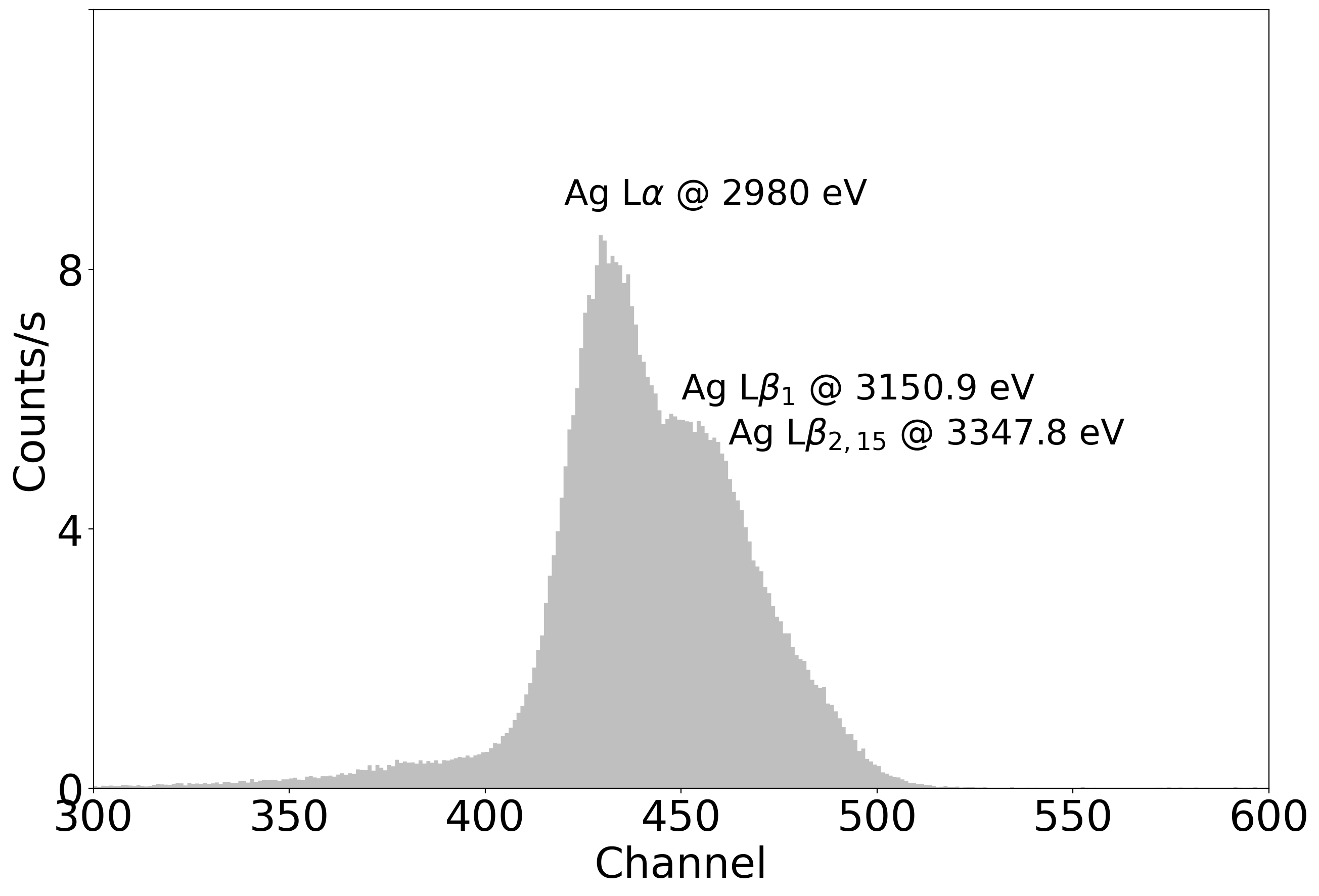}
\includegraphics[width=0.48\textwidth]{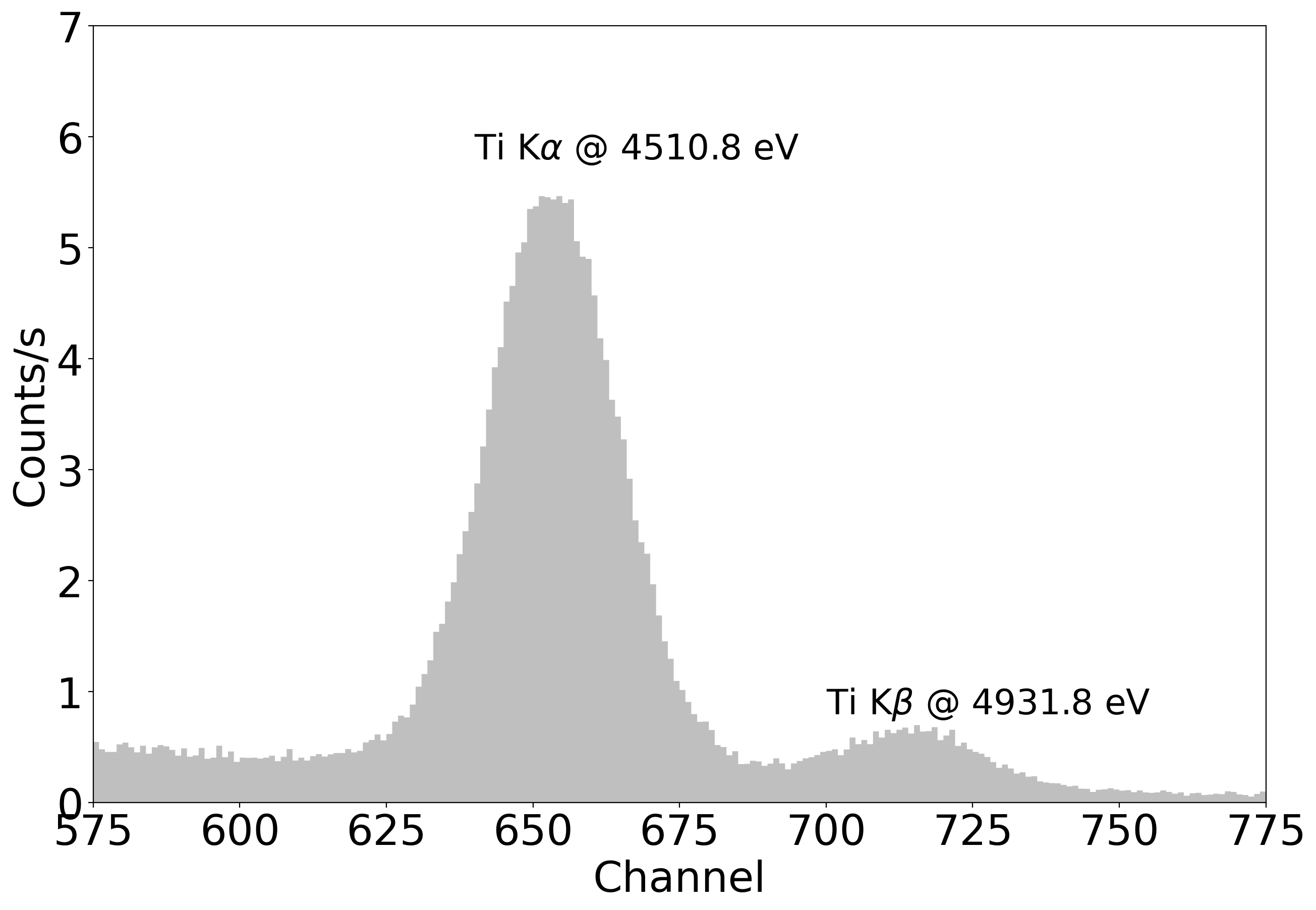}
\caption{\label{fig:cmos_spectra}
The X-ray spectra measured at the center of CMOS 4 onboard FM1 using different targets. The characteristic X-ray emissions lines are marked with theoretical line energies.}
\end{figure*}

The 12 CMOS sensors onboard FM1, FM5 and FM11, along with other 36 sensors were firstly tested at NAOC for defects (bad pixels, bad columns, bad rows and bad clusters), readout noise, bias, Energy-Channel (EC) relation and energy response under different temperatures. These results have been summarized and will be presented in a forthcoming paper (Ling et al. in preparation). After these comprehensive tests at NAOC, the CMOS sensors were shipped to SITP, integrated into detector assemblies and mounted at the focal plane of each MA.
During the calibration campaign at 100XF, the performance of the 12 CMOS sensors were calibrated under the in-flight working condition. We note that the calibration for the energy response of the four CMOS detectors onboard FM11 (CMOS 41--44) is somehow limited due primarily to the insufficient number of targets being used. The properties of other eight detectors (CMOS 1--4, CMOS 17--20) were thoroughly investigated, as detailed below.

\begin{figure*}
    \centering
    \includegraphics[width=0.45\textwidth]{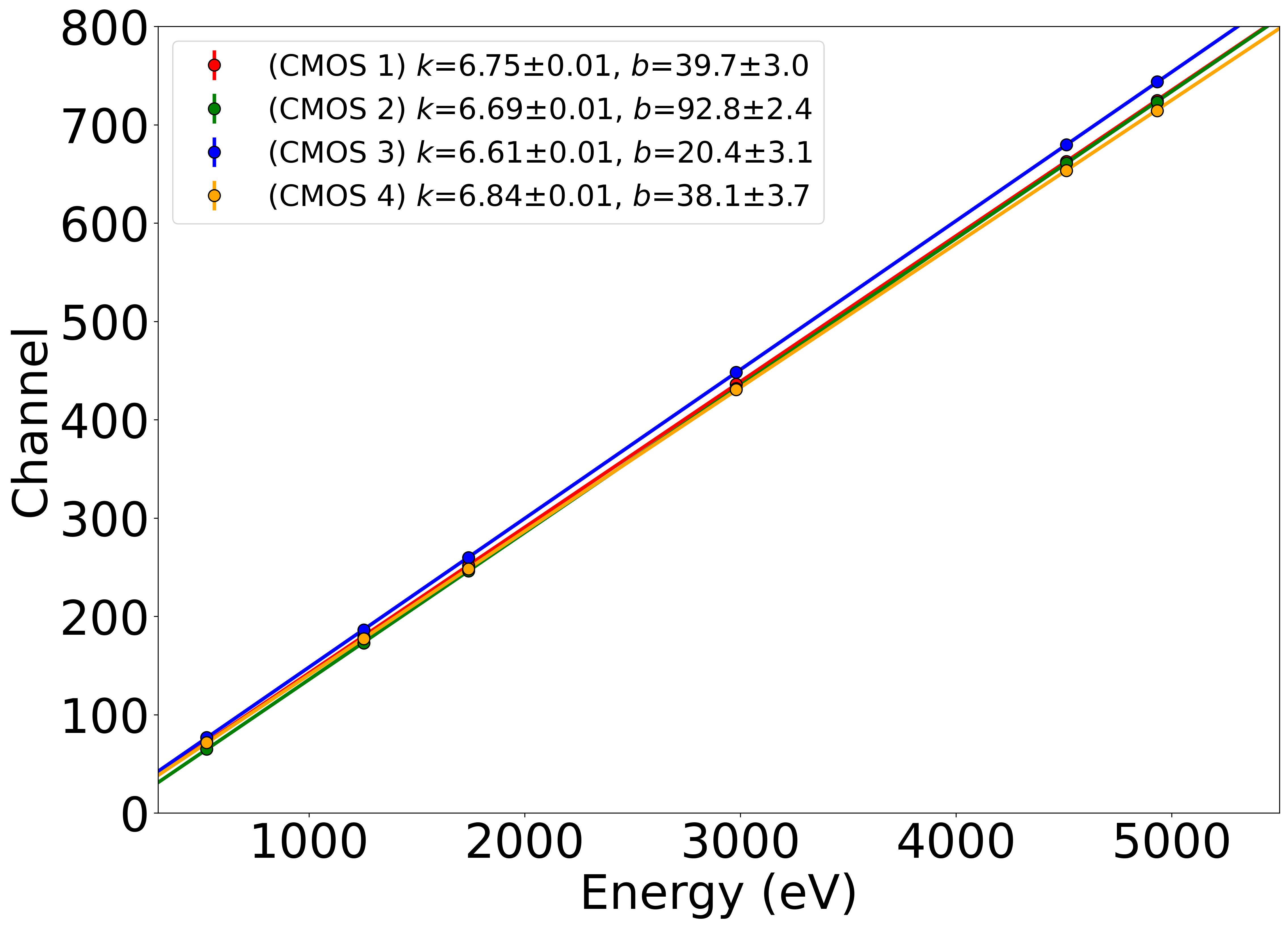}
    \includegraphics[width=0.45\textwidth]{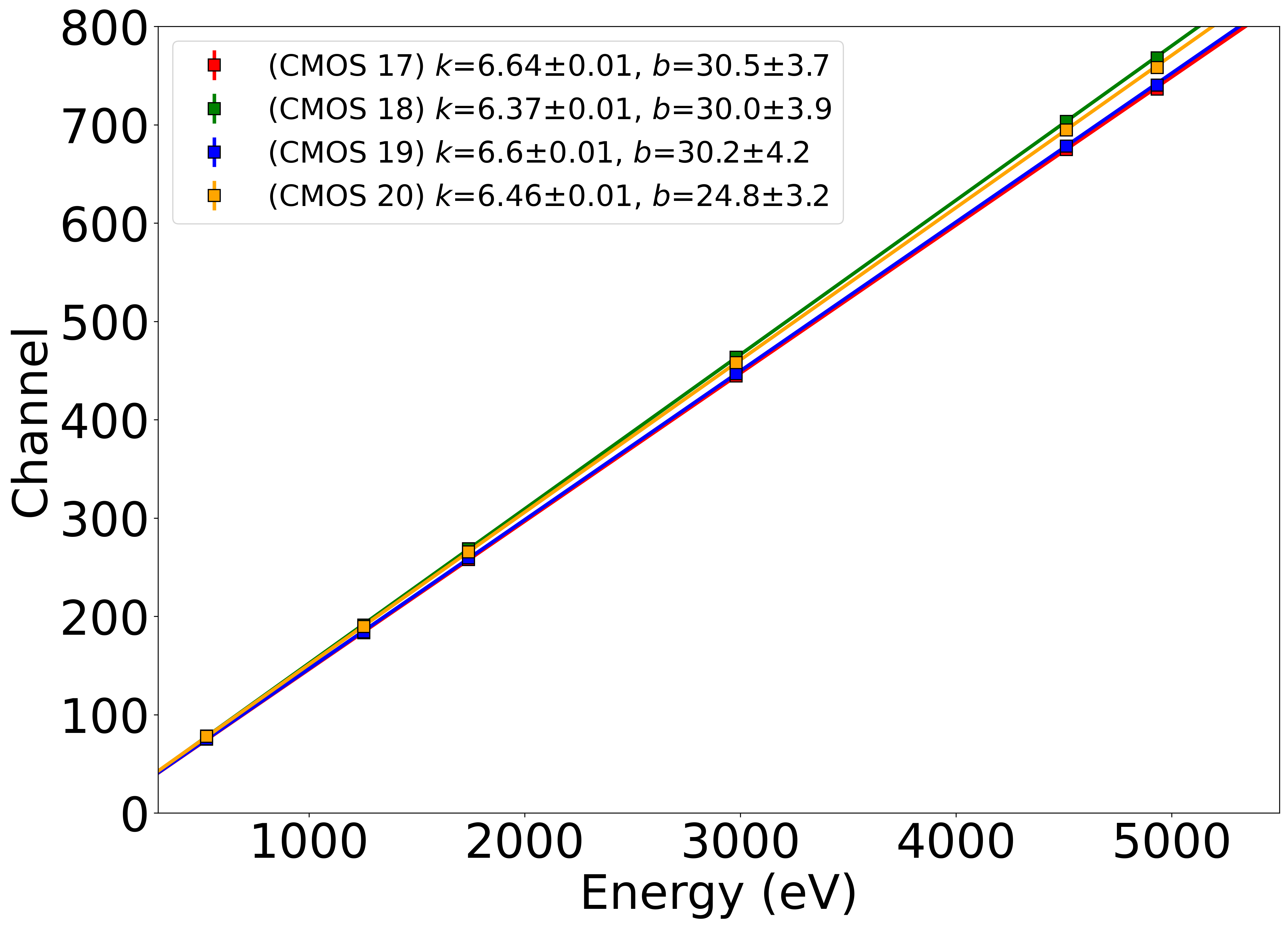}    
    \caption{EC relation measured in the direction along the center of the eight CMOS sensors aboard FM1 ({\em Left} panel) and FM5 ({\em Right} panel). Different colors and symbols refer to different sensors.}
    \label{fig:cmos_gain}
\end{figure*}

With the five (for FM1) and six (for FM5) targets employed in the experiment (see Table \ref{tab:epwxt_schedule}), the CMOS response spectra with a total of seven characteristic X-ray emission lines are obtained at various incident angles.
As an example, in Fig. \ref{fig:cmos_spectra} we present the measured X-ray spectra for these emission lines, obtained at the center of CMOS 4 aboard FM1.
The results for other directions and CMOS sensors are similar.
For each of the emission line, a Gaussian fit is performed in order to obtain the central channel (in PHA unit) as well as the dispersion of the line. Note that for Mg K$\alpha$ line an additional Gaussian component is introduced to fit the weak Aluminum K$\alpha$ line ($1.49$ keV) induced by the Al filter. On the other hand, we adopt three Gaussian components when fitting the Ag L triplets (Ag L$\alpha$, Ag L$\beta_1$ and Ag L$\beta_{2,15}$).
The Cu L line is not included in our subsequent analysis, due primarily to its asymmetric profile potentially induced by an impurity of the target material which is difficult to be precisely quantified.
With the PHA channels and the energies of the six emission lines (O K$\alpha$, Mg K$\alpha$, Si K$\alpha$, Ag L$\alpha$, Ti K$\alpha$ and Ti K$\beta$), we establish the relation between the energy ($E$) and PHA channel ($C$) by fitting the two variables with a linear function
\begin{equation}
    \label{eq:gain}
    E = k\times C + b,
\end{equation}
where $k$ signifies the gain coefficient and $b$ is a constant.
For all the eight CMOS detectors, a strong linear correlation between the energy and PHA channel is found, as shown in Fig. \ref{fig:cmos_gain}.
By comparing the gain values at different locations of the detector, we observe a very slight non-uniformity of less than $1$ percent.
The gain coefficients for the eight detectors are found to be in range of $6.4-6.9$ eV/DN. 
We note that the measured gain coefficients align well with those obtained at NAOC (Ling et al. in preparation).
This indicates that the energy response of the CMOS detectors remains stable after the integration of the module.

With the FWHM (in units of DN) of the Gaussian fit, and the energy-independent gain value, we may derive the spectral resolution (in unit of eV) at different energies of X-ray emission lines.
It is found that, except for CMOS 18, the spectral resolutions at different energies are consistent among various CMOS sensors. 
Specifically, the resolution is found to be $\sim120$ eV at $0.53$ keV, $\sim130$ eV at $1.25$ keV, $\sim140$ eV at $1.74$ keV, $\sim160$ eV at $2.98$ keV and $\sim180$ eV at $4.51$ keV.
On the other hand, CMOS 18 exhibits a slightly poorer energy resolution than other detectors, with $\sim136\pm5$ eV at $0.53$ keV, $\sim144\pm4$ eV at $1.25$ keV, $\sim153\pm4$ eV at $1.74$ keV, $\sim170\pm4$ eV at $2.98$ keV and $\sim191\pm4$ eV at $4.51$ keV.
The spectral resolution is also found to be largely in-dependent of the CMOS position, with a slight fluctuation of $\lesssim5$ percent.

\begin{figure*}
    \centering
    \includegraphics[width=0.45\textwidth]{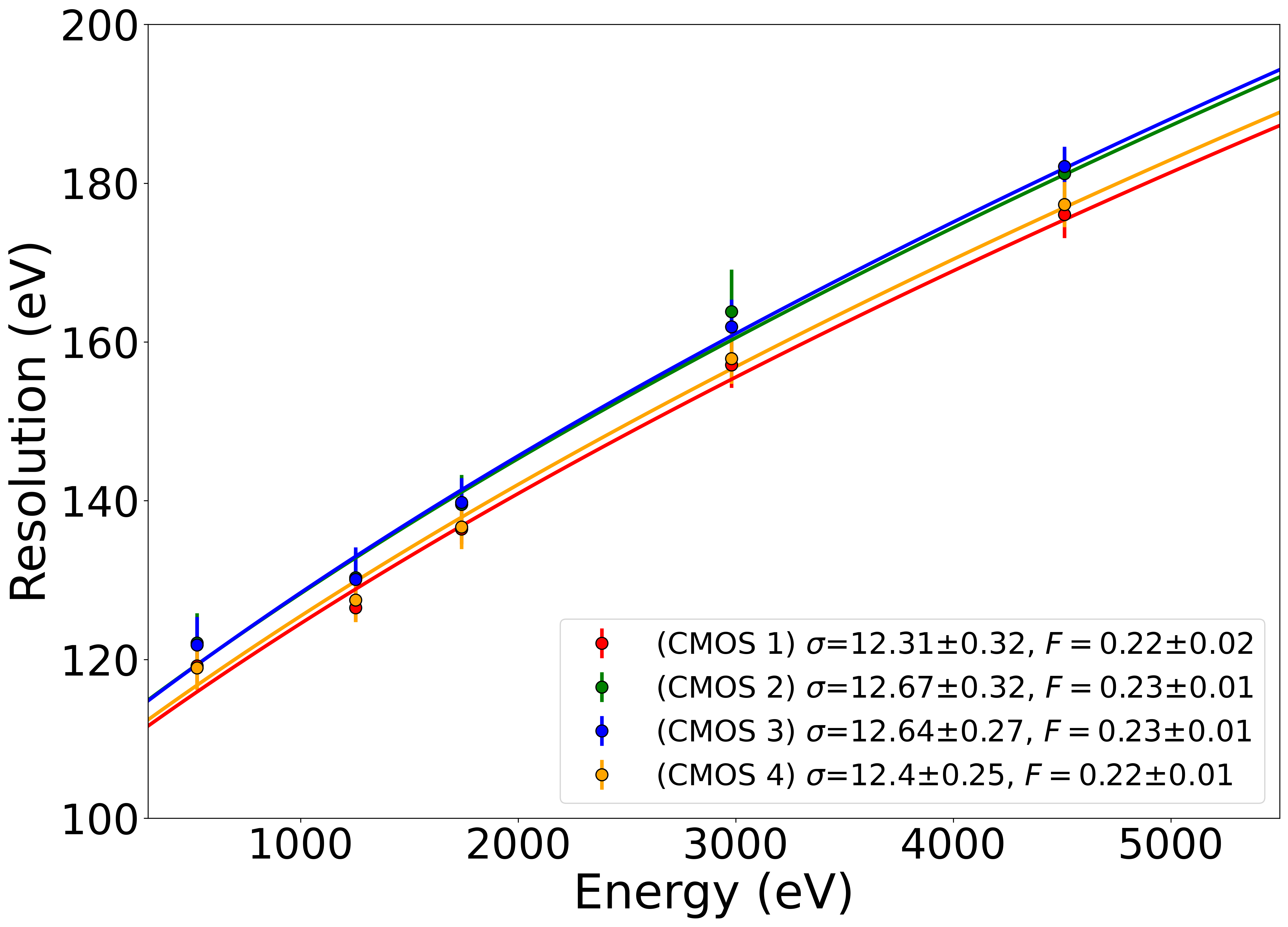}
    \includegraphics[width=0.45\textwidth]{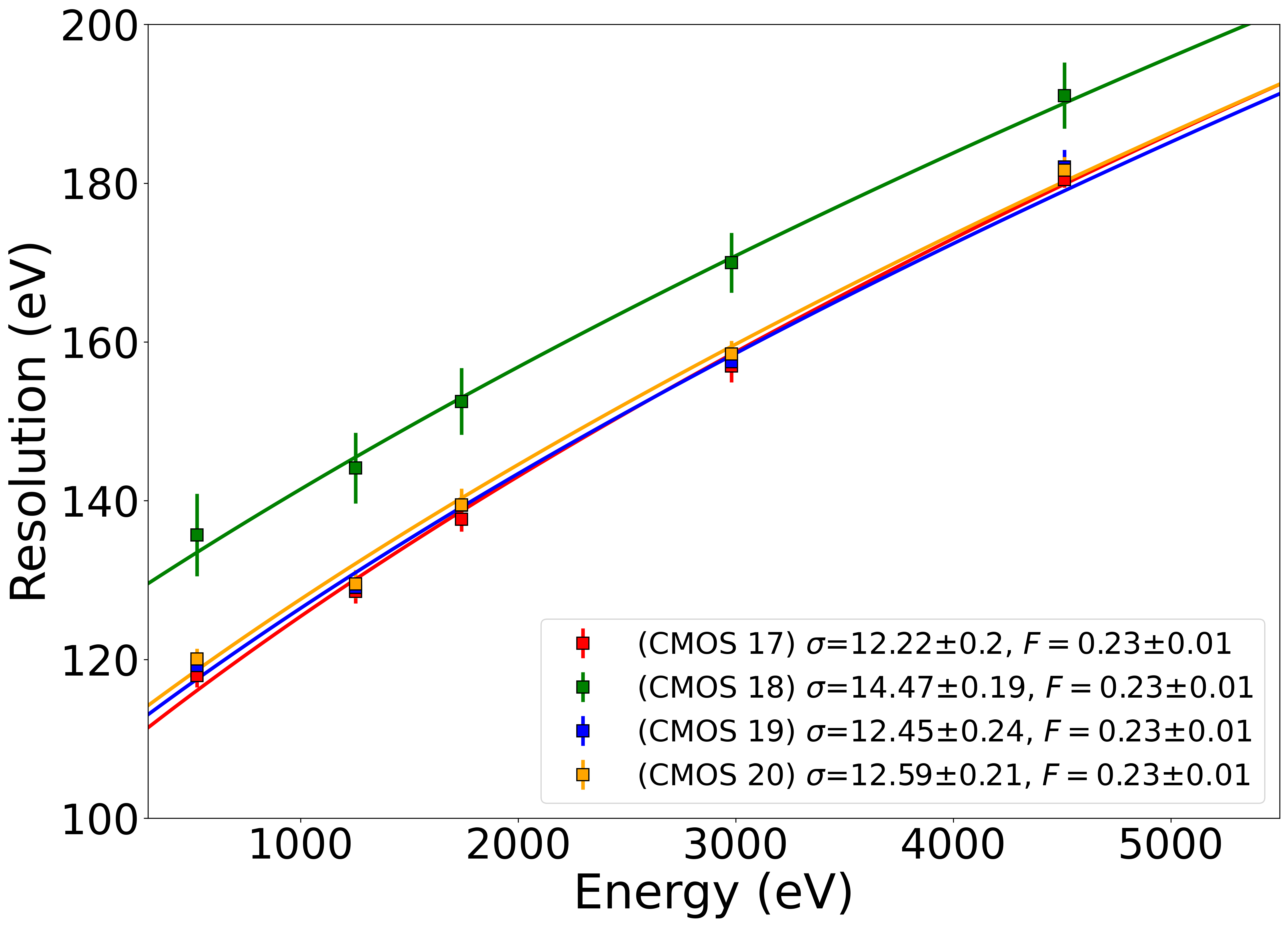}    
    \caption{Energy resolution as a function of the photon energy for the eight CMOS detectors aboard FM1 ({\em Left} panel) and FM5 ({\em Right} panel). Different detectors are denoted by different colors and symbols. The resolution measurements are taken from the mean and standard deviation of the whole detector. The best-fit relations of Equation \ref{eq:resolution} for different detectors are plotted as solid lines.}
    \label{fig:cmos_resolution_vs_energy}
\end{figure*}

The relation between the energy resolution ($FWHM_{\rm E}$, in units of eV) and photon energy (in units of eV) can be formulated as \cite{Holland2013}
\begin{equation}
    \label{eq:resolution}
    FWHM_{\rm E} = 2.35~\omega~(\sigma^2+FE/\omega)^{1/2}
\end{equation}
where $E$ represents the photon energy, $\omega$ represents the average ionization energy and is set to $3.65$, $F$ represents the Fano factor \citep{Fano1947} and $\sigma$ represents the equivalent noise charge.
We fit the Equation \ref{eq:resolution} to the data for each CMOS detector (see Fig. \ref{fig:cmos_resolution_vs_energy}).
The equivalent noise charge is in range of $\sim12-13$ for seven CMOS sensors except for CMOS 8, which has a larger value of $\sim14.5$. 
The Fano factor is found to be $\sim0.2$.

By analyzing the bias map and dark images (photon events without X-ray illumination), we obtained other properties of the CMOS detectors, including the readout noise and bad pixels (including the single bad pixels, bad columns and bad rows, clusters of bad pixels). 
Specifically, the readout noises of the twelve detectors are about $\sim3-4~{\rm e^{-}}$. 
The bad pixels are also marked and excluded from being triggered onboard and analyzed in on-ground data pipeline.
No significant large-scale spatial clustering of bad pixels was observed for these detectors.
In Table \ref{tab:cmos} we summarize the specifications of the 12 CMOS sensors. 

We note that the energy response properties of the CMOS sensors onboard WXT flight model modules show a close proximity to those aboard \textit{LEIA}, indicating their nearly equivalent capability in distinguishing photons of different energies. A detailed summary on the performance of all 48 CMOS sensors onboard the \textit{EP} satellite, including their energy response, defects, backgrounds and thermally induced variations will be presented in a forthcoming paper (Ling et al. in preparation).

\begin{table}
\centering
\caption{Characteristics of the twelve CMOS sensors aboard FM1, FM5 and FM11.}
\begin{tabular}{ccccc}
\toprule
Instrument & &  FM1 & & \\ \hline
 & CMOS 1 & CMOS 2 & CMOS 3 & CMOS 4 \\ \hline
Gain & $6.71\pm0.04$ & $6.68\pm0.02$ & $6.61\pm0.02$ & $6.86\pm0.03$ \\
Gain (NAOC) & $6.65\pm0.02$ & $6.66\pm0.02$ & $6.59\pm0.02$ & $6.81\pm0.02$ \\
Energy resolution /eV (0.53 keV) & $119 \pm 2$ & $122 \pm 4$ & $122 \pm 4$ & $119 \pm 3$ \\
Energy resolution /eV (1.25 keV) & $127 \pm 2$ & $130 \pm 3$ & $130 \pm 4$ & $127 \pm 3$  \\
Energy resolution /eV (1.74 keV) & $136 \pm 2$ & $140 \pm 4$ & $140 \pm 3$ & $137 \pm 3$ \\
Energy resolution /eV (2.98 keV) & $157 \pm 3$ & $164 \pm 5$ & $162 \pm 3$ & $158 \pm 3$ \\
Energy resolution /eV (4.51 keV) & $176 \pm 3$ & $181 \pm 3$ & $182 \pm 2$ & $177 \pm 3$  \\
Readout Noise (e$^{-}$)  & 3.3 & 3.4 & 3.4 & 3.4   \\
\botrule
Instrument & &  FM5 & & \\ \hline
 & CMOS 17 & CMOS 18 & CMOS 19 & CMOS 20 \\ \hline
Gain & $6.65\pm0.03$ & $6.36\pm0.02$ & $6.59\pm0.02$ & $6.45\pm0.01$ \\
Gain (NAOC) & $6.64\pm0.02$ & $6.31\pm0.02$ & $6.57\pm0.02$ & $6.41\pm0.02$ \\
Energy resolution /eV (0.53 keV) & $118 \pm 1$ & $136 \pm 5$ & $119 \pm 1$ & $120 \pm 1$ \\
Energy resolution /eV (1.25 keV) & $129 \pm 3$ & $144 \pm 4$ & $129 \pm 1$ & $129 \pm 2$  \\
Energy resolution /eV (1.74 keV) & $138 \pm 2$ & $153 \pm 4$ & $139 \pm 1$ & $139 \pm 2$ \\
Energy resolution /eV (2.98 keV) & $157 \pm 2$ & $170 \pm 4$ & $158 \pm 1$ & $159 \pm 2$ \\
Energy resolution /eV (4.51 keV) & $180 \pm 1$ & $191 \pm 4$ & $182 \pm 2$ & $182 \pm 2$  \\
Readout Noise (e$^{-}$) & 3.5 & 4.2 & 3.5 &  3.5  \\
\botrule
Instrument & &  FM11 & & \\ \hline
 & CMOS 41 & CMOS 42 & CMOS 43 & CMOS 44 \\ \hline
Gain (NAOC) & $6.60\pm0.02$ & $6.78\pm0.02$ & $6.37\pm0.02$ & $6.62\pm0.02$ \\
Readout Noise (e$^{-}$) &3.3 & 3.6 &3.3 & 3.3   \\
\botrule
\end{tabular}
\label{tab:cmos}
\end{table}

\subsection{Thermal test}
\label{sec:thermal_test_fm1}

\begin{figure*}
    \centering
    \includegraphics[width=0.48\textwidth]{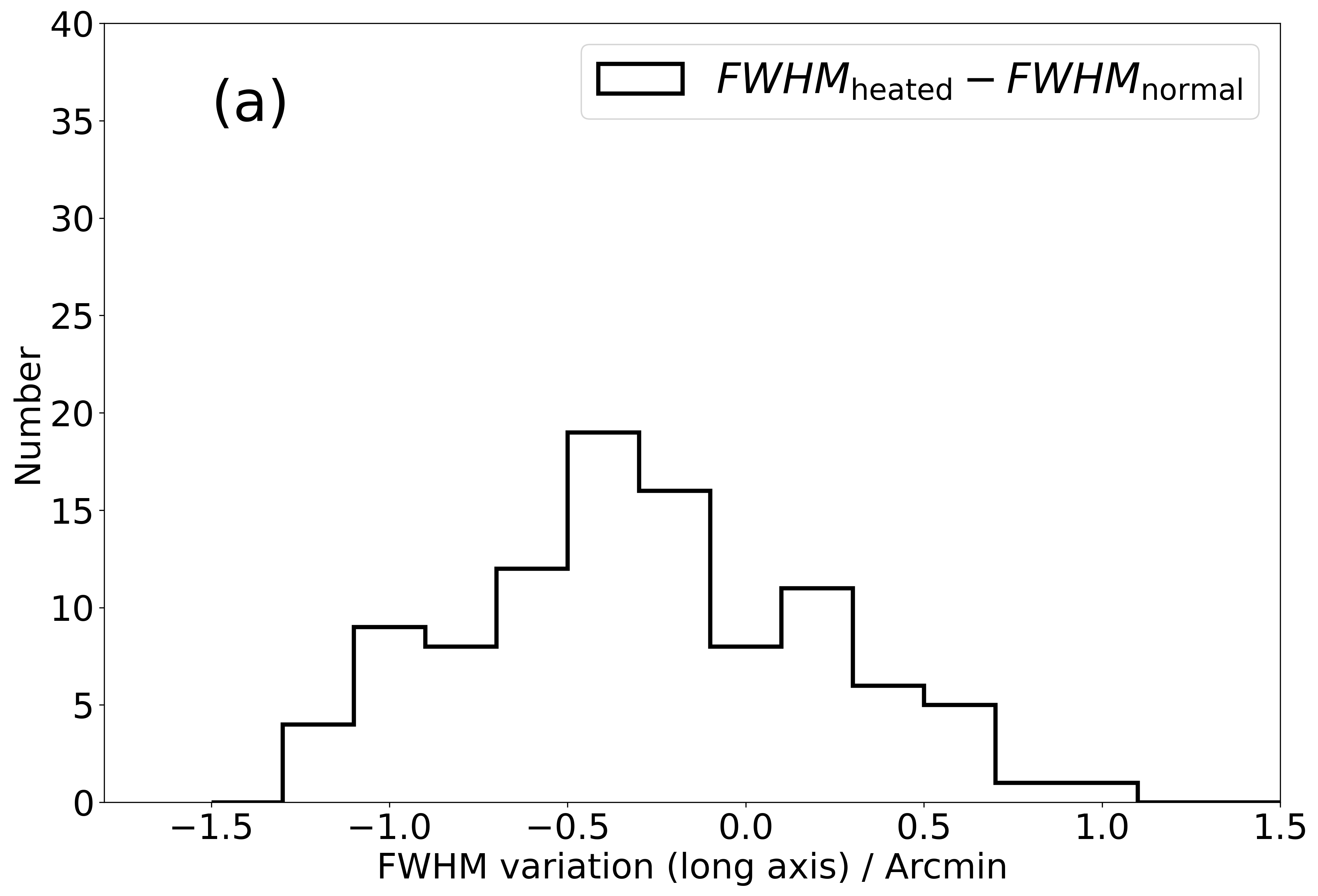}
    \includegraphics[width=0.48\textwidth]{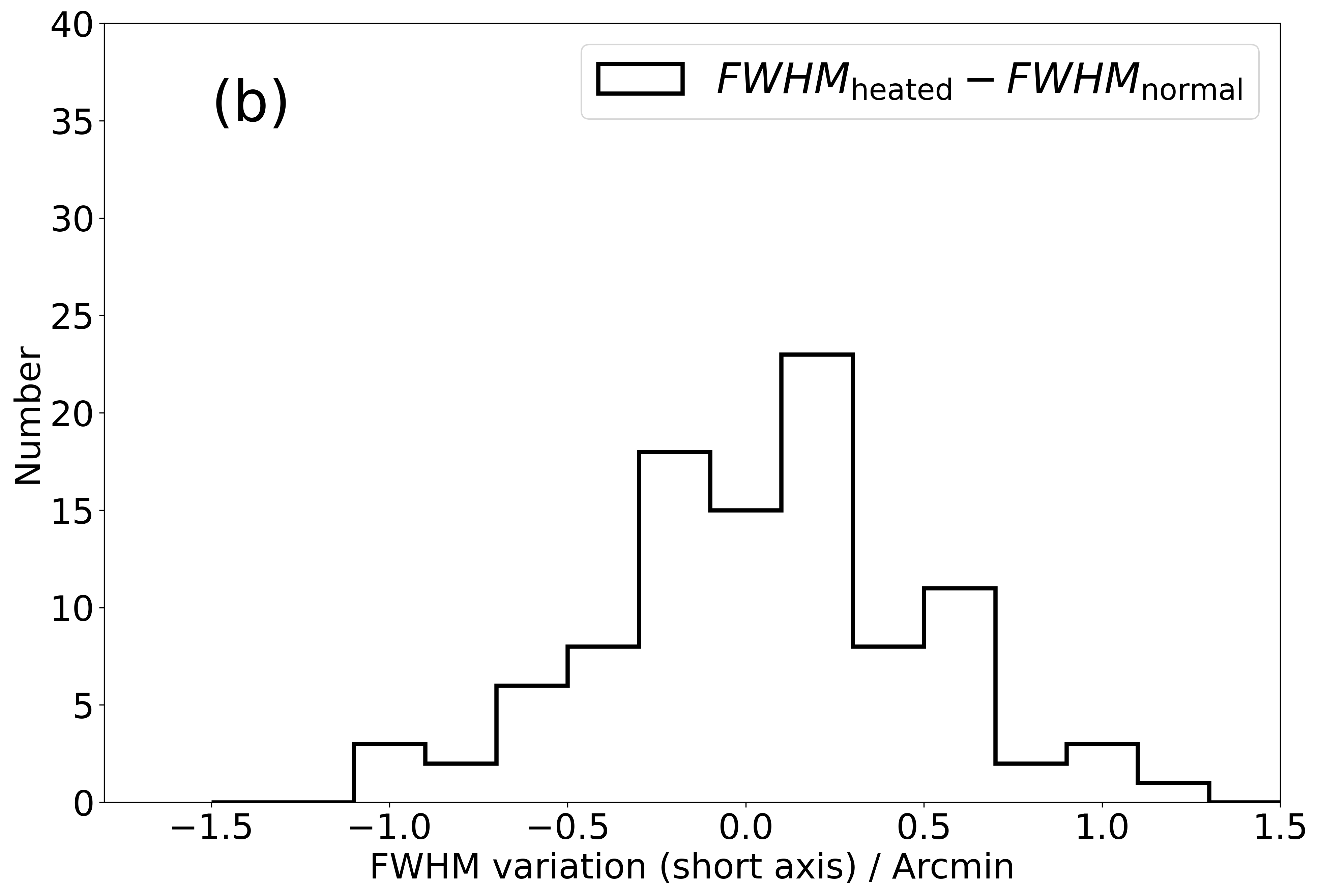}
    \includegraphics[width=0.48\textwidth]{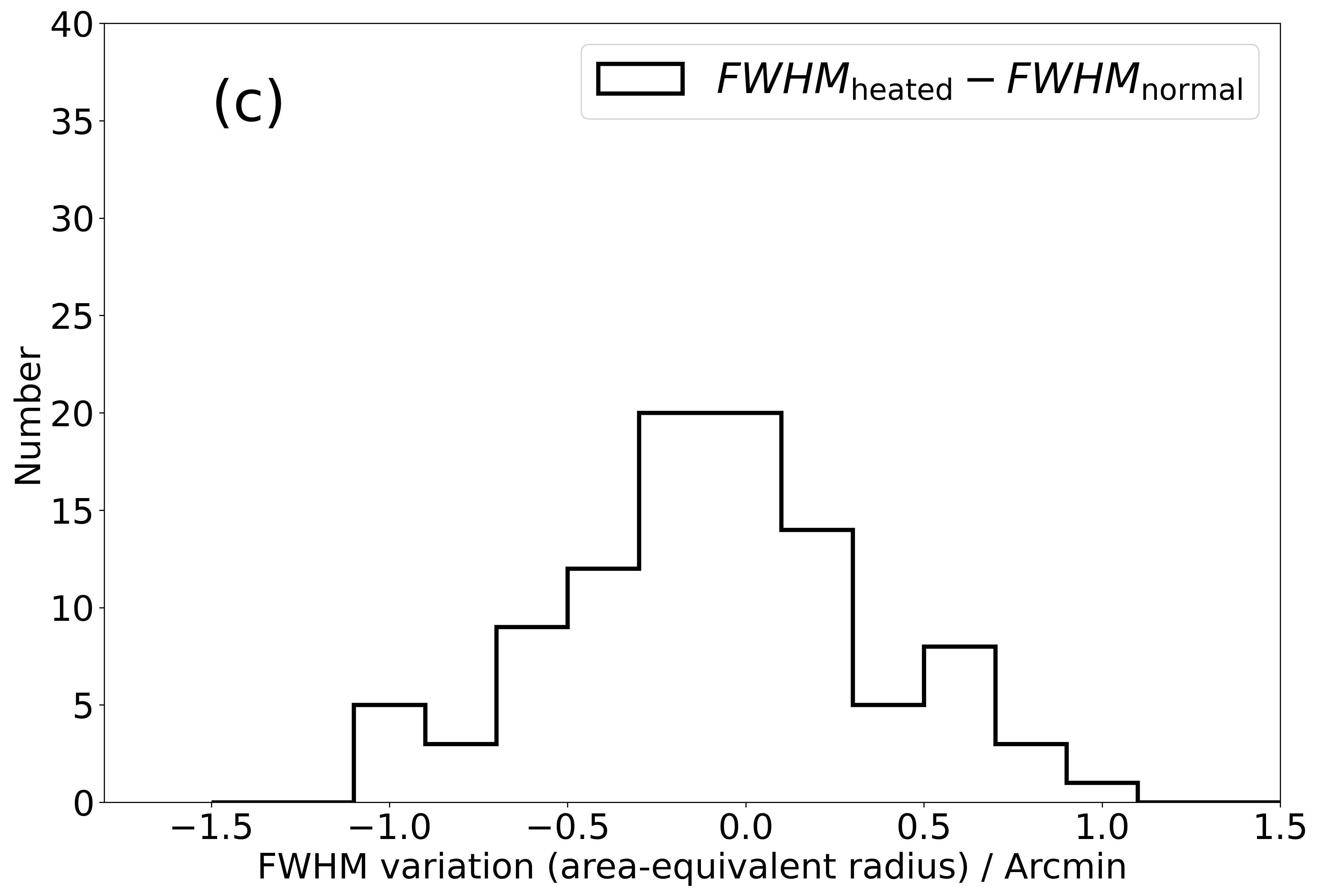}
    \includegraphics[width=0.48\textwidth]{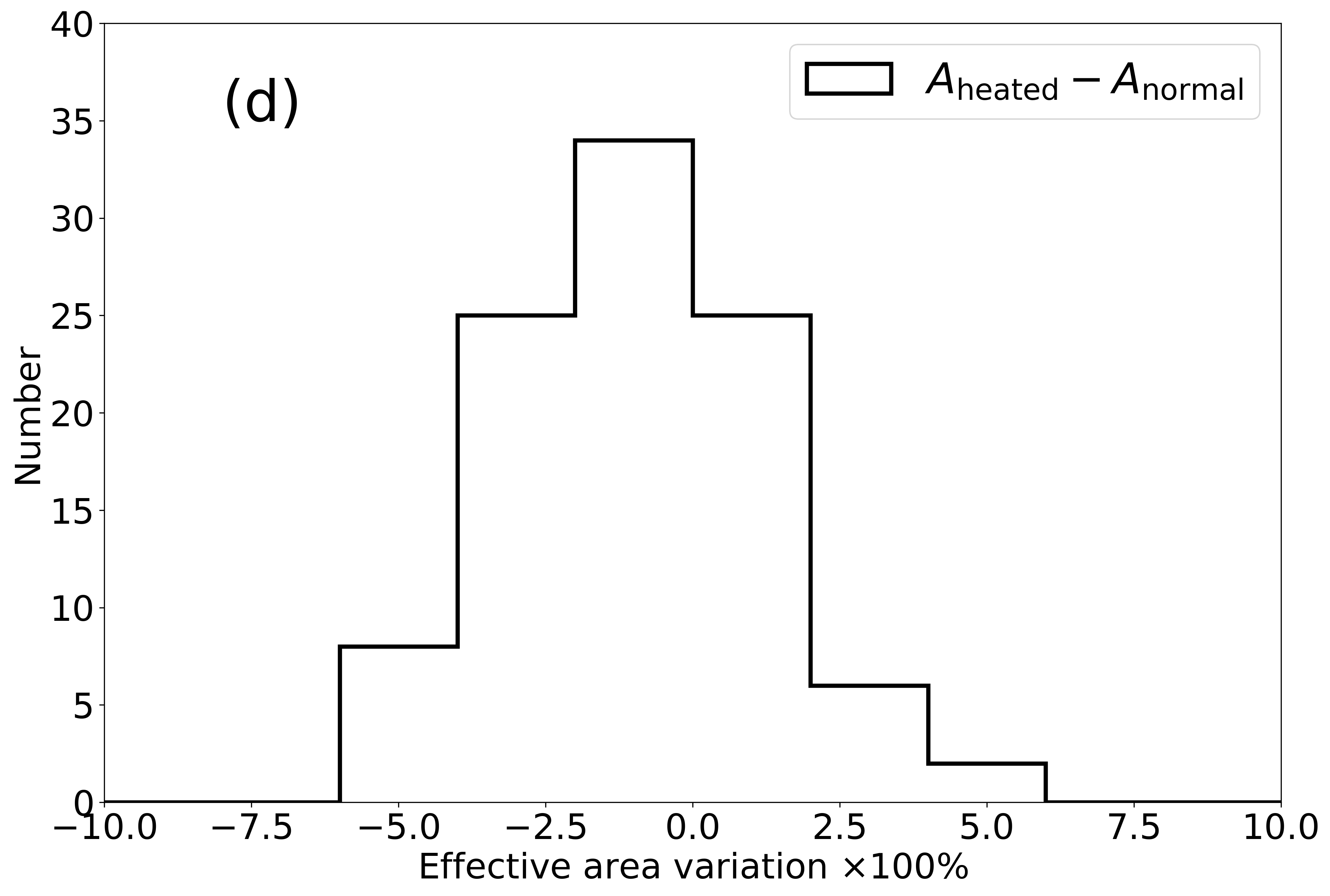}
    \caption{Panels (a), (b) and (c): distributions of the variations of all three measures of the FWHM (long axis, short axis and area-equivalent radius) at 1.25 keV, before and after the heating of the MA. Panel (d): distribution of the variation of the effective area at 1.25 kV, before and after the heating of the MA.}
    \label{fig:mpo_compare_thermaltest}
\end{figure*}

On the final day of the FM1 calibration experiment (2022 August 26th), we conducted a dedicated thermal test to characterize the temperature dependence of the instrument’s optical performance, aiming to quantify thermally induced variations in the PSF and effective area. 
While maintaining the detector at its nominal working temperature of $\sim-30\deg$C, we elevated the temperature of the MA from $\sim16\deg$C (nominal working temperature) to $\sim28\deg$C, and illuminate the instrument using Mg target.
For each of the FoV quadrant, a sub-grid of $5\times5$ points (`chip center' array + `chip edge' array) is utilized.
We then compare the measures of the PSF and effective area with those obtained under nominal working temperature, as shown in Fig. \ref{fig:mpo_compare_thermaltest}.
In Panels (a)-(c), we plot the histograms of the differences in the measured FWHM of the PSF focal spot before and after MA heating.
The variation in the imaging quality is found to be mild (typically within $1$ arcmin). 
The distribution of the variations in the effective area is shown in the Panel (d) of Fig. \ref{fig:mpo_compare_thermaltest}. 
The variations in the effective area fall within $-5\%$ to $5\%$, with a median of $-0.7\%$. 
These deviations are comparable to the measurement errors.
Our findings align well with previously obtained results by two separate qualification models, QM1 and QM2. 
For QM1, the thermal testing conducted at the University of Leicester as a collaborative work \cite{2020charly_wxt_LU} showed that the imaging quality remains quite stable despite a thermal gradient of $\sim9\deg$ across the optic (private communication, Feldman Charlotee).
Consistent results were obtained for QM2 during a similar calibration experiment performed also at 100XF. 
Specifically, when the MA was uniformly heated from $20\deg$C to $25\deg$C, no noticeable performance variations were observed.
In summary, the minimal observed variations in spatial resolution and effective area demonstrate that the optical performance remains stable when the MA is uniformly heated to $\sim28\deg$C,, confirming its robust thermal stability under operational conditions.

\section{Summary}
\label{sec:summary}

We report the results of the end-to-end calibration of three flight model modules (FM1, FM5 and FM11) for the Wide-field X-ray Telescope (WXT) instrument onboard the \textit{Einstein Probe}, performed at the IHEP 100-m X-ray Test Facility. Systematic measurements of PSF characteristics, effective area, and detector energy response were conducted across a wide range of the FoV using various characteristic emission lines. FM1 and FM5 modules were comprehensively calibrated using a $22\times22$ sampling grid across the FoV at the energy of Mg K$\alpha$ (1.25 keV), supplemented by $6\times6$ sub-grid measurements at O K$\alpha$ (0.53 keV), Si K$\alpha$ (1.74 keV), Ag L (2.98 keV), and Ti K$\alpha$ (4.51 keV). The properties at Cu L (930 eV) were also measured along the center of the each detector. FM11 characterization was limited to Mg K$\alpha$ due to the tight calibration schedule.

The PSFs exhibit largely uniform cruciform morphology across the FoV, consistent with lobster-eye optics predictions, underscoring both the superior imaging quality of the MPO plates and the precise mounting of the MA. The observed elliptical shape of the PSF and moderately broad distribution of the PSF FWHM originate from imperfections in micro-pore channel fabrication and assembling. 
Higher-energy photons produce more compact cruciform structures with attenuated higher-order components, and exhibit smaller values of FWHM, due essentially to the decreased critical angle of grazing incidence. There is an improvement in the imaging quality compared to \textit{LEIA} as evidenced by much smaller values of FWHM (in range of 3--7 arcminutes with a median of $\sim4.2$ arcmin).

The effective area is found to be energy-dependent, with a peak value of $\sim3~{\rm cm^2}$ at $\sim1$ keV.
The WXT optics is found to be almost free from vignetting, as verified by the roughly uniform distribution of the effective area within the FoV, which aligns well with the theoretical prediction of lobster eye optics. 
Meanwhile, due to the blocking of the incident X-ray photons by the mounting frame between the individual MPO plates, for some directions the effective areas are smaller than the nominal value of $\sim3~{\rm cm^2}$.
Both the energy dependence and spatial distribution pattern can be reproduced by Monte Carlo simulations.

The CMOS detectors demonstrate a reasonable spatial uniformity in both energy-independent gain coefficients (6.4-6.9 eV/DN) and spectral resolution for a given sensor. The measured spectral resolution is $\sim120$ eV at $0.53$ keV, $\sim130$ eV at $1.25$ keV (meeting design requirements of $170$ eV), $\sim140$ eV at $1.74$ keV, $\sim160$ eV at $2.98$ keV and $\sim180$ eV at $4.51$ keV for most detectors, except for CMOS 18 which exhibits a slightly poorer spectral resolution ($\Delta FWHM_{\rm E}\sim+10$ eV). The Fano factor is estimated to be $\sim0.2$, and the equivalent noise charge is in range of $12-13$ (except for a larger value of $\sim14.5$ for CMOS 18). The readout noises are found to be $\sim3-4~{\rm e^{-}}$ and no noticeable large-scale bad pixel clusters are identified.

As part of the FM1 calibration campaign, we evaluated the instrumental thermal stability by elevating the temperature of the mirror assembly by approximately 10 degrees. Quantitative analysis reveals no statistically significant variations in either the spatial resolution or the effective area ($\Delta FWHM_{\rm PSF}<1$ arcmin, $\Delta A_{\rm eff}< 5\%$), confirming the optical system's robustness against thermal perturbations.

The calibration of these three FM modules at the 100XF provides a crucial baseline for characterizing the instrumental performances and operational status prior to the launch of the \textit{Einstein Probe}.
The calibration results have been fully incorporated into the mission's calibration database and are currently implemented in the WXT science data processing pipeline.

\ 

\

\bmhead{Acknowledgments}
We thank all the members of the EP team and the EP consortium. The CAS team acknowledge contribution from ESA for calibration of the mirror assembly and tests of part of the devices. The Leicester and MPE teams acknowledge funding by ESA. 

\bmhead{Author’s Contribution}
H.-Q.C. and C.Z. led the paper writing. C.Z. led the calibration experiment design and data analysis. Z.-X.L., C.Z., W.-X.W., Y.-F.D, D.-H.Z., Z.-Q.J., X.-J.S., S.-L.S., Y.-F.C., Z.-W.C., W.F., Y.-X.H., J.-F.L., Z.-D.L., X.-H.M., Y.-L.X., A.-L.Y., Q.Z., L.-H.L., G.J. and W.-M.Y. contributed to the design and development the instrument. H.-Q.C., Z.-X.L., C.Z.,D.-H.Z., Y.-F.D., Z.-Q.J., X.-J.S., Y.-F.C., Z.-W.C., Y.-X.H., J.-F.L., Z.-D.L., X.-H.M., Y.-L.X. participated in the calibration experiment. H.-Q.C., Z.-X.L., C.Z., Y.L., H.-W.P., W.-X.W., D.-H.Z. contributed to the analysis of calibration data. Y.-S.W. is responsible for the construction of the test facility. X.-T.Y. and Z.-J.Z. led the experimental setup and data acquisition. All authors read and approved the final manuscript.

\bmhead{Funding}
This work is supported by the Einstein Probe project, a space mission supported by Strategic Priority Program on Space Science of Chinese Academy of Sciences, in collaboration with ESA, MPE and CNES. 
This work is supported by the National Natural Science Foundation of China (grant no. 12173055, 12173057, 12173056, 12203071). 

\bmhead{Availability of data and material}
The data supports the findings of the study are not openly available due to the intellectual property belong to the partners and are available from the corresponding author upon reasonable request.

\section*{Declarations}

\bmhead{Conflicts of interest}
The authors declare that they have no conflict of interest.

%%===========================================================================================%%
%% If you are submitting to one of the Nature Portfolio journals, using the eJP submission   %%
%% system, please include the references within the manuscript file itself. You may do this  %%
%% by copying the reference list from your .bbl file, paste it into the main manuscript .tex %%
%% file, and delete the associated \verb+\bibliography+ commands.                            %%
%%===========================================================================================%%

\bibliography{wxt_calib}% common bib file

\begin{thebibliography}{10}
\expandafter\ifx\csname url\endcsname\relax
  \def\url#1{\burl{#1}}\fi
\expandafter\ifx\csname urlprefix\endcsname\relax\def\urlprefix{URL }\fi
\providecommand{\bibinfo}[2]{#2}
\providecommand{\eprint}[2][]{\url{#2}}
\providecommand{\doi}[1]{\url{https://doi.org/#1}}
\bibcommenthead

\bibitem{1979ApJ...233..364A}
\bibinfo{author}{{Angel}, J.~R.~P.}
\newblock \bibinfo{title}{{Lobster eyes as X-ray telescopes.}}
\newblock \emph{\bibinfo{journal}{\apj}} \textbf{\bibinfo{volume}{233}},
  \bibinfo{pages}{364--373} (\bibinfo{year}{1979}).

\bibitem{1992SPIE.1546...41F}
\bibinfo{author}{{Fraser}, G.~W.}, \bibinfo{author}{{Lees}, J.~E.},
  \bibinfo{author}{{Pearson}, J.~F.}, \bibinfo{author}{{Sims}, M.~R.} \&
  \bibinfo{author}{{Roxburgh}, K.}
\newblock \bibinfo{editor}{{Hoover}, R.~B.} (ed.) \emph{\bibinfo{title}{{X-ray
  focusing using microchannel plates}}}.
\newblock (ed.\bibinfo{editor}{{Hoover}, R.~B.})
  \emph{\bibinfo{booktitle}{Multilayer and Grazing Incidence X-Ray/EUV
  Optics}}, Vol. \bibinfo{volume}{1546} of \emph{\bibinfo{series}{Society of
  Photo-Optical Instrumentation Engineers (SPIE) Conference Series}},
  \bibinfo{pages}{41--52} (\bibinfo{year}{1992}).

\bibitem{1998ExA.....8..281W}
\bibinfo{author}{{Willingale}, R.}, \bibinfo{author}{{Fraser}, G.~W.},
  \bibinfo{author}{{Brunton}, A.~N.} \& \bibinfo{author}{{Martin}, A.~P.}
\newblock \bibinfo{title}{{Hard X-ray imaging with microchannel plate optics}}.
\newblock \emph{\bibinfo{journal}{Experimental Astronomy}}
  \textbf{\bibinfo{volume}{8}}, \bibinfo{pages}{281--296}
  (\bibinfo{year}{1998}).

\bibitem{2002SPIE.4497..115F}
\bibinfo{author}{{Fraser}, G.~W.} \emph{et~al.}
\newblock \bibinfo{editor}{{Flanagan}, K.~A.} \& \bibinfo{editor}{{Siegmund},
  O. H.~W.} (eds) \emph{\bibinfo{title}{{LOBSTER-ISS: an imaging x-ray all-sky
  monitor for the International Space Station}}}.
\newblock (eds \bibinfo{editor}{{Flanagan}, K.~A.} \&
  \bibinfo{editor}{{Siegmund}, O. H.~W.}) \emph{\bibinfo{booktitle}{X-Ray and
  Gamma-Ray Instrumentation for Astronomy XII}}, Vol. \bibinfo{volume}{4497} of
  \emph{\bibinfo{series}{Society of Photo-Optical Instrumentation Engineers
  (SPIE) Conference Series}}, \bibinfo{pages}{115--126} (\bibinfo{year}{2002}).

\bibitem{2016SPIE.9905E..1YW}
\bibinfo{author}{{Willingale}, R.} \emph{et~al.}
\newblock \bibinfo{editor}{{den Herder}, J.-W.~A.},
  \bibinfo{editor}{{Takahashi}, T.} \& \bibinfo{editor}{{Bautz}, M.} (eds)
  \emph{\bibinfo{title}{{Aberrations in square pore micro-channel optics used
  for x-ray lobster eye telescopes}}}.
\newblock (eds \bibinfo{editor}{{den Herder}, J.-W.~A.},
  \bibinfo{editor}{{Takahashi}, T.} \& \bibinfo{editor}{{Bautz}, M.})
  \emph{\bibinfo{booktitle}{Space Telescopes and Instrumentation 2016:
  Ultraviolet to Gamma Ray}}, Vol. \bibinfo{volume}{9905} of
  \emph{\bibinfo{series}{Society of Photo-Optical Instrumentation Engineers
  (SPIE) Conference Series}}, \bibinfo{pages}{99051Y} (\bibinfo{year}{2016}).

\bibitem{Yuan2015}
\bibinfo{author}{{Yuan}, W.} \emph{et~al.}
\newblock \bibinfo{title}{{Einstein Probe - a small mission to monitor and
  explore the dynamic X-ray Universe}}.
\newblock \emph{\bibinfo{journal}{arXiv e-prints}}
  \bibinfo{pages}{arXiv:1506.07735} (\bibinfo{year}{2015}).

\bibitem{Yuan2018}
\bibinfo{author}{{Yuan}, W.} \emph{et~al.}
\newblock \bibinfo{editor}{{den Herder}, J.-W.~A.}, \bibinfo{editor}{{Nikzad},
  S.} \& \bibinfo{editor}{{Nakazawa}, K.} (eds) \emph{\bibinfo{title}{{Einstein
  Probe: a lobster-eye telescope for monitoring the x-ray sky}}}.
\newblock (eds \bibinfo{editor}{{den Herder}, J.-W.~A.},
  \bibinfo{editor}{{Nikzad}, S.} \& \bibinfo{editor}{{Nakazawa}, K.})
  \emph{\bibinfo{booktitle}{Space Telescopes and Instrumentation 2018:
  Ultraviolet to Gamma Ray}}, Vol. \bibinfo{volume}{10699} of
  \emph{\bibinfo{series}{Society of Photo-Optical Instrumentation Engineers
  (SPIE) Conference Series}}, \bibinfo{pages}{1069925} (\bibinfo{year}{2018}).

\bibitem{Yuan2022}
\bibinfo{author}{Yuan, W.}, \bibinfo{author}{Zhang, C.}, \bibinfo{author}{Chen,
  Y.} \& \bibinfo{author}{Ling, Z.}
\newblock \emph{\bibinfo{title}{The Einstein Probe Mission}},
  \bibinfo{pages}{1--30} (\bibinfo{publisher}{Springer Nature Singapore},
  \bibinfo{address}{Singapore}, \bibinfo{year}{2022}).
\newblock \urlprefix\url{https://doi.org/10.1007/978-981-16-4544-0_151-1}.

\bibitem{Yuan2025}
\bibinfo{author}{Yuan, W.} \emph{et~al.}
\newblock \bibinfo{title}{Science objectives of the einstein probe mission}.
\newblock \emph{\bibinfo{journal}{Science China Physics, Mechanics, and
  Astronomy}} \textbf{\bibinfo{volume}{68}} (\bibinfo{year}{2025}).
\newblock \urlprefix\url{http://dx.doi.org/10.1007/s11433-024-2600-3}.

\bibitem{Yin2024}
\bibinfo{author}{{Yin}, Y.-H.~I.} \emph{et~al.}
\newblock \bibinfo{title}{{Triggering the Untriggered: The First Einstein
  Probe-detected Gamma-Ray Burst 240219A and Its Implications}}.
\newblock \emph{\bibinfo{journal}{\apjl}} \textbf{\bibinfo{volume}{975}},
  \bibinfo{pages}{L27} (\bibinfo{year}{2024}).

\bibitem{LiuY2025}
\bibinfo{author}{{Liu}, Y.} \emph{et~al.}
\newblock \bibinfo{title}{{Soft X-ray prompt emission from the high-redshift
  gamma-ray burst EP240315a}}.
\newblock \emph{\bibinfo{journal}{Nature Astronomy}}
  \textbf{\bibinfo{volume}{9}}, \bibinfo{pages}{564--576}
  (\bibinfo{year}{2025}).

\bibitem{2024HSun}
\bibinfo{author}{{Sun}, H.} \emph{et~al.}
\newblock \bibinfo{title}{{Extragalactic fast X-ray transient from a weak
  relativistic jet associated with a Type Ic-BL supernova}}.
\newblock \emph{\bibinfo{journal}{arXiv e-prints}}
  \bibinfo{pages}{arXiv:2410.02315} (\bibinfo{year}{2024}).

\bibitem{Jiang2025}
\bibinfo{author}{{Jiang}, S.-Q.} \emph{et~al.}
\newblock \bibinfo{title}{{EP240801a/XRF 240801B: An X-ray Flash Detected by
  the Einstein Probe and Implications of its Multiband Afterglow}}.
\newblock \emph{\bibinfo{journal}{arXiv e-prints}}
  \bibinfo{pages}{arXiv:2503.04306} (\bibinfo{year}{2025}).

\bibitem{2025Jin}
\bibinfo{author}{{Jin}, C.~C.} \emph{et~al.}
\newblock \bibinfo{title}{{An Intermediate-mass Black Hole Lurking in A
  Galactic Halo Caught Alive during Outburst}}.
\newblock \emph{\bibinfo{journal}{arXiv e-prints}}
  \bibinfo{pages}{arXiv:2501.09580} (\bibinfo{year}{2025}).

\bibitem{Marino2025}
\bibinfo{author}{{Marino}, A.} \emph{et~al.}
\newblock \bibinfo{title}{{Einstein Probe Discovery of EP J005245.1‑722843: A
  Rare Be{\textendash}White Dwarf Binary in the Small Magellanic Cloud?}}
\newblock \emph{\bibinfo{journal}{\apjl}} \textbf{\bibinfo{volume}{980}},
  \bibinfo{pages}{L36} (\bibinfo{year}{2025}).

\bibitem{2025YangBeXRB}
\bibinfo{author}{{Yang}, H.~N.} \emph{et~al.}
\newblock \bibinfo{title}{{Broad-band study of the Be X-ray binary RX
  J0520.5-6932 during its outburst in 2024}}.
\newblock \emph{\bibinfo{journal}{\mnras}} \textbf{\bibinfo{volume}{536}},
  \bibinfo{pages}{1357--1373} (\bibinfo{year}{2025}).

\bibitem{LiuMJ_AGN}
\bibinfo{author}{{Liu}, M.} \emph{et~al.}
\newblock \bibinfo{title}{{Detection of an Orphan X-Ray Flare from a Blazar
  Candidate EP240709a with the Einstein Probe}}.
\newblock \emph{\bibinfo{journal}{\apj}} \textbf{\bibinfo{volume}{984}},
  \bibinfo{pages}{5} (\bibinfo{year}{2025}).

\bibitem{ZhangWD2025}
\bibinfo{author}{{Zhang}, W.} \emph{et~al.}
\newblock \bibinfo{title}{{Einstein Probe discovery of EP240408a: A peculiar
  X-ray transient with an intermediate timescale}}.
\newblock \emph{\bibinfo{journal}{Science China Physics, Mechanics, and
  Astronomy}} \textbf{\bibinfo{volume}{68}}, \bibinfo{pages}{219511}
  (\bibinfo{year}{2025}).

\bibitem{2022WangWXa}
\bibinfo{author}{{Wang}, W.} \emph{et~al.}
\newblock \bibinfo{editor}{{Holland}, A.~D.} \& \bibinfo{editor}{{Beletic}, J.}
  (eds) \emph{\bibinfo{title}{{Design and test results of scientific x-ray CMOS
  cameras}}}.
\newblock (eds \bibinfo{editor}{{Holland}, A.~D.} \&
  \bibinfo{editor}{{Beletic}, J.}) \emph{\bibinfo{booktitle}{X-Ray, Optical,
  and Infrared Detectors for Astronomy X}}, Vol. \bibinfo{volume}{12191} of
  \emph{\bibinfo{series}{Society of Photo-Optical Instrumentation Engineers
  (SPIE) Conference Series}}, \bibinfo{pages}{121910L} (\bibinfo{year}{2022}).
\newblock \eprint{2209.13163}.

\bibitem{2022WangWXb}
\bibinfo{author}{{Wang}, W.~X.}, \bibinfo{author}{{Ling}, Z.~X.},
  \bibinfo{author}{{Zhang}, C.}, \bibinfo{author}{{Yuan}, W.~M.} \&
  \bibinfo{author}{{Zhang}, S.~N.}
\newblock \bibinfo{title}{{Design and test results of different aluminum
  coating layers on the sCMOS sensors for soft X-ray detection}}.
\newblock \emph{\bibinfo{journal}{Journal of Instrumentation}}
  \textbf{\bibinfo{volume}{17}}, \bibinfo{pages}{P12016}
  (\bibinfo{year}{2022}).

\bibitem{Wu2022}
\bibinfo{author}{{Wu}, Q.} \emph{et~al.}
\newblock \bibinfo{title}{{X-Ray Performance of a Customized Large-format
  Scientific CMOS Detector}}.
\newblock \emph{\bibinfo{journal}{\pasp}} \textbf{\bibinfo{volume}{134}},
  \bibinfo{pages}{035006} (\bibinfo{year}{2022}).

\bibitem{Wu2023a}
\bibinfo{author}{Wu, Q.}, \bibinfo{author}{Ling, Z.}, \bibinfo{author}{Zhang,
  C.}, \bibinfo{author}{Zhang, S.-N.} \& \bibinfo{author}{Yuan, W.}
\newblock \bibinfo{title}{An aluminum-coated scmos sensor for x-ray astronomy}.
\newblock \emph{\bibinfo{journal}{Publications of the Astronomical Society of
  the Pacific}} \textbf{\bibinfo{volume}{135}}, \bibinfo{pages}{115002}
  (\bibinfo{year}{2023}).
\newblock \urlprefix\url{https://dx.doi.org/10.1088/1538-3873/ad03d7}.

\bibitem{Wu2023b}
\bibinfo{author}{{Wu}, Q.} \emph{et~al.}
\newblock \bibinfo{title}{{Investigating the image lag of a scientific CMOS
  sensor in X-ray detection}}.
\newblock \emph{\bibinfo{journal}{Nuclear Instruments and Methods in Physics
  Research A}} \textbf{\bibinfo{volume}{1050}}, \bibinfo{pages}{168180}
  (\bibinfo{year}{2023}).

\bibitem{LiuMJ2023}
\bibinfo{author}{{Liu}, M.} \emph{et~al.}
\newblock \bibinfo{title}{{Radiation effects on scientific CMOS sensors for
  X-ray astronomy: I. Proton irradiation}}.
\newblock \emph{\bibinfo{journal}{Journal of Astronomical Telescopes,
  Instruments, and Systems}} \textbf{\bibinfo{volume}{9}},
  \bibinfo{pages}{046003} (\bibinfo{year}{2023}).

\bibitem{ChenMX2024}
\bibinfo{author}{Chen, M.} \emph{et~al.}
\newblock \bibinfo{title}{{Radiation effects on scientific complementary
  metal-oxide-semiconductor detectors for x-ray astronomy: II. Total ionizing
  dose irradiation}}.
\newblock \emph{\bibinfo{journal}{Journal of Astronomical Telescopes,
  Instruments, and Systems}} \textbf{\bibinfo{volume}{10}},
  \bibinfo{pages}{026001} (\bibinfo{year}{2024}).
\newblock \urlprefix\url{https://doi.org/10.1117/1.JATIS.10.2.026001}.

\bibitem{LiuMJ2025}
\bibinfo{author}{{Liu}, M.} \emph{et~al.}
\newblock \bibinfo{title}{{Long-term stability of scientific X-ray CMOS
  detectors}}.
\newblock \emph{\bibinfo{journal}{Nuclear Instruments and Methods in Physics
  Research A}} \textbf{\bibinfo{volume}{1072}}, \bibinfo{pages}{170146}
  (\bibinfo{year}{2025}).

\bibitem{2022ChenYifan}
\bibinfo{author}{{Chen}, Y.} \emph{et~al.}
\newblock \bibinfo{title}{{Detection system of the lobster eye telescope with
  large field of view}}.
\newblock \emph{\bibinfo{journal}{Applied Optics}}
  \textbf{\bibinfo{volume}{61}}, \bibinfo{pages}{8813} (\bibinfo{year}{2022}).

\bibitem{2012SPIE.8443E..3XZ}
\bibinfo{author}{{Zhang}, C.}, \bibinfo{author}{{Ling}, Z.} \&
  \bibinfo{author}{{Zhang}, S.-N.}
\newblock \bibinfo{editor}{{Takahashi}, T.}, \bibinfo{editor}{{Murray}, S.~S.}
  \& \bibinfo{editor}{{den Herder}, J.-W.~A.} (eds)
  \emph{\bibinfo{title}{{Development of the super-high angular resolution
  principle for x-ray imaging: experimental demonstrations}}}.
\newblock (eds \bibinfo{editor}{{Takahashi}, T.}, \bibinfo{editor}{{Murray},
  S.~S.} \& \bibinfo{editor}{{den Herder}, J.-W.~A.})
  \emph{\bibinfo{booktitle}{Space Telescopes and Instrumentation 2012:
  Ultraviolet to Gamma Ray}}, Vol. \bibinfo{volume}{8443} of
  \emph{\bibinfo{series}{Society of Photo-Optical Instrumentation Engineers
  (SPIE) Conference Series}}, \bibinfo{pages}{84433X} (\bibinfo{year}{2012}).

\bibitem{2005ExA....20..405F}
\bibinfo{author}{{Freyberg}, M.~J.} \emph{et~al.}
\newblock \bibinfo{title}{{The MPE X-ray test facility PANTER: Calibration of
  hard X-ray (15-50 kev) optics}}.
\newblock \emph{\bibinfo{journal}{Experimental Astronomy}}
  \textbf{\bibinfo{volume}{20}}, \bibinfo{pages}{405--412}
  (\bibinfo{year}{2005}).

\bibitem{2019SPIE11119E..16B}
\bibinfo{author}{{Bradshaw}, M.} \emph{et~al.}
\newblock \bibinfo{editor}{{O'Dell}, S.~L.} \& \bibinfo{editor}{{Pareschi}, G.}
  (eds) \emph{\bibinfo{title}{{Developments in testing x-ray optics at MPE's
  PANTER facility}}}.
\newblock (eds \bibinfo{editor}{{O'Dell}, S.~L.} \&
  \bibinfo{editor}{{Pareschi}, G.}) \emph{\bibinfo{booktitle}{Optics for EUV,
  X-Ray, and Gamma-Ray Astronomy IX}}, Vol. \bibinfo{volume}{11119} of
  \emph{\bibinfo{series}{Society of Photo-Optical Instrumentation Engineers
  (SPIE) Conference Series}}, \bibinfo{pages}{1111916} (\bibinfo{year}{2019}).

\bibitem{EPWXT_Panter2023}
\bibinfo{author}{{Rukdee}, S.} \emph{et~al.}
\newblock \bibinfo{editor}{{O'Dell}, S.~L.}, \bibinfo{editor}{{Gaskin}, J.~A.},
  \bibinfo{editor}{{Pareschi}, G.} \& \bibinfo{editor}{{Spiga}, D.} (eds)
  \emph{\bibinfo{title}{{Einstein probe wide field telescope flight mirror
  module characterization at PANTER test facility}}}.
\newblock (eds \bibinfo{editor}{{O'Dell}, S.~L.}, \bibinfo{editor}{{Gaskin},
  J.~A.}, \bibinfo{editor}{{Pareschi}, G.} \& \bibinfo{editor}{{Spiga}, D.})
  \emph{\bibinfo{booktitle}{Optics for EUV, X-Ray, and Gamma-Ray Astronomy
  XI}}, Vol. \bibinfo{volume}{12679} of \emph{\bibinfo{series}{Society of
  Photo-Optical Instrumentation Engineers (SPIE) Conference Series}},
  \bibinfo{pages}{1267910} (\bibinfo{year}{2023}).

\bibitem{2023ExA....55..427W}
\bibinfo{author}{{Wang}, Y.} \emph{et~al.}
\newblock \bibinfo{title}{{The 100-m X-ray test facility at IHEP}}.
\newblock \emph{\bibinfo{journal}{Experimental Astronomy}}
  \textbf{\bibinfo{volume}{55}}, \bibinfo{pages}{427--445}
  (\bibinfo{year}{2023}).

\bibitem{2023LingZXRAA}
\bibinfo{author}{{Ling}, Z.~X.} \emph{et~al.}
\newblock \bibinfo{title}{{The Lobster Eye Imager for Astronomy Onboard the
  SATech-01 Satellite}}.
\newblock \emph{\bibinfo{journal}{Research in Astronomy and Astrophysics}}
  \textbf{\bibinfo{volume}{23}}, \bibinfo{pages}{095007}
  (\bibinfo{year}{2023}).

\bibitem{2022ZhangChenApJL}
\bibinfo{author}{{Zhang}, C.} \emph{et~al.}
\newblock \bibinfo{title}{{First Wide Field-of-view X-Ray Observations by a
  Lobster-eye Focusing Telescope in Orbit}}.
\newblock \emph{\bibinfo{journal}{\apjl}} \textbf{\bibinfo{volume}{941}},
  \bibinfo{pages}{L2} (\bibinfo{year}{2022}).

\bibitem{Cheng2024}
\bibinfo{author}{{Cheng}, H.} \emph{et~al.}
\newblock \bibinfo{title}{{Ground calibration result of the Lobster Eye Imager
  for Astronomy}}.
\newblock \emph{\bibinfo{journal}{Experimental Astronomy}}
  \textbf{\bibinfo{volume}{57}}, \bibinfo{pages}{10} (\bibinfo{year}{2024}).

\bibitem{2014SPIE.9144E..4EZ}
\bibinfo{author}{{Zhao}, D.} \emph{et~al.}
\newblock \bibinfo{editor}{{Takahashi}, T.}, \bibinfo{editor}{{den Herder},
  J.-W.~A.} \& \bibinfo{editor}{{Bautz}, M.} (eds) \emph{\bibinfo{title}{{Ray
  tracing simulations for the wide-field x-ray telescope of the Einstein Probe
  mission based on Geant4 and XRTG4}}}.
\newblock (eds \bibinfo{editor}{{Takahashi}, T.}, \bibinfo{editor}{{den
  Herder}, J.-W.~A.} \& \bibinfo{editor}{{Bautz}, M.})
  \emph{\bibinfo{booktitle}{Space Telescopes and Instrumentation 2014:
  Ultraviolet to Gamma Ray}}, Vol. \bibinfo{volume}{9144} of
  \emph{\bibinfo{series}{Society of Photo-Optical Instrumentation Engineers
  (SPIE) Conference Series}}, \bibinfo{pages}{91444E} (\bibinfo{year}{2014}).

\bibitem{2021OptCo.48326656L}
\bibinfo{author}{{Li}, L.} \emph{et~al.}
\newblock \bibinfo{title}{{Study on the optical properties of Angel Lobster eye
  X-ray flat micro pore optical device}}.
\newblock \emph{\bibinfo{journal}{Optics Communications}}
  \textbf{\bibinfo{volume}{483}}, \bibinfo{pages}{126656}
  (\bibinfo{year}{2021}).

\bibitem{2022PASP..134k5002L}
\bibinfo{author}{{Li}, L.} \emph{et~al.}
\newblock \bibinfo{title}{{Fabrication and Performance of Lobster Eye X-Ray
  Micro Pore Optics with the Ultra-high Aspect Ratio}}.
\newblock \emph{\bibinfo{journal}{\pasp}} \textbf{\bibinfo{volume}{134}},
  \bibinfo{pages}{115002} (\bibinfo{year}{2022}).

\bibitem{2017ZhaoDH_simulation}
\bibinfo{author}{{Zhao}, D.} \emph{et~al.}
\newblock \bibinfo{title}{{Geant4 simulations of a wide-angle x-ray focusing
  telescope}}.
\newblock \emph{\bibinfo{journal}{Experimental Astronomy}}
  \textbf{\bibinfo{volume}{43}}, \bibinfo{pages}{267--283}
  (\bibinfo{year}{2017}).

\bibitem{Holland2013}
\bibinfo{author}{Holland, A.}
\newblock \emph{\bibinfo{title}{X-ray CCDs}}, \bibinfo{pages}{443--453}
  (\bibinfo{publisher}{Springer New York}, \bibinfo{address}{New York, NY},
  \bibinfo{year}{2013}).
\newblock \urlprefix\url{https://doi.org/10.1007/978-1-4614-7804-1_24}.

\bibitem{Fano1947}
\bibinfo{author}{Fano, U.}
\newblock \bibinfo{title}{Ionization yield of radiations. ii. the fluctuations
  of the number of ions}.
\newblock \emph{\bibinfo{journal}{Phys. Rev.}} \textbf{\bibinfo{volume}{72}},
  \bibinfo{pages}{26--29} (\bibinfo{year}{1947}).
\newblock \urlprefix\url{https://link.aps.org/doi/10.1103/PhysRev.72.26}.

\bibitem{2020charly_wxt_LU}
\bibinfo{author}{Feldman, C.} \emph{et~al.}
\newblock \bibinfo{editor}{den Herder, J.-W.~A.}, \bibinfo{editor}{Nikzad, S.}
  \& \bibinfo{editor}{Nakazawa, K.} (eds) \emph{\bibinfo{title}{{Testing of the
  WXT optics at the University of Leicester}}}.
\newblock (eds \bibinfo{editor}{den Herder, J.-W.~A.}, \bibinfo{editor}{Nikzad,
  S.} \& \bibinfo{editor}{Nakazawa, K.}) \emph{\bibinfo{booktitle}{Space
  Telescopes and Instrumentation 2020: Ultraviolet to Gamma Ray}}, Vol.
  \bibinfo{volume}{11444}, \bibinfo{pages}{114447R}.
  \bibinfo{organization}{International Society for Optics and Photonics}
  (\bibinfo{publisher}{SPIE}, \bibinfo{year}{2020}).
\newblock \urlprefix\url{https://doi.org/10.1117/12.2562194}.

\end{thebibliography}
%% if required, the content of .bbl file can be included here once bbl is generated
%%\input sn-article.bbl

\end{document}